\documentclass[superscriptaddress,dvipsnames,nofootinbib,amsmath,amssymb,prd,notitlepage]{revtex4-1}

\usepackage[english]{babel}

\usepackage[letterpaper,top=2cm,bottom=2cm,left=3cm,right=3cm,marginparwidth=1.75cm]{geometry}

\usepackage[normalem]{ulem}

\usepackage{amsmath,amssymb}
\usepackage{graphicx}

\usepackage{ragged2e}

\usepackage[colorlinks=true, allcolors=blue]{hyperref}
\usepackage{comment}

\newcommand{\be}{\begin{equation}}
\newcommand{\ee}{\end{equation}}
\newcommand{\bea}{\begin{eqnarray}}
\newcommand{\eea}{\end{eqnarray}}
\newcommand{\dv}{\ensuremath{\mathrm{d}}}
\newcommand{\LCDM}{\ensuremath{\Lambda}CDM}
\newcommand{\Scal}{\ensuremath{\mathcal{S}}}
\newcommand{\Lm}{\ensuremath{\mathcal L_m}}
\newcommand{\al}{\ensuremath{\alpha}}
\newcommand{\bt}{\ensuremath{\beta}}
\newcommand{\gm}{\ensuremath{\gamma}}
\newcommand{\ka}{\ensuremath{\kappa}}
\newcommand{\Hc}{\ensuremath{\mathcal{H}}}
\newcommand{\Om}{\ensuremath{\Omega}}
\newcommand{\OmQ}{\ensuremath{{\Omega_{Q}}_0}}
\newcommand{\Omm}{\ensuremath{{\Omega_{m}}_0}}
\newcommand{\Omb}{\ensuremath{{\Omega_{b}}_0}}
\newcommand{\Omc}{\ensuremath{{\Omega_{c}}_0}}
\newcommand{\Omr}{\ensuremath{{\Omega_{r}}_0}}

\newcommand{\Omde}{\ensuremath{{\Omega_{de}}_0}}
\newcommand{\OmDE}{\ensuremath{{\Omega_{DE}}_0}}

\newcommand{\rhoi}{\ensuremath{{\rho_{i}}_0}}
\newcommand{\OmL}{\ensuremath{{\Omega_{\Lambda}}_0}}
\newcommand{\weff}{\ensuremath{w_{\text{eff}}}}

\def\l{\left}
\def\r{\right}

\newcommand{\lsim}{\raisebox{-0.13cm}{~\shortstack{$<$ \\[-0.07cm]
      $\sim$}}~}

\begin{document}

\title{Cosmological study of a symmetric teleparallel gravity model}
\author{Tiago B. Gon\c{c}alves}
\email{tbgoncalves@ciencias.ulisboa.pt}
 \author{Lu\'is Atayde}
 \email{luisbbatayde@gmail.com}
\affiliation{Instituto de Astrof\'{i}sica e Ci\^{e}ncias do Espa\c{c}o, Faculdade de Ci\^{e}ncias da Universidade de Lisboa, Edif\'{i}cio C8, Campo Grande, P-1749-016 Lisbon, Portugal}
 \affiliation{Departamento de F\'{i}sica, Faculdade de Ci\^{e}ncias da Universidade de Lisboa, Edif\'{i}cio C8, Campo Grande, P-1749-016 Lisbon, Portugal}

\author{Noemi Frusciante}
\email{noemi.frusciante@unina.it}
\affiliation{Dipartimento di Fisica ``E. Pancini", Universit\`a degli Studi  di Napoli  ``Federico II", Compl. Univ. di Monte S. Angelo, Edificio G, Via Cinthia, I-80126, Napoli, Italy}


\begin{abstract}
We study a Symmetric Teleparallel Gravity with a Lagrangian of Logarithmic form. The full model leads to an accelerated Universe and for specific values of the free parameters the Hubble rate reduces to the well known DGP model, though the evolution of the gravitational potentials are different. We consider different branches of the Logarithmic model, among which sDGP and nDGP. The phenomenology of both the background and linear perturbations is discussed, including all the relevant effects on Cosmic Microwave Background radiation (CMB) angular power spectrum, lensing and matter power spectra.  To this purpose, we modified  the Einstein--Boltzmann code \texttt{MGCAMB}. 
Finally, we derive bounds on the free parameters which are in agreement with early dark energy constraint from CMB and Big Bang Nucleosynthesis  constraint on the helium abundance.
\end{abstract}
\date{\today}
\maketitle
\tableofcontents
\section{Introduction}

In the  geometrical interpretation of gravity, the gravitational potential is described by  the metric, $g_{\mu\nu}$, and the gravitational field  by the affine connection, $\Gamma^\al_{\phantom{x}\mu\nu}$. In General Relativity (GR),  the affine connection  is torsion--free (${\Gamma^\al_{\phantom{x}\mu\nu} = \Gamma^\al_{\phantom{x}\nu\mu}}$) and  metric--compatible (${\nabla_\al g_{\mu\nu} = 0}$). 
In this case,  the metric and the connection are not independent, and gravity is described by the Riemann curvature tensor $R^\al_{\phantom{x}\bt\mu\nu}$.
One can give up these features \cite{BeltranJimenez:2017tkd, BeltranJimenez:2018vdo} and   introduce the \emph{torsion}  tensor, $T^\al_{\phantom{x}\mu\nu} = \Gamma^\al_{\phantom{x}\mu\nu} - \Gamma^\al_{\phantom{x}\nu\mu}$, and the \emph{non-metricity} tensor, $Q_{\al\mu\nu} = \nabla_a g_{\mu\nu}$.
A theory  with no curvature, metric--compatible, and which has only torsion can result in the same field equations as GR---this is the so--called Teleparallel Equivalent of GR (TEGR) (see \cite{Maluf:2013gaa} for a review).
Similarly, considering a metric incompatible theory  with vanishing torsion and curvature  also results in the same field equations as GR---the Symmetric Teleparallel Equivalent of GR (STEGR) \cite{Nester:1998mp, BeltranJimenez:2017tkd, BeltranJimenez:2019tme}. 
The actions in GR, TEGR and STEGR  linearly depend on the scalars constructed from these tensors, respectively, the Ricci scalar ($R$), the torsion scalar ($T$) and the non-metricity scalar ($Q$). 

Observational evidences show a Universe that currently is in an accelerated expansion phase. This phenomenon can be explained within GR, and equivalents, with the inclusion of a cosmological constant ($\Lambda$), resulting in the $\Lambda$CDM model.
However, theoretical long--lasting issues \cite{Weinberg:1988cp,Carroll:2000fy,Velten:2014nra,Joyce:2014kja}  and recent observational tensions \cite{Riess:2019cxk,Wong:2019kwg,Freedman:2019jwv,DiValentino:2020zio,Kuijken:2015vca,deJong:2015wca,Hildebrandt:2016iqg,DiValentino:2020vvd} are putting into question the standard cosmological model, and motivating theories that modify the gravitational interaction \cite{Clifton:2011jh, Capozziello:2011et, Avelino:2016lpj,CANTATA:2021ktz}. 
Generalisation of GR, by taking a Lagrangian dependent on a general function of curvature scalar, $f(R)$, has been widely studied~\cite{ Sotiriou:2008rp, DeFelice:2010aj}. By analogy, there has been exploration of modified gravity (MG) theories of the form $f(T)$ \cite{Ferraro:2006jd, Ferraro:2008ey, Bengochea:2008gz} (see \cite{Cai:2015emx} for a review) and of the form $f(Q)$ \cite{BeltranJimenez:2017tkd, Lazkoz:2019sjl, Dialektopoulos:2019mtr, Lu:2019hra, Barros:2020bgg, Esposito:2021ect, Frusciante:2021sio, Anagnostopoulos:2021ydo, Atayde:2021pgb, Albuquerque:2022eac, Ferreira:2022jcd, Anagnostopoulos:2022gej, Khyllep:2022spx, Atayde:2023aoj, Ferreira:2023awf, Shabani:2023xfn, De:2023xua}. 

In this paper we focus on a specific model of $f(Q)$. We select a  Logarithmic form of the $f(Q)$ function introduced in \cite{Ayuso:2021vtj}, which  resembles the background of DGP (Dvali--Gabadadze--Porrati) braneworld cosmologies \cite{Dvali:2000hr}. 
The DGP theory considers the 4D spacetime as a brane embedded in a 5D Minkowski space, in which the extra dimension is of infinite size.  
Interestingly, the 4D spacetime in the DGP model has two cosmological branches~\cite{Deffayet:2000uy}: one branch can present self-accelerating behaviour (sDGP),  where the late--time acceleration is sourced by the graviton itself and does not resort to $\Lambda$; and the other is the so--called normal branch (nDGP), in which case to realize the late--time expansion the presence of a cosmological constant or a dark energy (DE) component is required. The sDGP branch, while being more attractive as an alternative to $\Lambda$, suffers from a ghost-like instability \cite{Luty:2003vm,Nicolis:2004qq,Charmousis:2006pn} and is incompatible with observations \cite{Fairbairn:2005ue,Maartens:2006yt,Fang:2008kc}. 
A first study \cite{Ayuso:2021vtj} of this Logarithmic $f(Q)$ model  focused  at the level of the cosmological background  behaviour  where a non-zero cosmological constant was allowed in both the nDGP and sDGP branches. Their results show that in the sDGP branch, the $f(Q)$  modifications  work in a way of reducing the effect of the cosmological constant. While observational constraints at background level, using Type Ia Supernovae (SNIa), cosmic clocks, Cosmic Microwave Background (CMB) shift and Baryon Acoustic Oscillations (BAO) data, in either branch, did not indicate a significant statistical preference as compared to \LCDM. 

In this work we  perform a thorough analysis of the cosmology of the logarithmic form in \cite{Ayuso:2021vtj}: firstly, we propose a different treatment of the background, by studying the general case as well as  the DGP-like behaviours, and we provide early dark energy (EDE) constraint from CMB and Big Bang Nucleosynthesis (BBN) constraint on the helium abundance; secondly, we move forward by providing theoretical predictions for cosmological observables  on the CMB temperature, lensing and matter power spectra. 
 The CMB, whose anisotropies mainly originate at the time of the last--scattering, is a good probe of large scales and early--time cosmology. 
And so, it is also a tool to study and constrain MG~\cite{Planck:2015bue} because MG can affect the expansion history, changing the distance to the last--scattering surface (shifting the angular power spectrum peaks) \cite{Hu:1996vq}; can change the evolution of gravitational potentials at early and late times, affecting the Integrated Sachs--Wolfe (ISW) effect \cite{Sachs:1967er, Kofman:1985fp}; can change the gravitational lensing \cite{Acquaviva:2005xz, Carbone:2013dna} and the growth of structures \cite{Peebles:1984ge, 1993MNRAS.262..717B}; among other effects. 
To this purpose, we modified the public Einstein-Boltzmann solver \texttt{MGCAMB}~\cite{Zhao:2008bn, Hojjati:2011ix, Zucca:2019xhg, Wang:2023tjj} to implement the Logarithmic $f(Q)$ model, which we aim to publicly release in the near future. 

Let us note that while this paper was in preparation, it has been raised the question that the  $f(Q)$ theories might be pathological \cite{Gomes:2023tur,Heisenberg:2023wgk}: either  they are  strongly coupled or one of the seven propagating  modes is a ghost. Our model should fall in the first case. In this respect we stress that strong coupling problems have been identified also in other alternative theories  for which subsequent works shed light on the actual viability of the questioned theories \cite{Deffayet:2005ys,Mukohyama:2010xz,Gabadadze:2019lld,Hu:2023juh}.  To this extent we provide a phenomenological analysis showing that at least at phenomenological level in our model there are no instabilities, while waiting for further studies. 

This paper is organized as follows. In Section \ref{sec:fQ} we review the $f(Q)$-gravity and we introduce the particular form of $f(Q)$ we investigate. In Section \ref{Sec:phenoback} we study the background evolution of the Logarithmic $f(Q)$ and some specific branches. For any of them, where applicable, we derive constraints on the model parameters by using
EDE constraint from CMB and BBN constraint on the helium abundance. We then describe the adopted approach to linear perturbations in Section \ref{Sec:linearpheno} and present the phenomenology of the cosmological observables. Finally, we conclude in Section \ref{Sec:Conclusion}.

\section{$f(Q)$-gravity}\label{sec:fQ}

In this Section we review the theoretical framework of the $f(Q)$-gravity and we introduce the form of the $f(Q)$ function we will analyse in this work.

\subsection{General framework}

The action in $f(Q)$ gravity, within the Palatini formalism \cite{BeltranJimenez:2018vdo} (metric--affine formalism, in which the metric $g_{\mu\nu}$ and the connection $\Gamma^\alpha_{\phantom{x}\mu\nu}$ are to be independent fields), can be written as \cite{BeltranJimenez:2017tkd, BeltranJimenez:2019tme}
\be
    \label{eq:action}
    \Scal = \int\sqrt{-g} d^4 x \left[-\frac{1}{2\ka^2} f(Q) + \Lm (g_{\mu\nu},\chi_i)\right] \, ,
\ee
where $g$ is the determinant of the metric $g_{\mu\nu}$; $\ka^2=8\pi G_N/c^4$, with $G_N$ being the Newton's gravitational constant and $c$ is the speed of light (we set $c=1$); $\Lm$ is the Lagrangian density of the matter fields, $\chi_i$; and  $f(Q)$ is a general function of the non-metricity scalar ${Q=-Q_{\al\mu\nu}P^{\al\mu\nu}}$. 
In the latter ${Q_{\al\mu\nu}\equiv\nabla_\al g_{\mu\nu}}$ is the non-metricity tensor ($\nabla_\al$ is the covariant derivative defined by the affine connection $\Gamma^\alpha_{\phantom{x}\mu\nu}$) and
$P^{\al}_{\phantom{\al}\mu\nu} = -\frac12 L^{\al}_{\phantom{\al}\mu\nu} + \frac14\l(Q^\al-\tilde{Q}^\al\r)g_{\mu\nu} - \frac14\delta^\al_{(\mu}Q^{\phantom{x}}_{\nu)}$ 
is the non-metricity conjugate, 
where ${L^\al_{\phantom{\al}\mu\nu}=\frac12(Q^\al_{\phantom{\al}\mu\nu}-Q_{(\mu\nu)}^{\phantom{(\mu\nu)} \al})}$ is the disformation tensor,
and $Q_\al=g^{\mu\nu}Q_{\al\mu\nu}$ 
and $\tilde{Q}_\al=g^{\mu\nu}Q_{\mu\al\nu}$ are the two independent contractions of the non-metricity tensor. 
This choice of the non-metricity conjugate, $P^{\al}_{\phantom{\al}\mu\nu}$, results in a non-metricity scalar, $Q$, such that ${f(Q)=Q}$ corresponds to the STEGR~\cite{BeltranJimenez:2017tkd}. 
Varying the action in Eq.~\eqref{eq:action} with respect to (w.r.t.) the metric yields the following field equations
\cite{BeltranJimenez:2017tkd, BeltranJimenez:2019tme, Dialektopoulos:2019mtr, Anagnostopoulos:2021ydo}: 
\be
    \label{eq:field-eqs}
    \frac{2}{\sqrt{-g}} \nabla_\al\left(\sqrt{-g}f_{Q}P^{\al\mu}{}_\nu\right) +  \frac{1}{2}\delta^\mu_{\phantom{x}\nu} f +  f_{Q} P^{\mu\al\bt}Q_{\nu\al\bt}   =   T^{\mu}{}_{\nu} \,,
\ee
where the subscript $Q$ denotes derivatives w.r.t. $Q$, i.e. $f_Q \equiv \dv f/\dv Q$; $\delta^\mu_{\phantom{x}\nu}$ is the Kronecker delta; and $T_{\mu\nu}$ is the stress energy tensor that we will consider having the form for a perfect fluid, i.e. ${T^{\mu}{}_{\nu}=\mathrm{diag}\left(-\rho,p,p,p\right)}$, where $\rho$ is the energy density and $p$ the isotropic pressure.
Furthermore, varying the action  in Eq.~\eqref{eq:action} w.r.t.~the connection yields the following field equations \cite{BeltranJimenez:2017tkd, BeltranJimenez:2019tme}: 
\be
    \label{eq:connection-eqs}
    \nabla_\mu\nabla_\nu\left(\sqrt{-g}f_Q P^{\mu\nu}_{\phantom{xx}\alpha}\right) = 0 \,,
\ee
which holds in the absence of hypermomentum ( ${\mathfrak{H}_{\lambda}^{\phantom{x}\mu\nu} \equiv -\frac12 \delta \left(\sqrt{-g}\Lm\right)/\delta \Gamma^\alpha_{\phantom{X}\mu\nu}=0}$, i.e. $\Lm$ is assumed to depend on the metric but not on the affine connection).

We now adopt the coordinate choice such that ${\Gamma^\al_{\phantom{\al}\mu\nu}=0}$, this is known as the coincident gauge  \cite{BeltranJimenez:2017tkd, BeltranJimenez:2019tme}.  
This choice of the coincident gauge is without loss of generality, since, in $f(Q)$ gravity the connection is assumed to be torsion--free and flat, leaving an inertial connection which may be removed by a coordinate gauge choice~\cite{BeltranJimenez:2018vdo}.

Then, we consider a flat Friedmann-Lema{\^i}tre-Robertson-Walker (FLRW) metric with line element~\footnote{
    Note that the lapse function $N^2(t)$, in $\dv s^2=-N^2(t)\dv t^2+a(t)^2\delta_{ij}\dv x^i \dv x^j$, has been set to $1$. In principle, this freedom would not exist, since the diffeomorphism gauge freedom has ben used to fix the coincident gauge. However, this choice can still be done because in the $f(Q)$ theories there is a time--reparameterisation invariance~\cite{BeltranJimenez:2017tkd, BeltranJimenez:2018vdo}.}
\be
    \label{eq:FLRW-flat}
    \dv s^2=-\dv t^2+a(t)^2\delta_{ij}\dv x^i \dv x^j\,,
\ee
where $a(t)$ is the time dependent scale factor, and the roman indices run over the spatial coordinates. 
With these choices, the non-metricity scalar  takes the form 
\be
    \label{eq:Q-flatFLRW}
    Q = 6H^2 \,,
\ee
where $H\equiv\dot a/a$ is the Hubble function, overdots  denoting derivatives w.r.t.~cosmic time $t$. If the matter fields are described by perfect fluids with energy density $\rho_i$ and pressure $p_i$, then the modified Friedmann and Raychaudhuri equations are, respectively,
\be
    \label{eq:Fried}
    6f_QH^2-\frac12f = \ka^2\rho \,,
\ee
\be
    \label{eq:Raych}
    \l(12H^2f_{QQ}+f_Q\r)\dot{H}=-\frac{\ka^2}{2} \l(\rho+p\r)\,,
\ee
where ${f_{QQ}\equiv\dv^2 f/\dv Q^2}$. 
By $\rho$ we denote the sum of the contributions to the energy density from radiation ($r$) and matter ($m$)---where matter ($m$) includes cold dark matter ($c$) and baryons ($b$). In other words, $\rho\equiv \Sigma_i \rho_i$, where $i$ runs over all ordinary species ($r, c, b$). Likewise, for the pressure, $p\equiv \sum_i p_i$. 
Each cosmological fluid is \emph{individually} conserved, i.e. 
they each obey the continuity equation
\be
    \label{eq:Continuity}
    \dot \rho_i = -3H\left(\rho_i +p_i\right) \,.
\ee
Then, the evolution of densities with the scale factor $a$ is given by
\be
    \label{eq:densities}
    \rho_i = \rhoi a^{-3\left(1+w_i\right)} \,,
\ee
assuming a constant equation of state for each of the fluids of the form $w_i=p_i/\rho_i$ (in detail, $w_{c,b}=0$ for cold dark matter and baryons and $w_r=1/3$ for radiation). We note that a variable with subscript $0$ (e.g. $\rhoi$) denotes the value of the function at present time (when $a=a_0\equiv 1$; or, in terms of redshift: when $z=0$, redshift being defined by ${z+1=a/a_0}$).
We also introduce the usual definition of density parameters: $\Om_i \equiv \ka^2\rho_i/(3 H^2)$. 

Note that the conservation laws of the theory can be derived from Eqs.~\eqref{eq:field-eqs} and \eqref{eq:connection-eqs}. These conservation laws,  in the absence hypermomentum, coincide with those of GR. 
Conversely, the conservation equation, Eq.~\eqref{eq:Continuity}, together with Eq.~\eqref{eq:field-eqs}
lead to the equivalent of the connection equation, Eq.~\eqref{eq:connection-eqs}, in the coincident gauge~\cite{BeltranJimenez:2018vdo}. 
Thus, we may drop Eq.~\eqref{eq:connection-eqs}, and work solely with the field equations, Eq.~\eqref{eq:field-eqs}, and the continuity equations, Eq.~\eqref{eq:Continuity}.

It is possible to recast Eqs.~\eqref{eq:Fried} and~\eqref{eq:Raych} in a form similar to the one of GR by introducing a `dark $Q$--fluid':
\bea
    \label{eq:Fried-fQ}
    &&3H^2=\ka^2(\rho+\rho_Q) \,,\\
    \label{eq:Raych-fQ}
    &&\dot{H}=-\frac{\ka^2}{2}\l(\rho+p+\rho_Q+p_Q\r) \,,
\eea
where the density $\rho_Q$ and the pressure $p_Q$ are defined as
\bea
    \label{eq:rhoQ}
    &&\rho_Q=\frac{1}{\ka^2}\l(3H^2+\frac{1}{2}f-6f_QH^2\r) \,,\\
    \label{eq:pQ}
    &&p_Q=\frac{1}{\ka^2}\l[2\dot{H}(12H^2f_{QQ}+f_Q-1)+H^2(6f_Q-3)-\frac{1}{2}f\r] \,.
\eea
The Raychaudhuri equation, Eq.~\eqref{eq:Raych-fQ}, can be obtained by taking the time derivative of the Friedmann equation, Eq.~\eqref{eq:Fried-fQ}, and using the continuity equation, Eq.~\eqref{eq:Continuity}, to replace $\dot{\rho}$ and $\dot\rho_Q$. Thus, only two of these three equations are independent, and we may work solely with the Friedmann and continuity equations.

The $f(Q)$ equivalent to classical GR (up to a boundary term), the STEGR, is given by $f(Q)=Q$. 
Of interest is also the class of models with ${f(Q)=Q+M\sqrt{Q}}$,
where $M$ is a constant parameter, understood as a mass scale \cite{BeltranJimenez:2017tkd, BeltranJimenez:2019tme}. This model can mimic the background evolution of $\Lambda$CDM, while it behaves differently at the level of perturbations \cite{Barros:2020bgg, Frusciante:2021sio,Atayde:2021pgb}. Many other forms of $f(Q)$ have been adopted in literature  \cite{BeltranJimenez:2019tme,Ayuso:2020dcu,Anagnostopoulos:2021ydo,Anagnostopoulos:2022gej} which have been explored in different contexts. In the next sub-section we will outline the form of $f(Q)$ adopted in this work.

\subsection{Logarithmic $f(Q)$}\label{sec:fQ-log}

The class of models we will investigate is based on the form of $f(Q)$ given by \cite{Ayuso:2021vtj} 
\be
    \label{eq:fQ-DGP}
    f(Q)= \dfrac{\al'}{2}\sqrt{Q}\ln{\left(\gm' Q\right)} + \bt Q \,,
\ee 
where $\al'$, $\bt$ and $\gm'$ are constant parameters~\footnote{
    \label{fn:gammaAyuso} 
    In Ref.~\cite{Ayuso:2021vtj} the parameter $\gm^\prime$ is not present. Eq.~\eqref{eq:fQ-DGP} matches the one in Ref.~\cite{Ayuso:2021vtj} for $\gm^\prime=1$. In Ref.~\cite{Ayuso:2021vtj} this parameter is not relevant  because  the authors are interested in studying only the background dynamics which is not affected by this parameter (see Eq.~\eqref{eq:H-sols}). In our analysis this parameter enters in the perturbation equations and therefore we include it.  The parameter $\gm'$ appears as an integration constant and for a detailed derivation of the $f(Q)$ form in Eq.~\eqref{eq:fQ-DGP}, see Appendix~\ref{appdx:fQderivation}.
}. 
Note that the GR equivalent is recovered for the choice $\al'=0$ and $\bt=1$.
While the constant $\bt$ is dimensionless, for the other two parameters the dimensions need to be such that $\left[\al'\right]=\left[Q\right]^{1/2}$ and $\left[\gm'\right]=\left[Q\right]^{-1}$. Since we will be preforming a numerical study, it is preferable to work with dimensionless quantities, therefore we define the following dimensionless constant parameters: 
\be
    \label{eq:parms_nodimension}
    \al\equiv \frac{\al'}{\sqrt{Q_0}}  \,, \qquad \gm\equiv \gm' Q_0  \,,
\ee
where  the constant $Q_0$ is the present value of $Q$ as defined in Eq.~\eqref{eq:Q-flatFLRW}, i.e. ${Q_0\equiv 6H_0^2}$. Equation~\eqref{eq:fQ-DGP}, thus, becomes 
\be
    \label{eq:fQ-DGP-new}
    f(Q)= \dfrac{\al}{2}\sqrt{Q_0 Q}\ln{\left(\gm\frac{Q}{Q_0}\right)} + \bt Q \,,
\ee 
which is the one we will use throughout this work.
Additionally,  $f_Q$ is identified as the effective Planck mass ($M_{\textrm{eff}}^2=f_Q G_N$) \cite{BeltranJimenez:2019tme, Frusciante:2021sio}, therefore it should remain positive, a condition we will impose in our analysis, to avoid ghost instability. 

 The modified Friedmann equation, Eq.~\eqref{eq:Fried}, and the Raychaudhuri equation, Eq.~\eqref{eq:Raych},
become
\be
    \label{eq:Fried-DGP-1}
    3\bt H^2 + 3\al H_0  H = \ka^2\rho  \,,
\ee 
and \be
    \label{eq:Raych-DGP}
    \left(\frac{\al H_0}{2 H } +\bt\right)\dot H = -\frac{\ka^2}{2} \left(\rho +p\right) \,,
\ee 
respectively. Equation  \eqref{eq:Fried-DGP-1} resembles the DGP-like behaviour \cite{Deffayet:2000uy, Lazkoz:2006gp, Ayuso:2021vtj} and has the following  solutions:
\be
    \label{eq:H-sols}
    H = \dfrac{\pm \sqrt{\frac43\bt \ka^2\rho + \al^2 H_0^2} - \al H_0}{2\bt} \,,
\ee
 for $\bt\neq 0$.
In an expanding universe, the Hubble function $H$ is positive. To guarantee $H>0$ we will use the solution with the ``plus'' sign ($+$), since it is the only one that always remains positive (considering $\bt>0$~\footnote{
    We consider $\bt>0$, since the GR equivalent is recovered for $\bt=1$, and we will study small deviations from this value.}).
So, the solution we take has the following form:
\be
    \label{eq:Hofa-sol}
    H = \frac{H_0}{2\bt}\left[\sqrt{4 \bt \left(\frac{\Omr}{a^4}+\frac{\Omm}{a^3}\right) + \al^2} - \al\right] \,.
\ee
Note that the parameter $\gm$ does not affect the evolution of the background.

For this class of $f(Q)$ models, Eq.~\eqref{eq:rhoQ} yields the following `dark $Q$-field' energy density:
\be
    \label{eq:rhoQ-Log}
    \rho_Q=\frac{3 H^2}{\ka^2}\l(1 -\bt -\dfrac{\al H_0 }{H} \r) \,. 
\ee
So, defining its density parameter in the standard way, $\Om_Q \equiv \ka^2\rho_Q/3H^2$, we have 
\be
    \label{eq:OmQ}
    \Om_Q = 1 -\bt -\dfrac{\al H_0 }{H} \,,
\ee
whose value at present time can be easily computed, yielding 
\be 
    \label{eq:OmQ0}
    \OmQ = 1 -\al -\bt \,.
\ee 
Moreover, the pressure of the $Q$-field, defined in Eq.~\eqref{eq:pQ}, is, then, given by
\be
    \label{eq:pQ-Log}
    p_Q=-\frac{1}{\ka^2}\l[\l(2 -2\bt-\frac{\al H_0}{H}\r)\dot H +3\l(1-\bt -\frac{\al H_0}{H}\r)  H^2 \r] \,.
\ee

In the next Section, we study,  analytically, the evolution of the background cosmology, taking particular attention to some limiting cases, which are approached by specific choices of the free parameters of the Logarithmic $f(Q)$ function.

\section{Phenomenology of the background evolution and viability} \label{Sec:phenoback}

We first study the background evolution of the Logarithmic modified cosmology, by considering different branches which will ensure the model is giving rise to late--time acceleration. Here we list the specific cases we analyze:
\begin{itemize}
    \item sLog model: we solve the Hubble evolution as that given by Eq.~\eqref{eq:Hofa-sol}. In principle, we have $\bt$ and $\al$ as free parameters, however by imposing the flatness condition we find that we can eliminate one parameter out of the two. We show that under the requirement $f_Q>0$  a viable expansion history with a late--time accelerated expansion can be achieved, where the $Q$-fluid works as an effective DE ($DE=Q$). We note that the case with $\beta=1$ resembles the DGP model at background, so we analyze it separately in the following. 
    For the sLog model, where in general $\bt\neq 1$, we work out BBN and EDE constraints. Under such conditions, we note that $\al<0$ and there is self-accelerated behaviour, hence the name sLog model.
    
    \item sDGP model: this is a sub-case of the sLog model. We solve the Hubble evolution given in Eq.~\eqref{eq:Hofa-sol} for $\beta=1$ and $\al<0$.  We note that $\alpha$ is constrained by the flatness condition.  For this case, the choice $\bt=1$ makes it DGP-like and we obtain a background that resembles the sDGP  branch \cite{Deffayet:2000uy}, hence the name. 
    We note that under the requirement $f_Q>0$ for a range of the $\gm$ parameter the model is viable.

    \item $\Lambda$--nDGP model: we solve the Hubble evolution  Eq. \eqref{eq:Hofa-sol} by considering $\bt=1$ and $\al>0$. In this case, the background behaves like the nDGP model, hence the name.  This model  does not give a late--time acceleration, and we follow a standard approach used for nDGP \cite{Dvali:2000hr, Lazkoz:2006gp}: in order to obtain a late--time accelerated expansion, we need to add an extra dark component (${DE=Q+de}$).   We assume the extra dark energy fluid to be  a cosmological constant (${DE=Q+\Lambda}$). This case is akin to what has already been discussed in Ref.~\cite{Ayuso:2021vtj}, but with some differences: in Ref.~\cite{Ayuso:2021vtj} the authors always included $\Lambda$ regardless of positive or negative $\al$, whereas in our work we limit to the case $\al>0$. That is because in the case $\al<0$ it is not necessary to introduce a cosmological constant since we fall back to the  sDGP case above, which is self-accelerating without the need of any additional component. Hereafter we refer to this model as $\Lambda$--nDGP.

\end{itemize}

We summarize the main features of the subcases we analyze in Tab.~\ref{Tab:Modelsummary}, where we anticipate some results of the following subsections. For illustrative purpose we fix the values of the cosmological parameters:
 ${\Omb = 0.0461}$, ${\Omc= 0.229}$, ${\Omr= 5\times 10^{-5}}$, $H_0=70$ km/s/Mpc,
in order to show concrete examples.

\begin{table}
\centering
\begin{tabular}{|l|l|l|l|l|l|}
\hline
\multicolumn{5}{|c|}{Summary of models features}  \\ \hline\hline
 Model name &  Extra parameters  &  $H(z)$ & Early DE constraint & BBN constraint \\ \hline \hline
sLog & 2\, ($\al\rightarrow \mbox{fixed}, \bt, \gm$) & Eq.~\eqref{eq:H-Exact} & $\Om_{Q}^e(z=50,\bt)$, Eq. \eqref{eq:eDEbeta}*  & $0.89\lesssim \bt \lesssim 1.14$\\
sDGP & 1 ($\al\rightarrow \mbox{fixed}, \bt=1, \gm$) & Eq.~\eqref{eq:H-Exact} with $\bt=1$ & satisfied ** & not applicable\\
$\Lambda$--nDGP &2 ($\al>0, \bt=1, \gm$)& Eq.~\eqref{eq:HLnDGP} & always satisfied & not applicable\\
\hline
\end{tabular}
\caption{A summary of the main features of each case we analyse. For early dark energy constraint, we refer to the bounds in Ref.~\cite{Planck:2015bue} and for the BBN constraints to Refs.~\cite{Olive:1994fe,Izotov:1998mj,OMeara:2000tmq}. 
For the cosmological parameter values we use in this work: (*) the constraint on the EDE leads to $\beta \gtrsim 0.98$, in the sLog model; (**) in the sDGP model this constraint is satisfied, as $\Om_{Q}^e(z=50)\approx 0.004$. }
\label{Tab:Modelsummary}
\end{table}

\subsection{sLog model}

In this subsection, we study the background evolution of the sLog model, whose Hubble function is that of Eq.~\eqref{eq:Hofa-sol}. In this case, the  energy density  and pressure of the effective DE correspond  to that of the $Q$-field (Eq.~\eqref{eq:rhoQ-Log} and  Eq.~\eqref{eq:pQ-Log}). We will show that the density of the $Q$-field can be negative at early time, therefore, we do not define an equation of state for $Q$ but instead we will only consider an effective equation of state for the system:
\be
w_{\rm eff}=-1-\frac23\frac{\dot{H}}{H^2}.
\ee

Additionally, we consider the  flatness constraint, i.e. ${\Om_Q +\Om_m + \Om_r =1}$, which can be obtained from the Friedmann equation. Evaluating this constraint at present time, we obtain the following relation 
\be
    \OmQ = 1  -\al -\bt = 1 - \Omr -\Omm \,,
\ee
 which  can be used  to reduce the number of free parameters of the model. We choose to write $\al$ as a function of the other parameters:
\be
    \label{eq:alpha-Exact}
    \al = \Omm +\Omr -\bt \,.
\ee 
The resulting model has then only one free parameter at background level: $\bt$,  which is positive defined. Let us note that  $\gm$ is another free parameter for the model when considering  perturbations.
We can now rewrite the Friedman equation only in terms of the independent parameters of the model, so the Hubble function, Eq.~\eqref{eq:Hofa-sol}, becomes
\be\label{eq:H-Exact}
    H=\frac{H_0}{2\bt}\l[\sqrt{ 4\bt \l(\frac{\Omr}{a^4} +\frac{\Omm}{a^3}\r) + \l(\Omm +\Omr -\bt\r)^2 } -\l(\Omm +\Omr -\bt\r)\r] \,.
\ee

Before moving ahead with the investigation of the background phenomenology, we want to consider the predictions of BBN and EDE constraints to give us an informative limit on the parameters of the sLog model:
\begin{itemize}
    \item BBN constraint: The BBN constraint on the helium abundance is $|\Delta Y_p| < 0.01$ \cite{Olive:1994fe,Izotov:1998mj,OMeara:2000tmq} and it can be related to the effective cosmological gravitational coupling, $G_c$. 
Such coupling in the case of the model under consideration is 
\be
    G_c=\frac{G_N}{\bt}\,.
\ee
It has been shown \cite{Chen:2000xxa} that the helium abundance can be also expressed in terms of a speed-up factor $\zeta\equiv H/\bar{H}$, where $\bar{H}$ is the cosmology of reference that we assume to be $\Lambda$CDM, and it reads:
\be\label{eq:BBN}
    |\Delta Y_p|=0.08(\zeta^2 - 1)\,.
\ee
 At BBN time when  matter components are dominant, we have that 
\be
    \zeta=\sqrt{\frac{G_c}{G_N}}=\sqrt{\frac{1}{\bt}}\,,
\ee
therefore by substituting it in Eq.~\eqref{eq:BBN} we obtain
\be
    \l|\frac{1}{\bt}-1\r|<\frac{1}{8}.
\ee
Since we are considering $\bt>0$, we obtain the following theoretical bounds on the $\bt$ parameter from BBN: 
\be
    \label{eq:BBN-beta-Exact}
    \frac{8}{9}<\bt <\frac{8}{7} \,,
\ee
i.e. $0.89\lsim \bt \lsim 1.14$ \footnote{BBN constraint on this model has also been obtained by \cite{Anagnostopoulos:2022gej}, with a different procedure and using a stronger bound on $|\Delta Y_p| < 10^{-4}$ than what we use here. In this regard, our bound is more conservative.}. 

\item EDE constraint: Another constraint one can apply is on the DE relative contribution to the early time cosmology ($z\sim 50$), which we denote by $\Om_{DE}^e$. 
From CMB data by Planck 2015 one has $\Om_{DE}^e<0.02$ at 95\% C.L.~\cite{Planck:2015bue} 
In the sLog model, the $f(Q)$ modification is the only contribution to DE, i.e. 
\begin{equation}
    \Om_{DE}(z) = \Om_{Q}(z) = 1 -\bt - \l(\Omm + \Omr - \bt\r)\frac{H_0}{H(z)} \,,
\end{equation}
obtained combining Eqs.~\eqref{eq:OmQ} and \eqref{eq:alpha-Exact}.
So, we apply the EDE constraint, ${\Om_{DE}^e<0.02}$, to ${\Om_{Q}(z=50)}$.  
Having ${\Om_{Q}(z=50)<0.02}$ translates on a bound on $\bt$, which depends on the values of the cosmological parameters $\Omm$ and $\Omr$. 
Solving for $\bt$; taking the limit $\Omr\rightarrow 0$ (because at $z\sim 50$ the Universe is already in the matter dominated era); and performing an expansion; we then obtain 
\be \label{eq:eDEbeta}
\beta \gtrsim 0.001 \sqrt{\frac{1}{\Omega _{\text{m0}}}}+0.98.
\ee
Thus, for the choice of cosmological parameters we have made in this work, it translates in:
\be
    \label{eq:EDE-beta-Exact}
    \bt \gtrsim 0.98 \,.
\ee
\end{itemize}
Combining both constraints, Eqs.~\eqref{eq:BBN-beta-Exact} and \eqref{eq:EDE-beta-Exact}, we estimate the following range of viable values of $\bt$:
\be
    \label{eq:constraint-beta-Exact}
    0.98 \lesssim \bt \lesssim 1.14 \,.
\ee

Then, we use the above results to provide some examples  on the viable phenomenology of the sLog model at background and highlight its differences w.r.t. \LCDM. Hereafter, we will consider $\gm\lesssim 10$ to satisfy the condition $f_Q>0$.

\begin{figure*}[t!]
    \begin{justify}
    \includegraphics[width=0.45\textwidth]{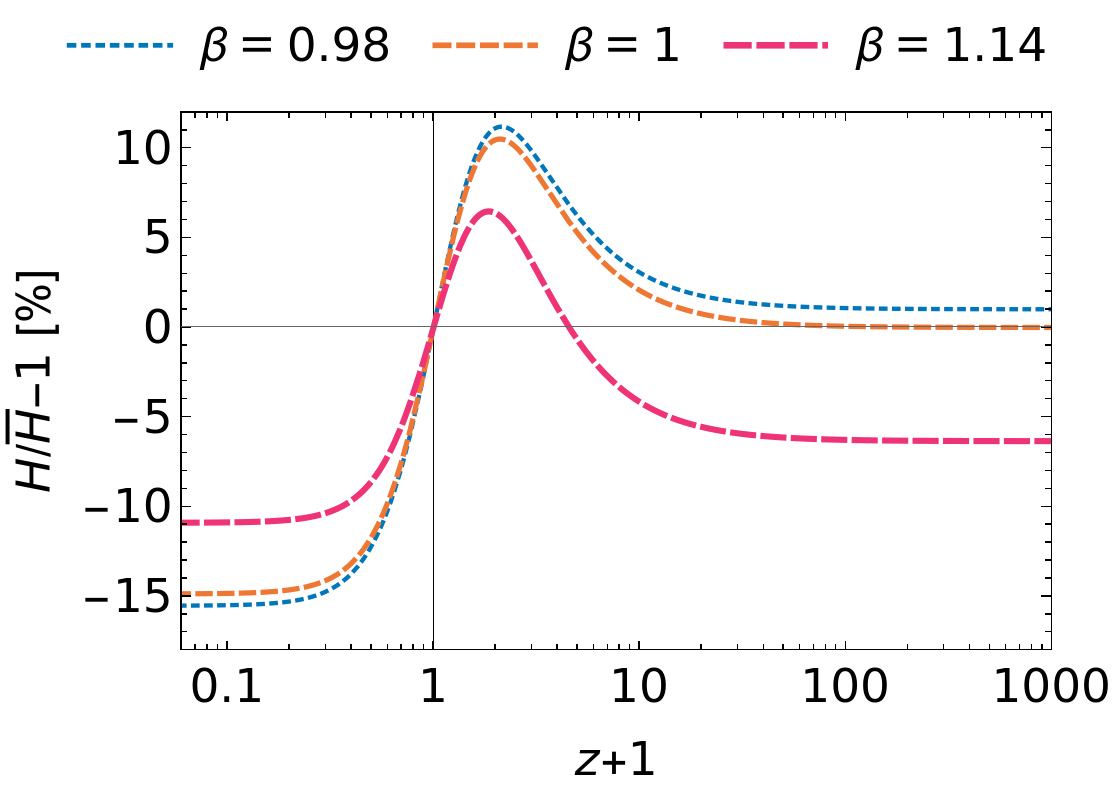}
    \hspace{15pt}
    \includegraphics[width=0.45\textwidth]{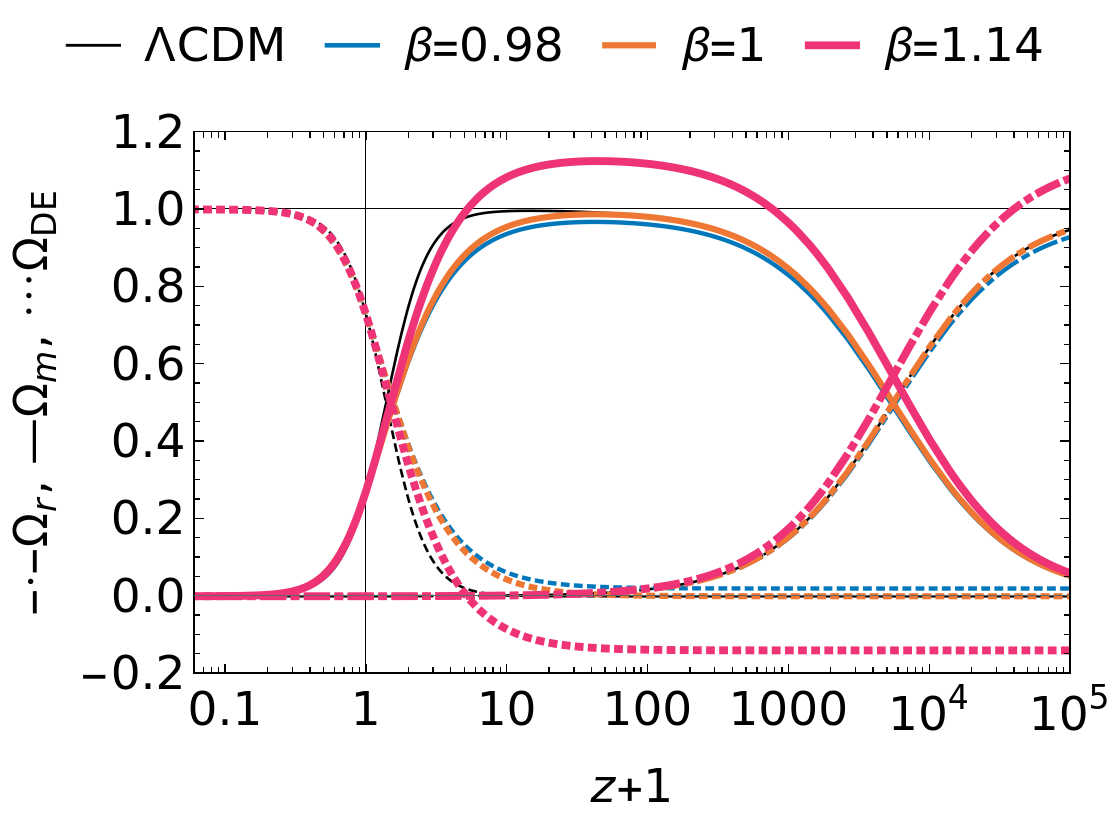}
    \justifying
    \caption{sLog model. \textit{Left panel:} Percentage relative difference of the Hubble evolution ($H$) w.r.t. \LCDM~($\bar H$) for different values of $\bt$ satisfying the BBN Constraint. \textit{Right panel:} Evolution with redshift of the density parameters.  $\Om_{DE}=\Om_\Lambda$ in the \LCDM~case and $\Om_{DE}=\Om_Q$ in the sLog model of $f(Q)$. The sLog model and the \LCDM~one share the same values of the cosmological parameters. The $\bt=1$ case corresponds to the sDGP model. The vertical line indicates the present time ($z=0$).}
    \label{fig:Exact-Hdiff-omegas}
    \end{justify}
\end{figure*}

Figure~\ref{fig:Exact-Hdiff-omegas} left panel shows the relative percentage difference of the Hubble function $H(z)$ of the sLog model relative to the standard cosmological scenario ($\bar H$).
We can notice that, due to a modified cosmological gravitational coupling constant, $G_c$, at large $z$, the rate of expansion of the MG model is larger than \LCDM~when $\bt<1$, while it is smaller when $\bt>1$. 
At small $z$ ($z+1\lesssim 5$), instead, the evolution of the expansion rate is  higher than in \LCDM, for all values of $\bt$ considered.

Figure~\ref{fig:Exact-Hdiff-omegas} right panel shows the evolution with redshift of the density parameters, $\Om_m(z)$, $\Om_r(z)$, and $\Om_Q(z)$. 
One can notice that for $\bt>1$, at some $z$ the matter components have density values that are  bigger than one while the $Q$ component has negative values. 
The opposite holds when $\bt<1$. 
That is because the expansion history changes (as we show in Fig.~\ref{fig:Exact-Hdiff-omegas} left panel) altering the time behaviour of the relative abundances. The $Q$-field balances this effect in order to respect the flatness condition. 
In the cases in which the density of the $Q$-field is negative the interpretation of the MG in terms of a fluid-like component is not well posed, representing instead a genuine geometrical modification of the gravitational sector as already found in other works on MG \cite{Amendola:2006we,Frusciante:2013zop,Frusciante:2015maa}. 
We also note that the time of equality between radiation and matter does not change, being the same as for \LCDM. On the other hand, the time of equality between matter and the $Q$-field is anticipated w.r.t. $\Lambda$CDM---the smaller $\bt$ is, the earlier this transition takes place.

\begin{figure}
    \centering
    \includegraphics[width=0.5\textwidth]{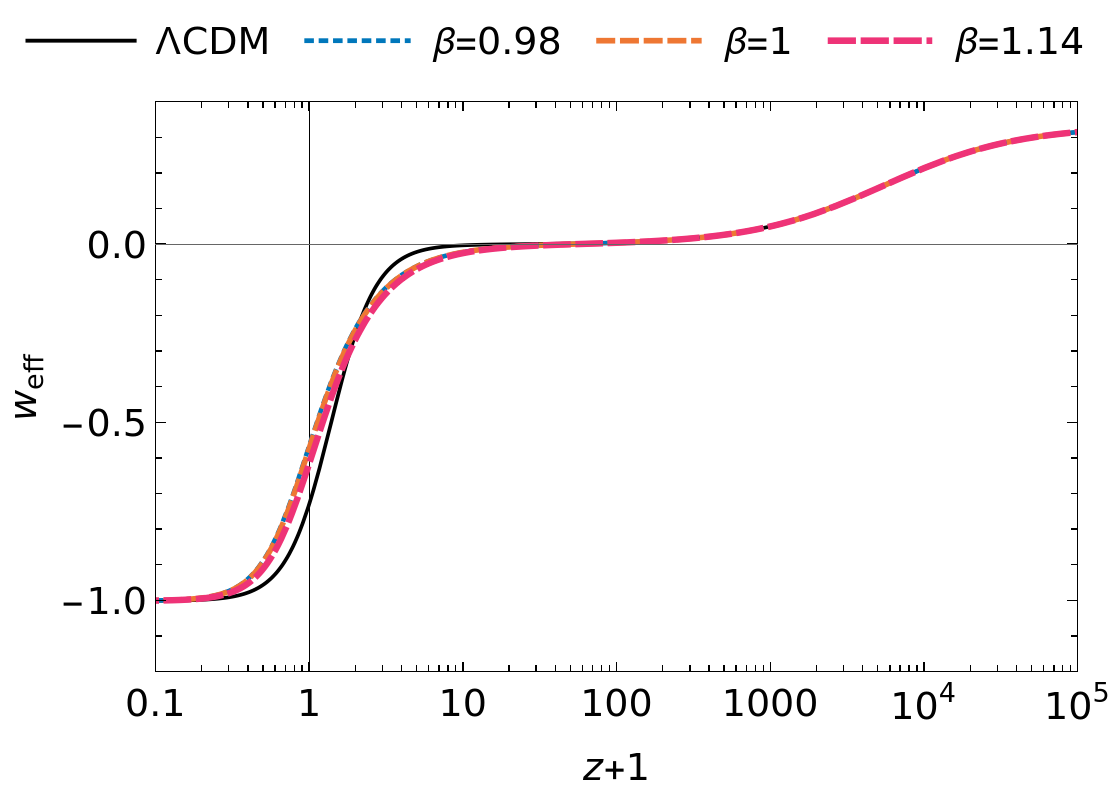}
    \hspace{10pt}
    \includegraphics[width=0.45\textwidth]{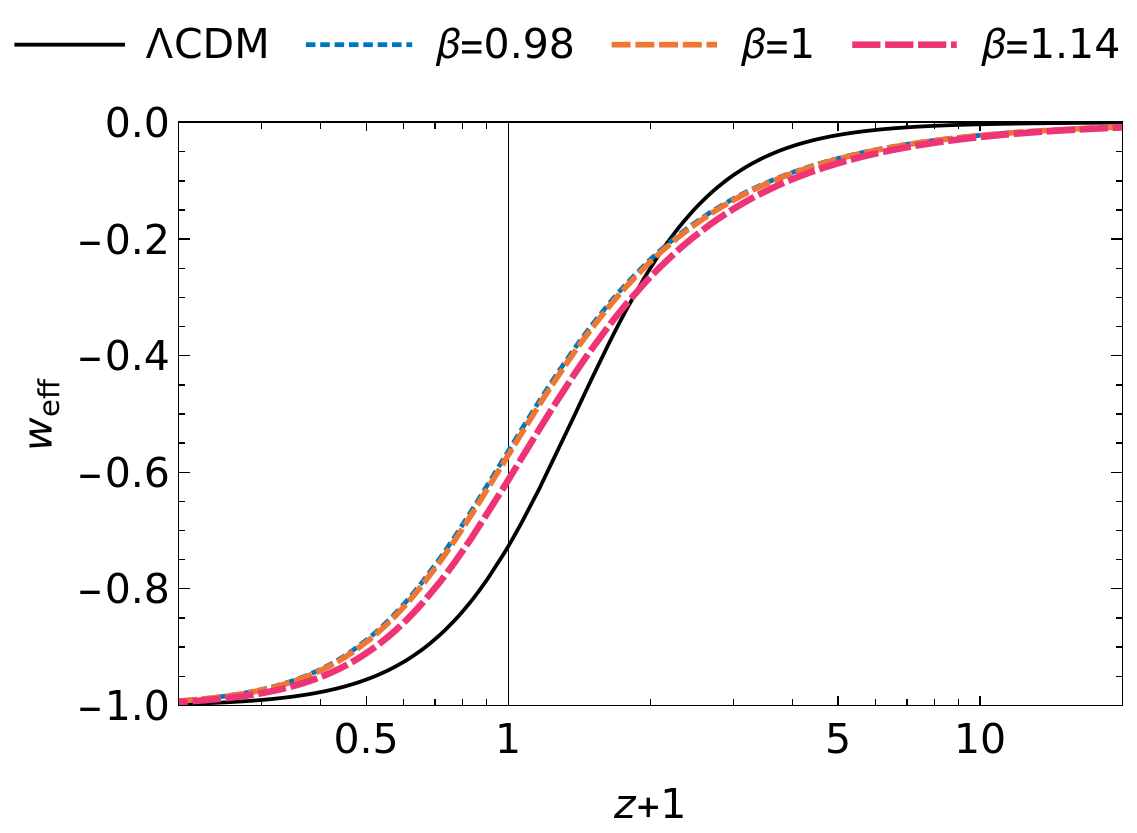}
    \caption{sLog model: effective equation of state ($\weff$) evolution with redshift ($z$), as compared to \LCDM. Significant differences only occur at late time ($z+1\lesssim 50$). The right panel shows a zoomed in version of the left one. The vertical line indicates the present time ($z=0$). The sLog model has the effect of smoothing the transition between matter domination ($\weff=0$) to dark energy domination ($\weff<-1/3$), as compared to \LCDM, i.e. the transition takes place over a longer period of time. The curve with the lowest value of $\bt$ ($\bt=0.98$) has the highest \weff~at present time (the transition takes longer).}
   \label{fig:weff_E}
\end{figure}

Finally, we  show the evolution of the effective equation of state in Fig.~\ref{fig:weff_E}. It follows the one of $\Lambda$CDM up $z+1\sim 50$, with deviations at later time. 
At late times, the $\weff$ for the sLog model also evolves from $\weff=0$ to $\weff=-1$, but this transition takes a longer time, starting earlier. 
At present time, the values of $\weff$ for the MG model are larger than the one for the standard scenario, the one for $\bt=0.98$ being the largest one.

\subsection{sDGP model}

In this Section, we explore the sDGP  scenario, which is a particular case of the sLog model, realized for the subcase $\bt=1$ (for which $\al<0$). The equation for the density  of the $Q$-field is
\be
    \rho_Q=-\frac{3}{\ka^2}\al H_0 H\,,
\ee
with a density parameter given by
\be
    \label{eq:OmQsDGP}
    \Om_Q = -\dfrac{\al H_0 }{H}\,.
\ee
The sDGP model has $\al<0$, so, $\Om_Q\geq 0$ always.  At  present time we have $\OmQ=-\al$ and considering the flatness condition
with $\OmQ=1-\Omm-\Omr$, we can then eliminate $\al$ as free parameter: 
\begin{equation}
    \label{eq:alpha-flat-sDGP}
    \al=\Omm+\Omr-1 \,.
\end{equation}
The sDGP case thus does not have any free parameter at background level.
The Friedmann equation follows from Eq.~\eqref{eq:H-Exact}, simply by setting $\bt=1$. 
Considering that $\beta=1$, the EDE constraint is satisfied (see Eq. \ref{eq:EDE-beta-Exact}), in particular we find that the density of the $Q$-field at early time is ${\Om_Q(z=50)\approx0.004}$.

\subsection{$\Lambda$--nDGP model}

Another case we will investigate is the one inspired by the nDGP scenario according to which   $\bt=1$ and $\al>0$.  For this nDGP model the late time acceleration phase cannot be achieved without an additional dark component, $\rho_{de}(a)$. We then have:
\be
     H^2+\al H_0 H = \dfrac{\ka^2}{3}(\rho+\rho_{de}) \, ,
\ee 
and the solution is

\be
    \label{eq:HnDGP}
    H= H_0 \left(\sqrt{ \frac{\Omr}{a^4} +\frac{\Omm}{a^3} + \Omde \,g(a) +\frac{\al^2}{4} }-\frac{\al}{2}\right)\,,
\ee
with $g(a)$ encoding the equation of state of the dark fluid component, $w_{de}$:
\begin{equation}
 g(a) = \exp{\l(-3\int_{1}^{a}\frac{1+w_{\rm de}(x)}{x}\mathrm{d}x\r)}\,.
\end{equation}
Now, we can define a new effective DE component, ${\rho_{DE} = \rho_Q + \rho_{de}}$, such that
\be
   H^2 = \dfrac{\ka^2}{3}(\rho+\rho_{DE})\, ,
\ee 
with
\be
    \rho_{DE}=-\frac{3}{\ka^2} \al H_0 H+\rho_{de}\,,
\ee
and 
\be
    \label{eq:OmDE-nDGP}
    \Om_{DE}=-\dfrac{\al H_0 }{H}+\Om_{de}.
\ee
As in previous sections we consider the flatness condition at present time:
\be
    \label{eq:flat-LnDGP}
    \OmDE\equiv-\al  +\Omde=1-\Omm-\Omr.
\ee

For the $\Lambda$--nDGP model, we select $w_{de}=-1$, such that the additional dark energy component is the cosmological constant. In this case we have that $g(a)=1$ and
\be
    \label{eq:OmL-LnDGP}
    \Omde\equiv\OmL=1-\Omm-\Omr+\al\,,
\ee
which we can use to constrain and replace $\OmL$. Thus, we are left with one extra free parameter at background: $\al$. 

The background evolution is given by
\be
    \label{eq:HLnDGP}
    H= H_0 \left(\sqrt{ \frac{\Omr}{a^4} +\frac{\Omm}{a^3} + 1-\Omm-\Omr+\al +\frac{\al^2}{4} }-\frac{\al}{2}\right)\,.
\ee

The effective DE in the $\Lambda$-nDGP case is given by Eq.~\eqref{eq:OmDE-nDGP}, but with ${\Om_{de}=\Om_\Lambda=\OmL H_0^2/H^2}$, so
\be
    \label{eq:OmDE-LnDGP}
    \Om_{DE}=-\dfrac{\al H_0 }{H}+\frac{\left(1-\Omm-\Omr+\al\right) H_0^2}{H^2}.
\ee
If we impose to Eq.~\eqref{eq:OmDE-LnDGP} the EDE bound, ${\Om_{DE}^e<0.02}$, this can be satisfied for any $\al>0$.

\begin{figure}[t!]
    \centering
    \includegraphics[width=0.45\textwidth]{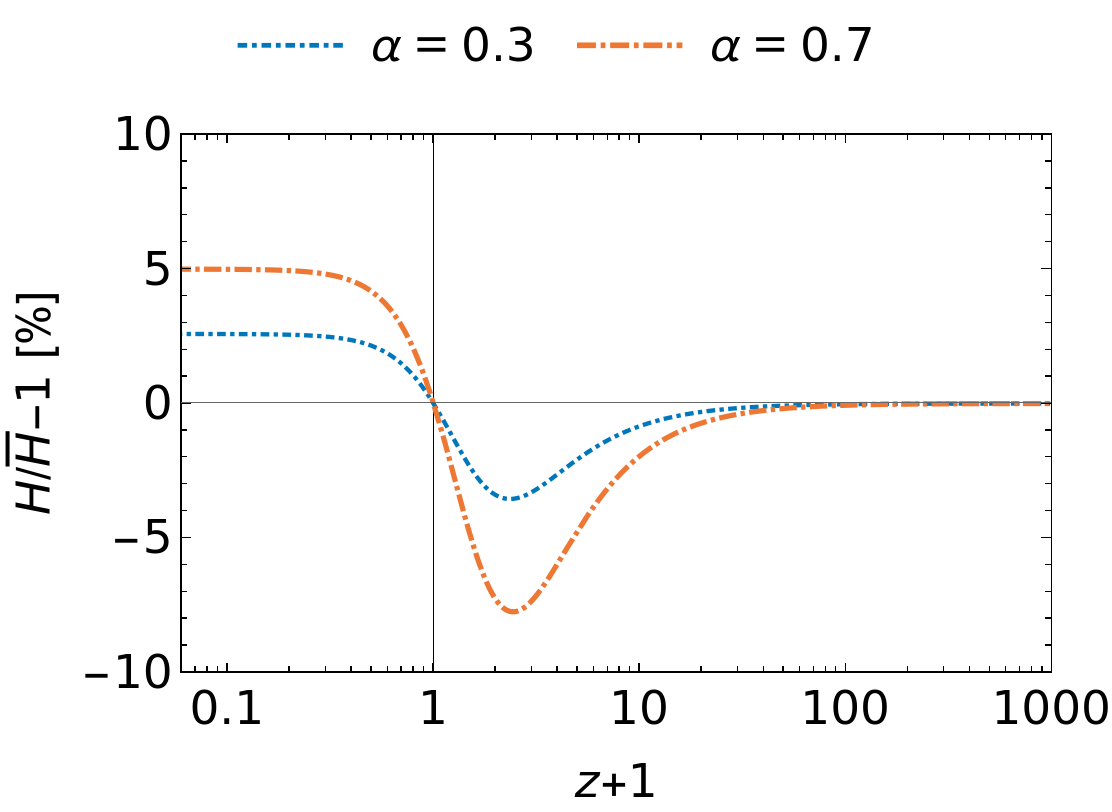}
    \hspace{15pt}
    \includegraphics[width=0.45\textwidth]{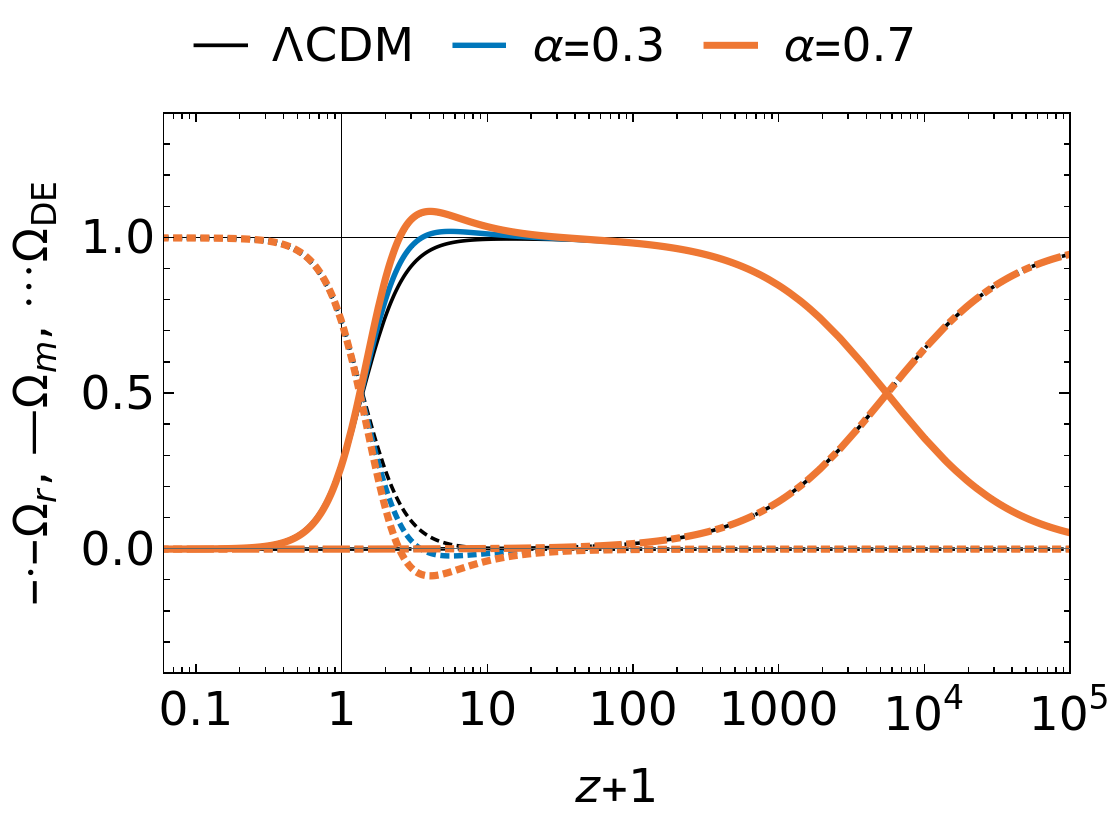}
    \caption{
    $\Lambda$-nDGP model ($\bt=1$, and $\gm$ does not affect the background).
    \textit{Left panel:} Percentage relative difference of the Hubble evolution ($H$) w.r.t. \LCDM~($\bar H$) for different values of $\al$. \textit{Right panel:} Evolution with redshift of the density parameters.  $\Om_{DE}=\Om_\Lambda$ in the \LCDM~case and $\Om_{DE}=\Om_\Lambda+\Om_Q$ in the $\Lambda$-nDGP model of $f(Q)$. The $\Lambda$-nDGP model and the \LCDM~one share the same values of the cosmological parameters for this plot. The vertical line indicates the present time ($z=0$).}
    \label{fig:Hdiff-Omegas-LnDGP}
\end{figure}

In Fig.~\ref{fig:Hdiff-Omegas-LnDGP} left panel, we show the relative percentage difference of the Hubble evolution for the $\Lambda$--nDGP model w.r.t.~\LCDM~for different values of $\al$, the only free parameter at background level.
We can notice that already with small values of $\al$ the relative difference in the expansion rate can reach the $\sim 8\%$.
In this model, the effect on the Hubble evolution only starts to become significant at ${z+1\lsim100}$. 
Furthermore, between ${0\lesssim z \lesssim 100}$, $H$ is smaller than in \LCDM, while in the future it is larger. 

Figure~\ref{fig:Hdiff-Omegas-LnDGP} right panel shows the evolution of the density parameters. 
Because the expansion history is unmodified w.r.t. LCDM at early time, it is indistinguishable from it, while for $z+1\lesssim100$, when the modifications appear, we notice significant differences:
 $\Omega_{DE}$ becomes negative, and, in order to maintain the flatness condition, $\Omega_{m}$ is larger than one (similarly to what we already saw for the sLog model).
The matter--DE equality happens later for the MG model than for \LCDM, with larger value of $\al$ moving the time of equality to smaller redshift.

\begin{figure}[ht!]
    \centering
    \includegraphics[width=0.5\textwidth]{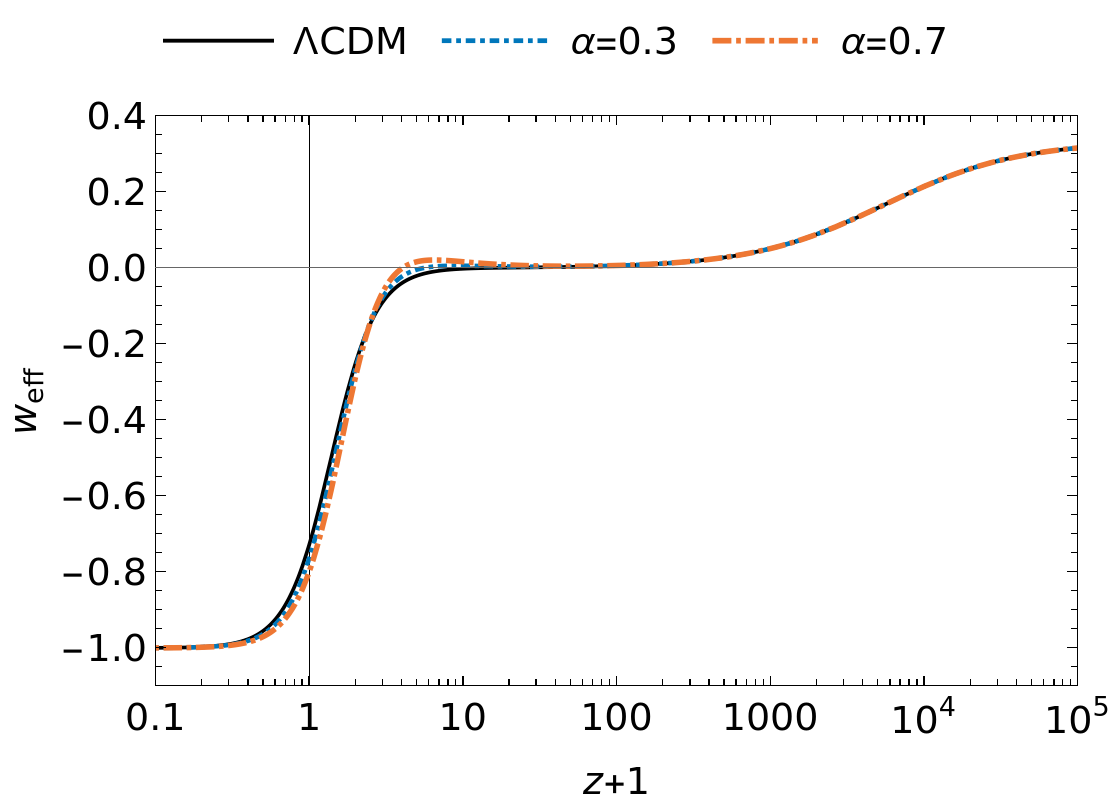}
    \hspace{10pt}
    \includegraphics[width=0.45\textwidth]{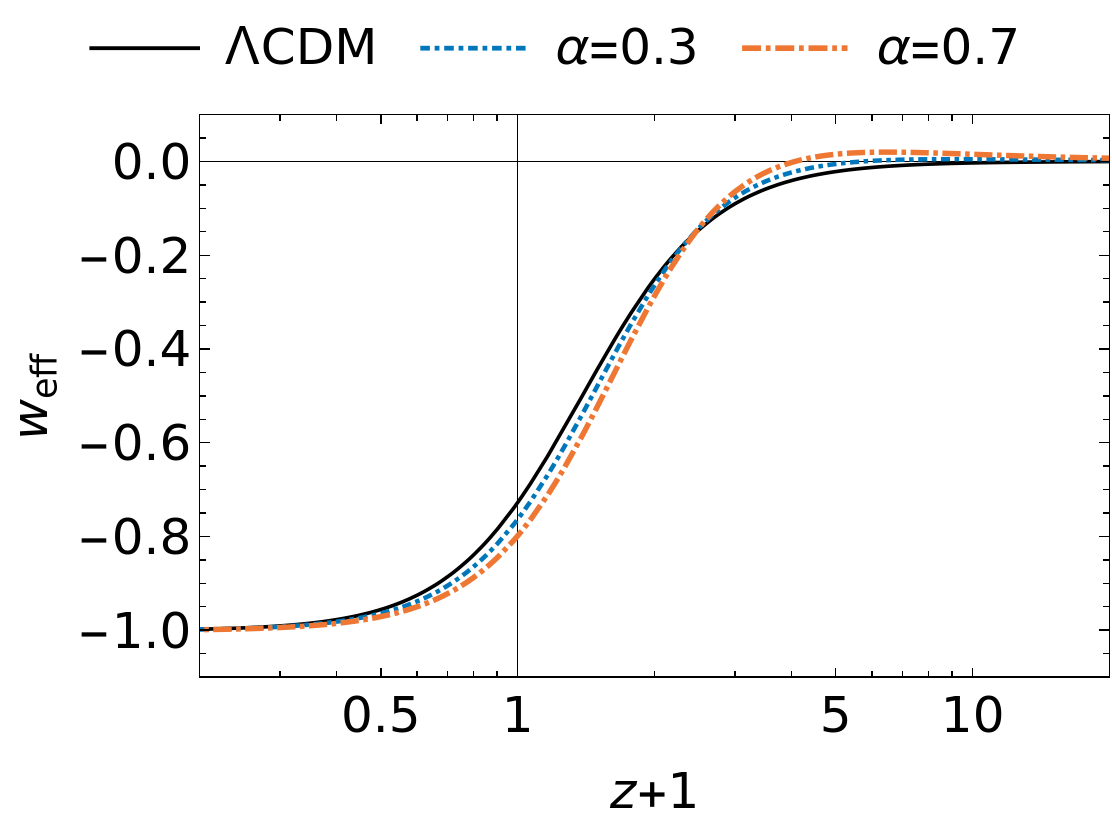}
    \caption{$\Lambda$-nDGP model: evolution of the effective equation of state ($\weff$) with redshift ($z$), as compared to \LCDM. The right panel is a zoomed in version of the left panel.
    In the $\Lambda$-nDGP model, there is a period around $z\sim 10$ when $\weff>0$ before transitioning to dark-energy domination.
    A higher value of $\al$ results in a steeper (quicker) matter--DE transition. The vertical line indicates the present time ($z=0$). The case with the larger $\al$ has the lower $\weff$ at present time.}
   \label{fig:weff-LnDGP}
\end{figure}

Figure~\ref{fig:weff-LnDGP} shows the evolution of the effective equation of state, $\weff$, where we can notice that there is a redshift range around ${z\sim 3 - 10}$ when $\weff$ becomes positive once again, an effect more pronounced for larger $\al$, before transitioning to dark-energy domination. In this case, this transition is faster than in the \LCDM~case, being steeper for a larger value of $\al$. This is connected with a larger value of $\al$ having matter domination era longer than \LCDM\, (see Fig.~\ref{fig:Exact-Hdiff-omegas}). 
We note that in this model it is easy to accommodate the requirement $f_Q>0$ given that the sign is driven by both $\alpha$ and $\gamma$, which are found to be degenerate among each other.

\section{Phenomenology of linear perturbations}\label{Sec:linearpheno}

In this Section, we review a model independent approach to MG models based on parametrizations of the gravitational potentials $\Psi(t,x_i)$ and $\Phi(t,x_i)$. 
The latter are introduced when considering  a perturbed flat FLRW metric which in Newtonian gauge reads 
 \be
    ds^2=-(1+2\Psi)dt^2+a^2(1-2\Phi)\delta_{i}^jdx^idx_j\,.
 \ee 
The parametrization of MG then consists of two coupling functions, ${\mu (a,k)}$ and ${\eta(a,k)}$, in general functions of the scale factor $a$ and the Fourier mode $k$, introduced at the level of linear perturbation equations \cite{Amendola:2007rr,PhysRevD.81.083534,Silvestri:2013ne,2010PhRvD..81j4023P,Amendola:2019laa}:  
\begin{eqnarray}
    \label{eq:Poisson}
    -\frac{k^2}{a^2}\Psi&=& 4 \pi G_N\,\mu (a,k)\,[\rho\Delta+3(\rho+p)\sigma] ,\\
    \frac{k^2}{a^2}[\Phi -\eta(a,k)\Psi] &=& 12\pi G_N\,\mu(a, k)\,(\rho + p) \sigma\,,
\end{eqnarray}
where  the gauge-invariant density contrast is defined as
 \begin{equation}
   \rho\Delta\equiv \rho\delta+3\frac{aH}{k^2}(\rho+p)v  \,,
 \end{equation}
being  $\delta$ the density contrast, $v$ the velocity, and $\sigma$  the matter anisotropic stress. 
The density contrast $\delta$ obeys the continuity equation, which in MG becomes
\begin{equation}
    \label{eq:continuity}
    \ddot\delta + 2H\dot\delta +   4\pi G_N a^2 \rho \,\mu(a,k)\, \delta = 0 \,.
\end{equation}

The function $\mu$ is the effective gravitational coupling, which in the GR limit reduces to $\mu_{GR}=1$. 
In MG we can then have a stronger (or weaker) gravitational interaction according to having $\mu>1$ (or $\mu<1$).
While  $\eta$~\footnote{
    Sometimes this coupling function is also denoted as $\gm$. Here in order to avoid confusion with the parameter $\gm$ of our model we choose the $\eta$ version.}
does not have a connection to observables, another coupling function can be introduced: $\Sigma(a,k)$ which accounts for the lensing effect. This is defined as~\cite{Amendola:2007rr,PhysRevD.81.083534,Silvestri:2013ne,2010PhRvD..81j4023P,Amendola:2019laa}:
\begin{eqnarray}
\frac{k^2}{a^2}(\Phi+\Psi)=-4\pi G_N\,\Sigma(a,k)\,[2\rho\Delta+3(\rho+p)\sigma].  \end{eqnarray}
The three coupling functions are connected in the case of negligible matter anisotropic stress as follows: 
\begin{equation}
  \Sigma=\frac{\mu}{2}(1+\eta)\,.  
\end{equation}
In the GR limit we also have $\Sigma_{GR}=1$ and $\eta_{GR}=1$ implying that $\Phi=\Psi$.

For $f(Q)$-gravity, it has been shown that the  equations up to linear order in perturbation~\cite{Jimenez:2019ovq}, when assuming the quasi-static approximation (QSA)~\footnote{
    In this approximation  the time derivatives of the perturbed quantities are neglected compared with their spatial derivatives and it is a valid  approximation for modes deep inside the Hubble radius \cite{Sawicki:2015zya}.
  },
can be cast in the above forms with $\eta=1$ (so, ${\Sigma=\mu}$), and 
\be
    \label{eq:mu}
    \mu(a)=\frac{1}{f_Q}\,,
\ee
which is  $k$-independent in this approximation.

\begin{figure}
\begin{minipage}[t]{0.49\linewidth}
\includegraphics[width=\textwidth]{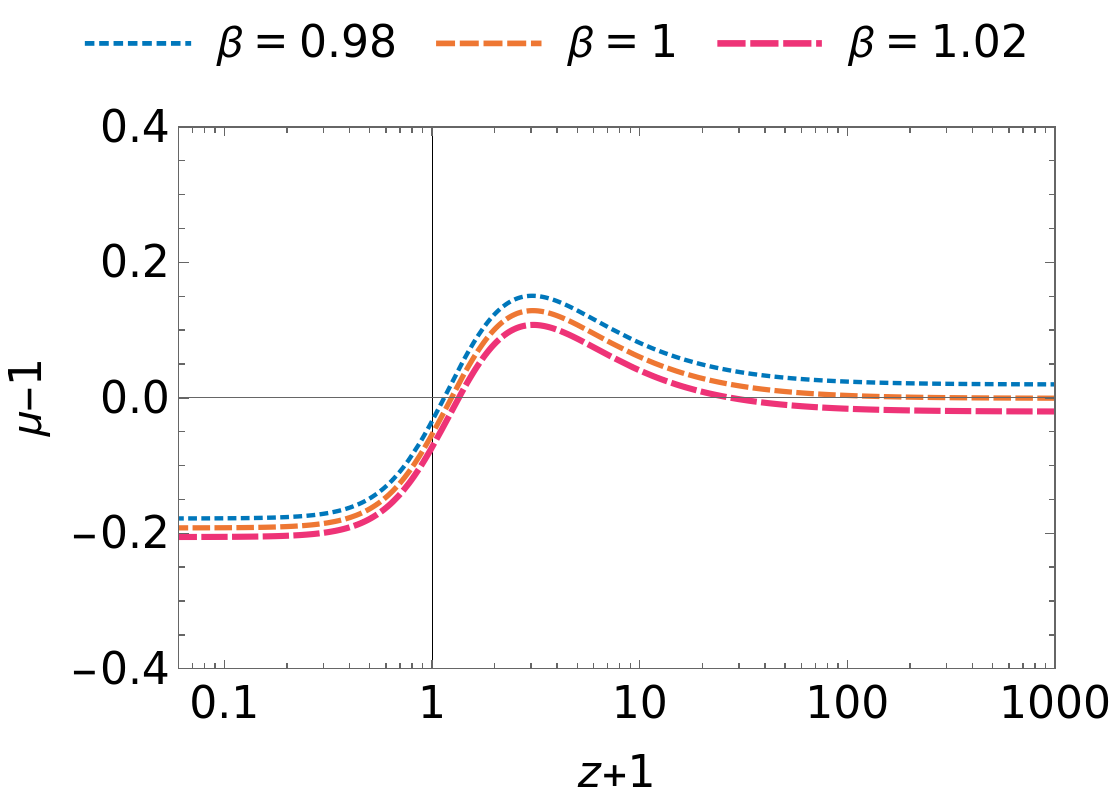}
        \centering
        \textbf{(a)} Varying $\bt$ (fixing $\gm=0.1$, $\al$ by Eq.~\eqref{eq:alpha-Exact}).
        \label{fig:mu_E_b}
\end{minipage}\hfill%
\begin{minipage}[t]{0.49\linewidth}
\includegraphics[width=\textwidth]{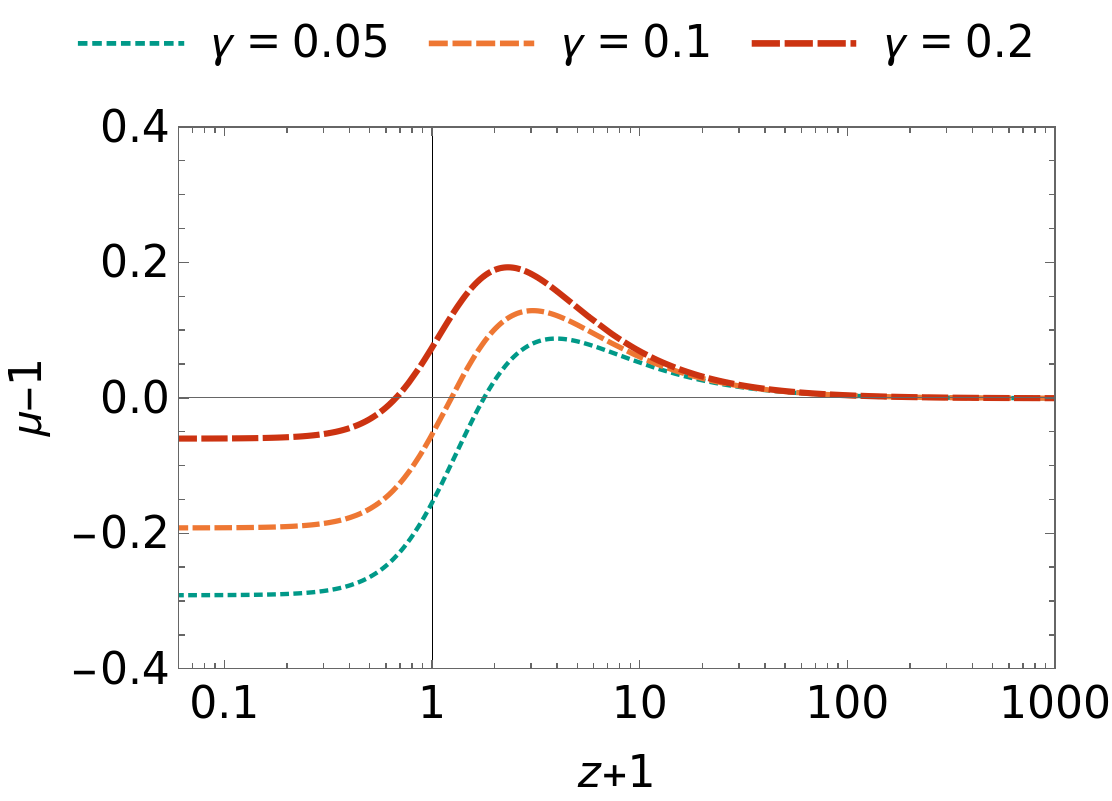}
\centering
        \textbf{(b)} Varying $\gm$ (fixing $\bt=1$, $\al$ by Eq.~\eqref{eq:alpha-Exact}).
        \label{fig:mu_E_g}
\end{minipage}%
    \caption{sLog model: difference between the effective gravitational coupling in the MG model and GR as a function of the redshift, i.e. $\mu(z)-1$, with $\mu$ as given in Eq.~\eqref{eq:mu}. 
    In general, $\bt$ is the only parameter with an effect at early times (high $z$), while $\gm$ has a stronger effect at late times (small $z$). 
    The curves with $\bt=1$ correspond to the subcase of the sDGP model. The vertical line indicates the present time ($z=0$).}
   \label{fig:mu_E}
\end{figure}

\begin{figure}
\begin{minipage}[t]{0.49\linewidth}
\includegraphics[width=\textwidth]{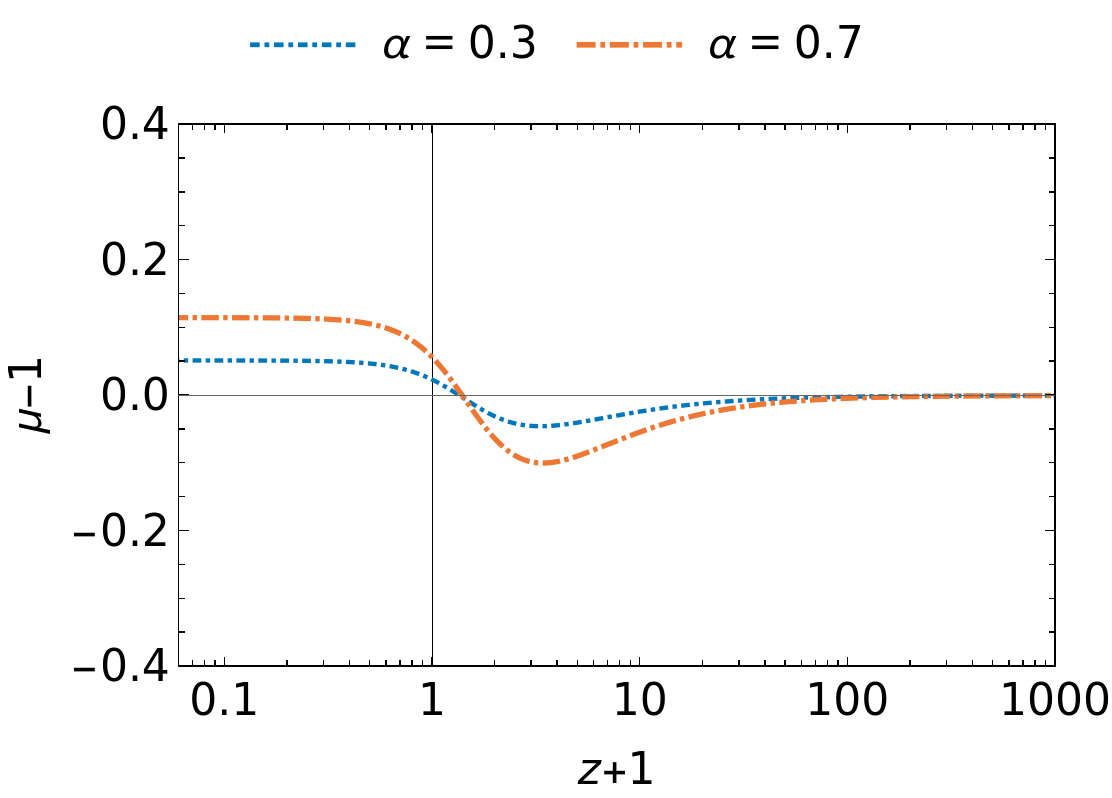}
        \centering
        \textbf{(a)} Varying $\al$ (fixing $\bt=1$, $\gm=0.1$).
        \label{fig:mu_L_a}
\end{minipage}\hfill%
\begin{minipage}[t]{0.49\linewidth}
\includegraphics[width=\textwidth]{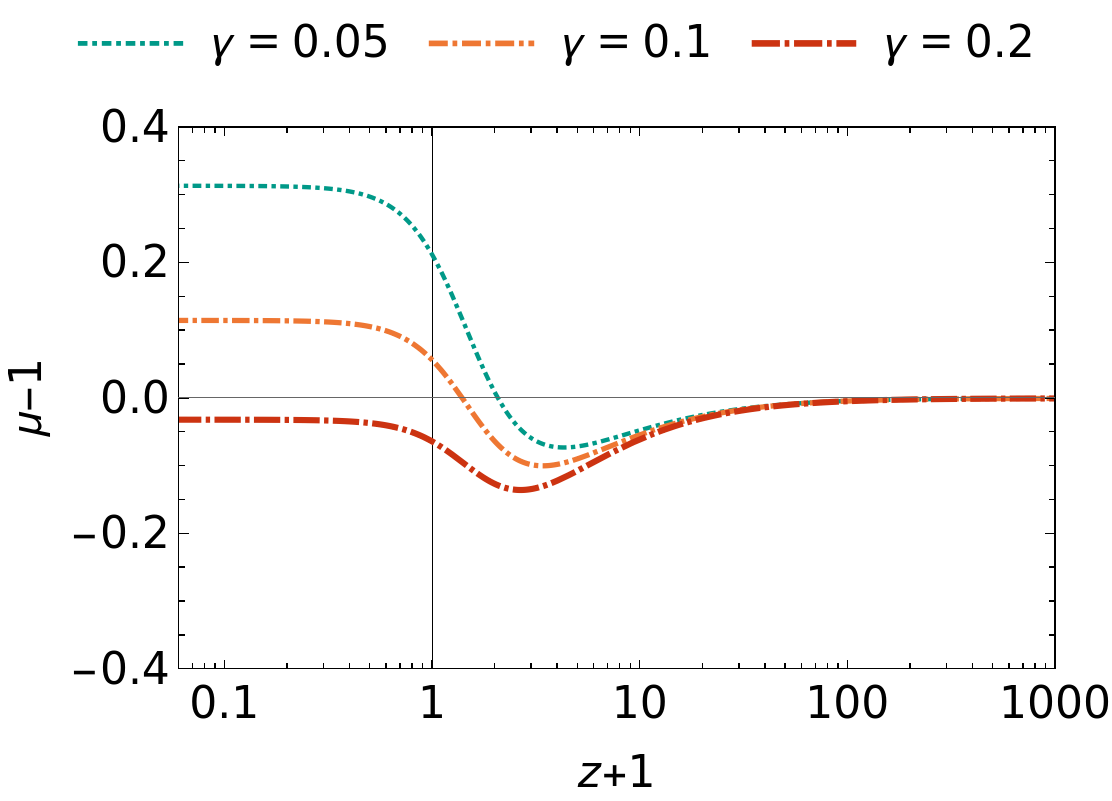}
\centering
        \textbf{(b)} Varying $\gm$ (fixing $\al=+0.7$, $\bt=1$).
        \label{fig:mu_L_g}
\end{minipage}%
    \caption{$\Lambda$-nDGP model: difference between the effective gravitational coupling  in the MG model and GR as a function of the redshift, i.e. $\mu(z)-1$, with $\mu$ as given in Eq.~\eqref{eq:mu}. 
    For this model, $\al>0$ and $\bt=1$ by definition. The vertical line indicates the present time ($z=0$).}
   \label{fig:mu_L}
\end{figure}

In  Figs.~\ref{fig:mu_E}--\ref{fig:mu_L}, we show the difference between the effective gravitational coupling $\mu$ and $\mu_{GR}=1$ as a function of the redshift ($\mu(z)-1$), for the different models we consider.  A  positive difference, ${\mu-1>0}$, means stronger gravity, and vice versa.  
These figures show the impact of the model parameters $\{\al,\bt,\gm\}$ on the gravitational coupling w.r.t. the standard scenario. 
For this analysis, we select a set of values for these parameters considering the  BBN and EDE constraints when applicable as discussed in Sec. \ref{Sec:phenoback} and we apply the requirement $f_Q>0$. 

As seen in Figs.~\ref{fig:mu_E}--\ref{fig:mu_L}, throughout the evolution of the universe, the impact of the different parameters may be  non-trivial. We
 will, however, attempt to draw some general conclusion: 
\begin{itemize}
    \item sLog: $\bt$ is the only parameter that affects the early time physics, with $\bt<1$ giving rise to stronger gravity, and vice versa.
    In general for $z+1\lesssim10$ the gravitational interaction is stronger than the \LCDM\, until close to present time when it can turn into a weaker interaction. The transition depending on both $\bt$ and $\gm$: the smaller $\gm$ is, the sooner the transition from stronger to weaker gravity happen.  

    \item $\Lambda$-nDGP: For $z+1\gtrsim 100$, we do not have any modification w.r.t.~\LCDM. For ${z+1\lesssim 100}$, the gravitational interaction becomes weaker, while close to present time the interaction can turn into stronger gravity w.r.t.~\LCDM, depending on the value of $\gamma$. The difference w.r.t. \LCDM\, is larger for larger values of $\al$. The $\gamma$ parameter can impact on the strength of the gravitational interaction at late times. 
\end{itemize}

In the next Section, we will investigate how the change in the gravitational interaction affects the cosmological observables such as  the CMB TT power spectrum, lensing potential auto-correlation power spectrum, and matter power spectrum.

\subsection{Impact on cosmological observables} \label{Sec:cosmoobservables}

In order to perform investigations of cosmological observables we have created a new patch to  the public~\footnote{
    \nolinkurl{https://github.com/sfu-cosmo/MGCAMB}. We have used the August 2018 version.
} Einstein--Boltzmann code \texttt{MGCAMB}~\cite{Zhao:2008bn, Hojjati:2011ix, Zucca:2019xhg, Wang:2023tjj}. We present in  Appendix~\ref{appdx:MGCAMB} the main modifications to the code. In the following, we provide a phenomenological study of the CMB temperature angular power spectrum, lensing power spectrum, and matter power spectrum. 
We expect a change in the
lensing potential given that $\Sigma=\mu\neq1$, a change in the growth of structure because $\mu$ can be different from 1 and modifies the  Poisson equation which sources the linear perturbations equation for matter density,
a modified shape of the temperature-temperature CMB power spectrum due to the integrated Sachs-Wolfe (ISW) sourced by $\dot{\Psi}+\dot{\Phi}$ and finally
 a shift of the high--$\ell$ peaks in the cases where the background expansion history is modified w.r.t. \LCDM~\cite{Planck:2015bue, Hu:1996vq, Sachs:1967er, Kofman:1985fp, Acquaviva:2005xz, Carbone:2013dna, Peebles:1984ge, 1993MNRAS.262..717B}.

\subsubsection{sLog \& sDGP models}

\begin{figure}
    \centering
    \makebox[\textwidth][c]{\includegraphics[width=1.15\textwidth]{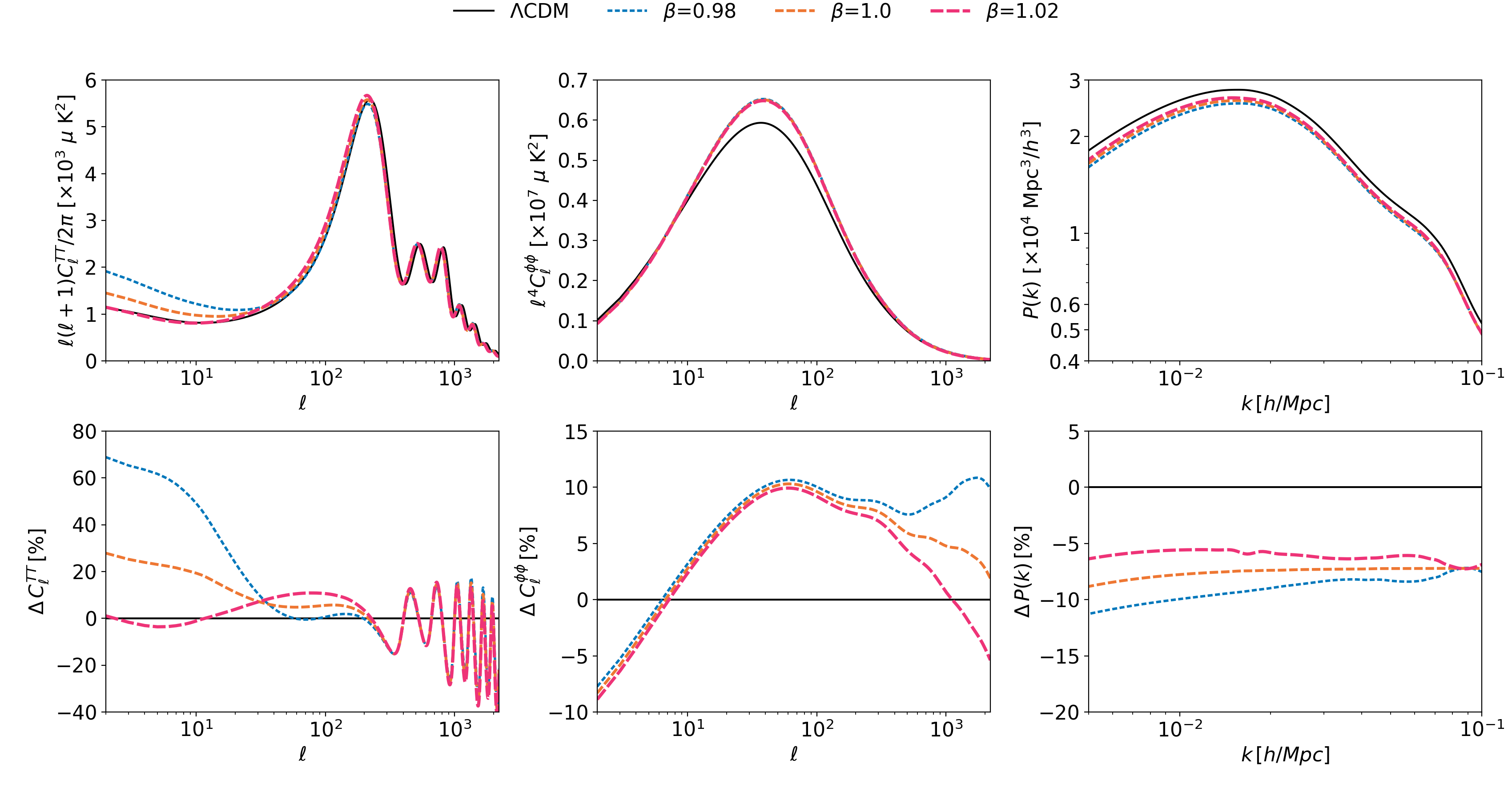}}
    \caption{sLog model: varying $\bt$, fixing $\gm=0.1$.
    \textit{Top panels:} CMB TT power spectrum (top left), lensing potential auto-correlation power spectrum (central panel) and the matter power spectrum (right panel).
    The case $\bt=1$ corresponds to the subcase of the sDGP model. 
    \textit{Bottom panels:} Relative percentage difference of the modified gravity (MG) w.r.t. to \LCDM~($\Delta=\mathrm{MG}/\textrm{\LCDM}-1$).}
    \label{fig:Perturb_E_b}
\end{figure}

\begin{figure}
    \centering
   \includegraphics[width=1.1\textwidth]{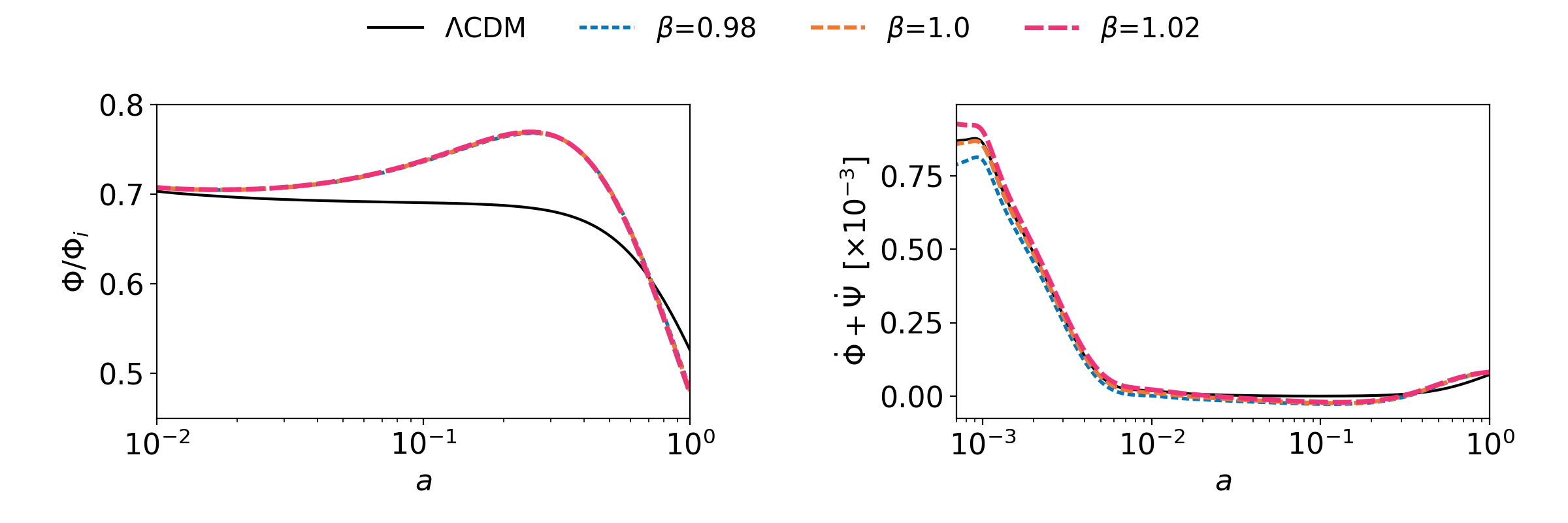}
    \caption{sLog model: varying $\bt$, fixing $\gm=0.1$.
    \textit{Left panel:} Evolution of the gravitational potential $\Phi$ normalized by its initial value $\Phi_i$ for the wavenumber $k =
0.01$ $h$Mpc$^-1$.
    \textit{Right panel:} Evolution of the time derivative $\dot{\Psi}+\dot{\Phi}$.}
    \label{fig:E_b_PhiPsi}
\end{figure}

\begin{figure}
    \centering
    \makebox[\textwidth][c]{\includegraphics[width=1.15\textwidth]{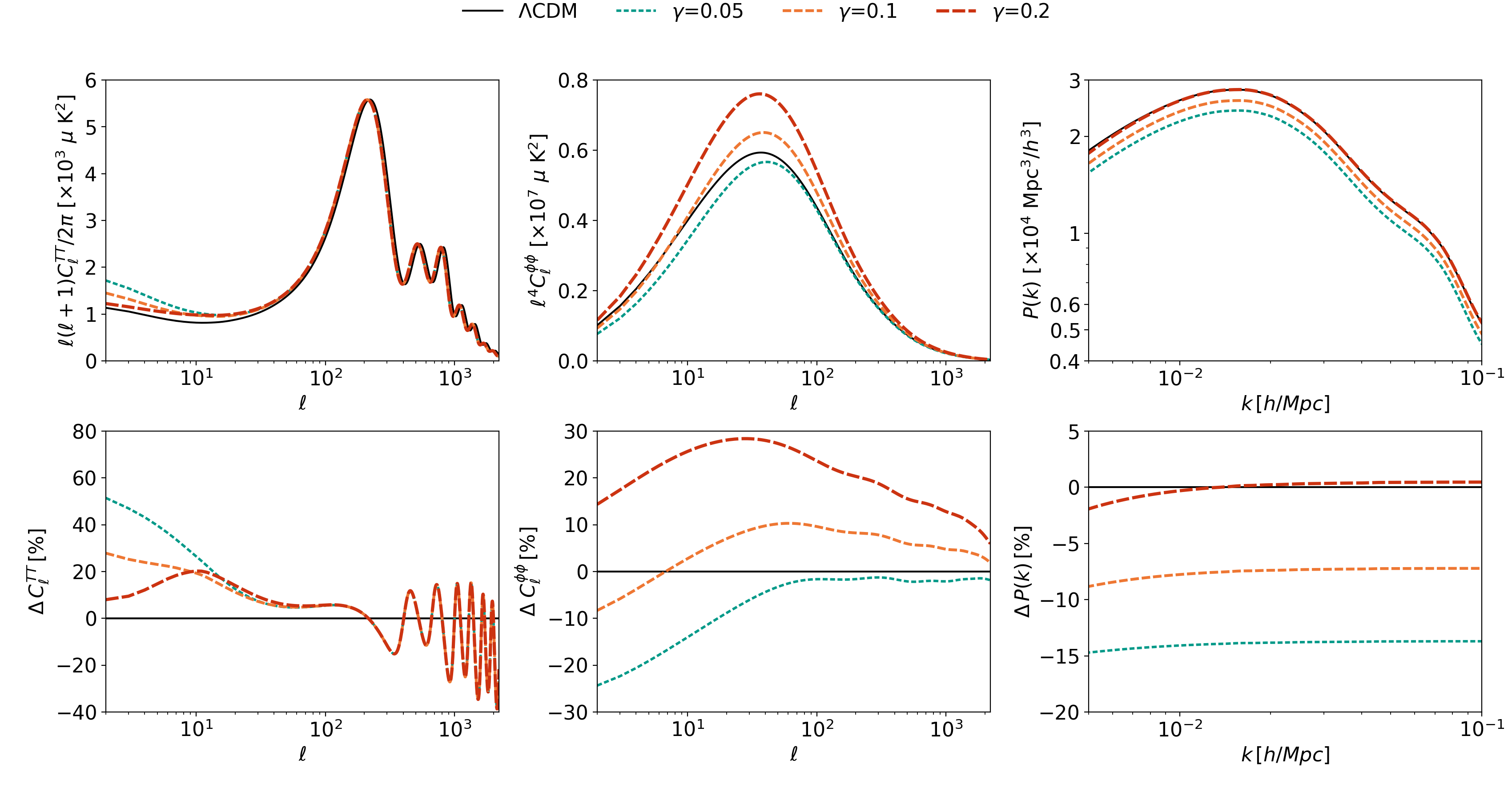}}
    \caption{sDGP model: varying $\gm$, fixing $\bt=1$.
    \textit{Top panels:} CMB TT power spectrum (left), lensing potential auto-correlation power spectrum (central panel) and the matter power spectrum (right panel). Since $\bt=1$, this also corresponds to the subcase of the sDGP model. 
    \textit{Bottom panels:} Relative percentage difference of the modified gravity (MG) w.r.t. to \LCDM~($\Delta=\mathrm{MG}/\textrm{\LCDM}-1$). }
    \label{fig:Perturb_E_g}
\end{figure}

\begin{figure}
    \centering
    \includegraphics[width=0.9\textwidth]{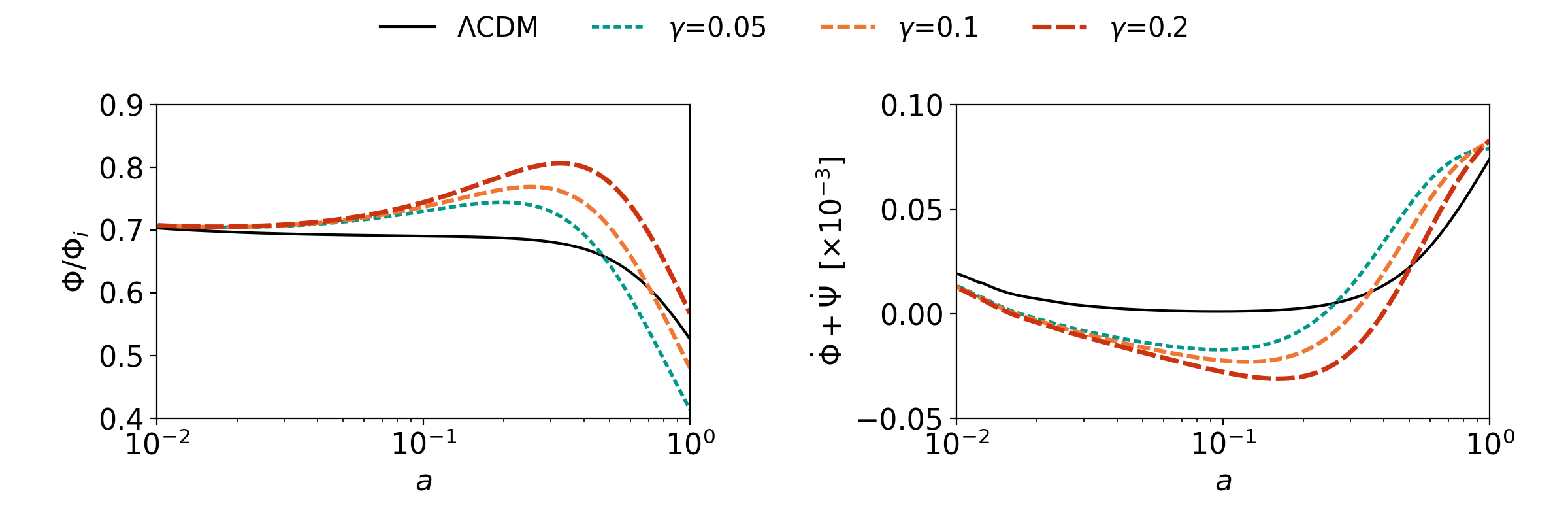}
    \caption{sDGP model: varying $\gm$, fixing $\bt=1$, choosing Fourier modes near ${k\sim 10^{-2}} \,h/\mathrm{Mpc}$ (each line is for a particular Fourier mode within ${k=9.5\pm 0.2\times 10^{-3}} \,h/\mathrm{Mpc}$).
    \textit{Left panel:} evolution of $\Phi$ with scale factor (normalized with the initial value $\Phi_i$). 
    \textit{Right panel:} Evolution of $\dot\Phi+\dot\Psi$ with scale factor.}
    \label{fig:E_g_PhiPsi}
\end{figure}

In this section we present the theoretical predictions for linear cosmological observables of the sLog model, which has two free parameters $\{\bt,\gm\}$ and of the sub-case  sDGP which has $\{\bt=1,\gm\}$.
The results are in Figs.~\ref{fig:Perturb_E_b} and \ref{fig:Perturb_E_g} where  we include the percentage relative difference w.r.t. the standard scenario. 
We also show the evolution of the gravitational potentials $\Phi$ and $\Psi$ in Figs.~\ref{fig:E_b_PhiPsi} and \ref{fig:E_g_PhiPsi}.  We select the following sets of parameters to show their phenomenology:  $\bt=0.98,1,1.02$ with $\gm=0.1$ and $\gamma=0.05, 0.1, 0.2$ with $\bt=1$. 

\paragraph*{TT power spectrum:} 
In Fig.~\ref{fig:Perturb_E_b} left panels, at high--$\ell$ we can notice a shift  in the acoustic peaks towards lower multipoles w.r.t. \LCDM. This is connected to a modified background expansion history (see Fig.~\ref{fig:Exact-Hdiff-omegas}).  
Additionally,  when $\bt>1$  the amplitude of the first acoustic peak is enhanced, while $\bt<1$ suppresses it. This is because $\bt$ impacts the early time ISW effect. Indeed, we notice a change in the evolution of the ISW source, $\dot{\Phi}+\dot{\Psi}$, in Fig.~\ref{fig:E_b_PhiPsi} right panel.   $\beta$ can be seen as a coupling parameter to the matter component, changing the strength of the gravitational interaction, and so affecting the amplitude of the acoustic oscillations. 
At low--$\ell$ in Fig.~\ref{fig:Perturb_E_b}, we notice that the ISW tail is enhanced in the sLog model w.r.t. \LCDM; and the smaller $\bt$ is, the larger is this enhancement.  This is expected given the modifications in the  evolution of the gravitational potentials at late time which impact the late--time ISW effect, see Fig.~\ref{fig:E_b_PhiPsi} right panel. 
Similarly, in Fig.~\ref{fig:Perturb_E_g} when $\beta=1$ and $\gamma$ is varied (which is specifically the case of sDGP), we notice that the smaller $\gm$ is, the more enhanced the power is in the low--$\ell$ tail. This is a consequence of the change due to $\gm$ in the rate of evolution  of the gravitational potentials as shown in Fig.~\ref{fig:E_g_PhiPsi} right panel. We note that the difference in the position of the acoustic peaks at high--$\ell$ is due to having a modified background expansion and it does not depend on $\gm$ since this parameter does not enter in the background evolution.

\paragraph*{Lensing power spectrum:} In Fig.~\ref{fig:Perturb_E_b} central panel when $\beta$ is varied, we notice that the lensing power spectrum is slightly enhanced compared to \LCDM\, with relative difference at most $\sim$10\%. The smaller the $\beta$ is, the more enhanced the lensing power spectrum is. 
When we fix $\beta=1$ and we vary $\gamma$,
see Fig.~\ref{fig:Perturb_E_g} central panel, we can see how the lensing power changes: smaller vales of $\gamma$ lead to a suppression w.r.t. \LCDM\, which might reach 20\%; values larger than $\gamma=0.1$ instead lead to an enhancement that can be $>15\%$. 

\paragraph*{Matter power spectrum.}
The matter power spectrum is suppressed  compared to \LCDM, when  $\beta$ is varied, see Fig.~\ref{fig:Perturb_E_b} right panel. 
The difference in power  can reach 10\%, with the smaller values of $\bt$ producing the lower power. 
When $\beta=1$  and $\gamma$ is varied, see Fig.~\ref{fig:Perturb_E_g}  right panel, we can notice that the smaller values of $\gamma$ produce a suppressed  matter power spectrum compared to \LCDM; however, when $\gamma=0.2$ the matter power spectrum is enhanced wrt \LCDM. 

In conclusion, we note that a degeneracy between $\bt$ and $\gm$ exists, as they can compensate the effects of each other. Nevertheless, the degeneracy might be broken since the parameter $\beta$ can be constrained at background level, where $\gm$ is not present, using background data (e.g. SNIa) but also from CMB data, given the impact this parameter has at high--$\ell$ in the TT power spectrum. We also expect a strong constraint on this parameter given the precision of the CMB data at those angular scales. We note that in this model it is possible to realize the situation in which given a set of values for the parameters the lensing power spectrum is enhanced and the matter power spectrum is suppressed. This property is very interesting because it might account for the excess of lensing power found in the CMB data by Planck \cite{Planck:2018vyg} and at the same time have a lower $\sigma_8$ as suggested by weak lensing measurements and redshift surveys
which prefer a weaker matter clustering than that expected from the standard cosmological model, thus helping in solving  the  $\sigma_8$ tension \cite{DiValentino:2020vvd}.

\subsubsection{$\Lambda$-nDGP model}

\begin{figure}[t!]
    \centering
    \makebox[\textwidth][c]{\includegraphics[width=1.15\textwidth]{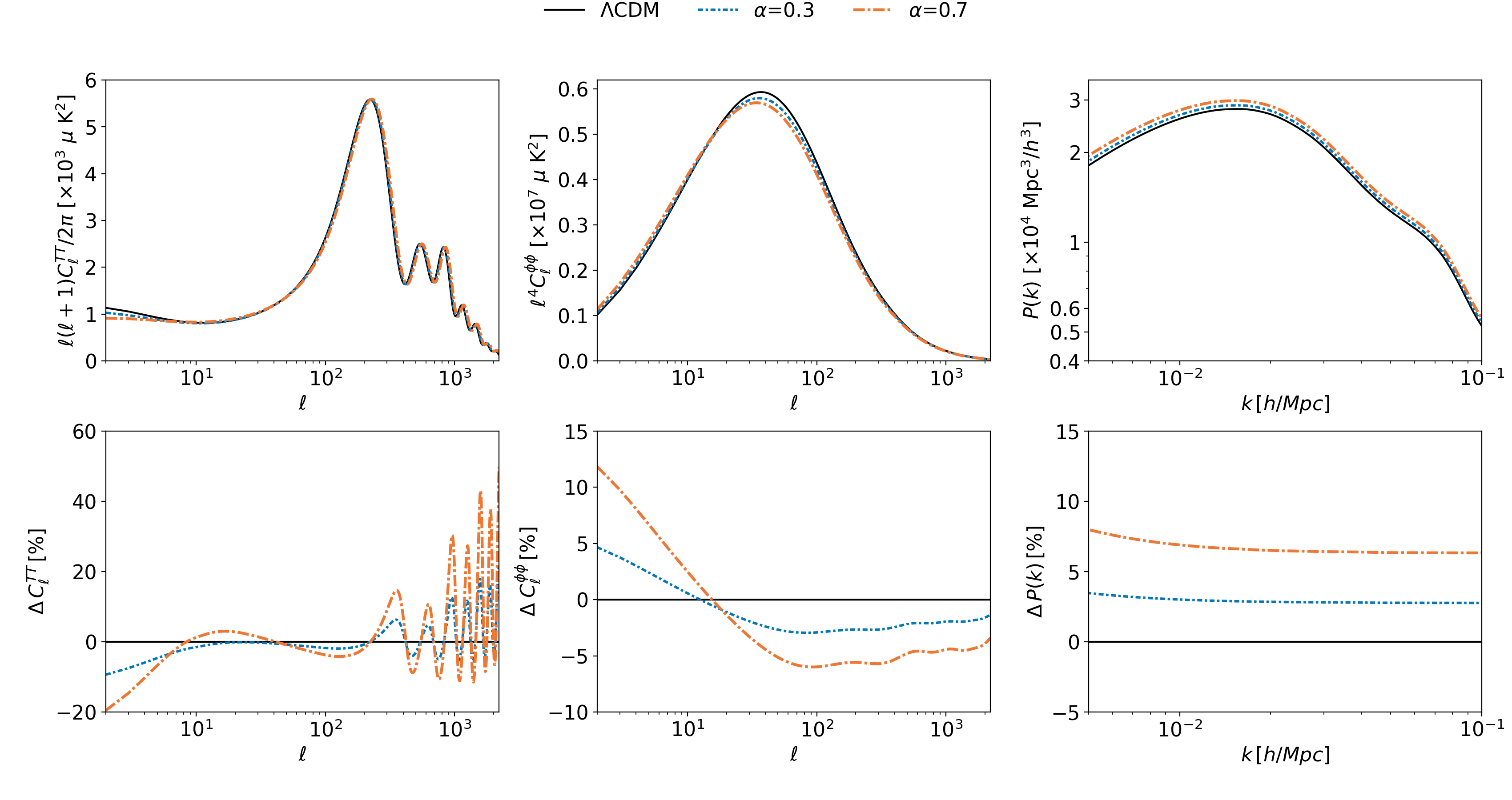}}
    \caption{
    $\Lambda$--nDGP model: varying $\al$, fixing $\bt=1$ and $\gm=0.1$.
    \textit{Top panels:} CMB TT power spectrum (top left), lensing potential auto-correlation power spectrum (central panel) and the matter power spectrum (right panel).
    \textit{Bottom panels:} Relative percentage difference of the modified gravity (MG) w.r.t. to \LCDM~($\Delta=\mathrm{MG}/\textrm{\LCDM}-1$).}
    \label{fig:Perturb_L_a}
\end{figure}

\begin{figure}
    \centering
    \includegraphics[width=0.9\textwidth]{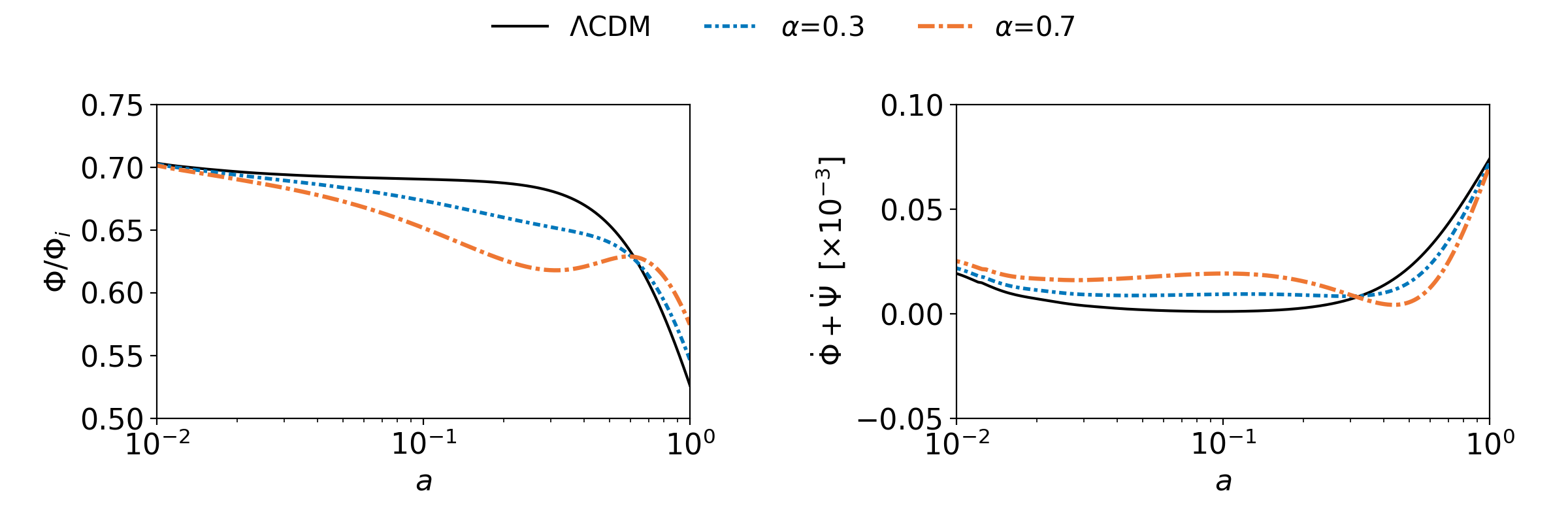}
    \caption{
    $\Lambda$--nDGP model: varying $\al$, fixing $\bt=1$ and $\gm=0.1$, choosing Fourier modes near ${k\sim 10^{-2}} \,h/\mathrm{Mpc}$ (each line is for a particular Fourier mode within ${k=9.6\pm 0.1\times 10^{-3}} \,h/\mathrm{Mpc}$).
    \textit{Left panel:} evolution of $\Phi$ with scale factor (normalized with the initial value $\Phi_i$). 
    \textit{Right panel:} Evolution of $\dot\Phi+\dot\Psi$ with scale factor.}
    \label{fig:L_a_PhiPsi}
\end{figure}

\begin{figure}[t!]
    \centering
    \makebox[\textwidth][c]{\includegraphics[width=1.15\textwidth]{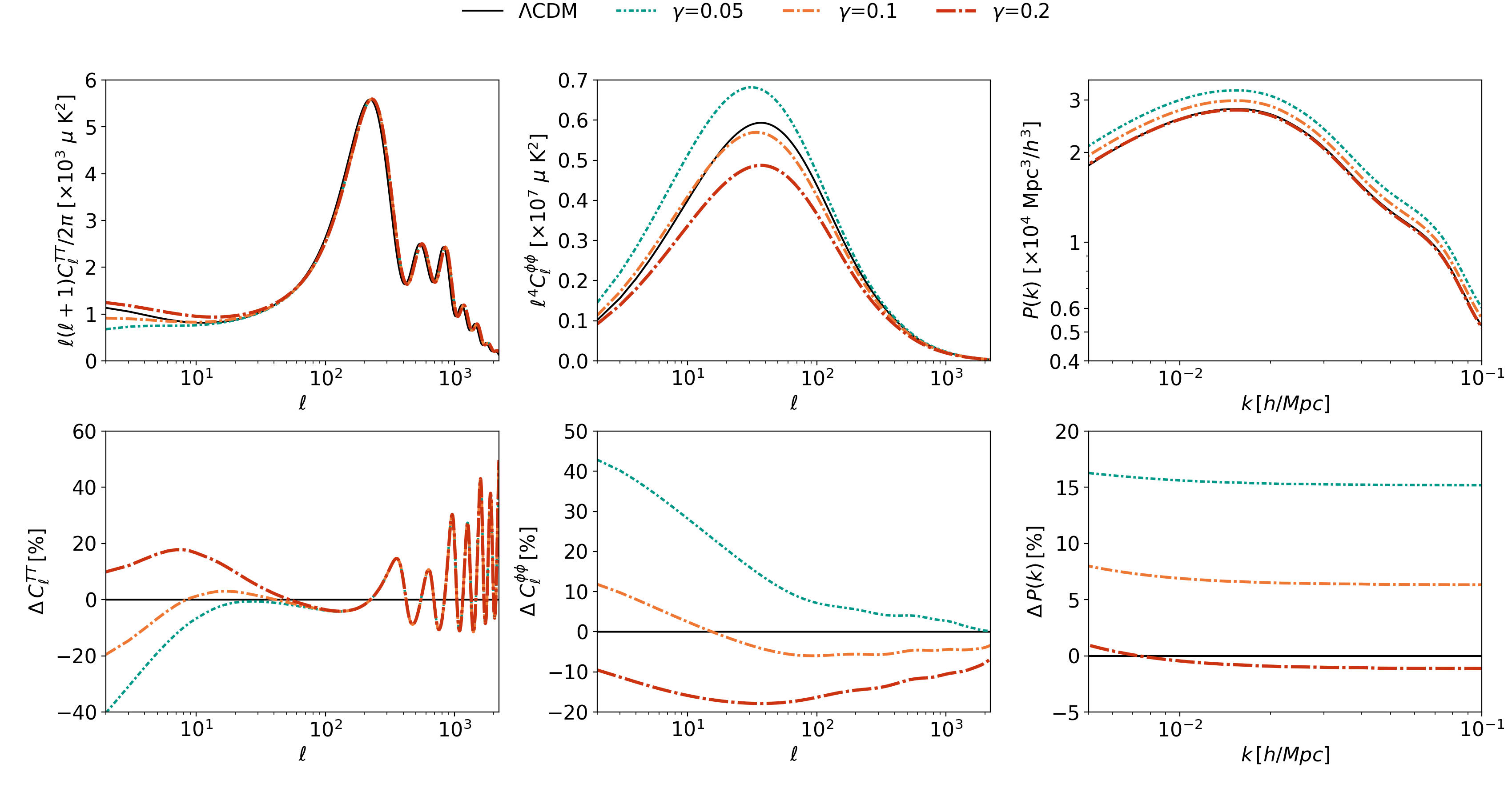}}
    \caption{
    $\Lambda$--nDGP model: varying $\gm$, fixing $\al=+0.7$ and $\bt=1$.
    \textit{Top panels:} CMB TT power spectrum (top left), lensing potential auto-correlation power spectrum (central panel) and the matter power spectrum (right panel).
    \textit{Bottom panels:} Relative percentage difference of the modified gravity (MG) w.r.t. to \LCDM~($\Delta=\mathrm{MG}/\textrm{\LCDM}-1$).}
    \label{fig:Perturb_L_g}
\end{figure}

\begin{figure}
    \centering
    \includegraphics[width=0.9\textwidth]{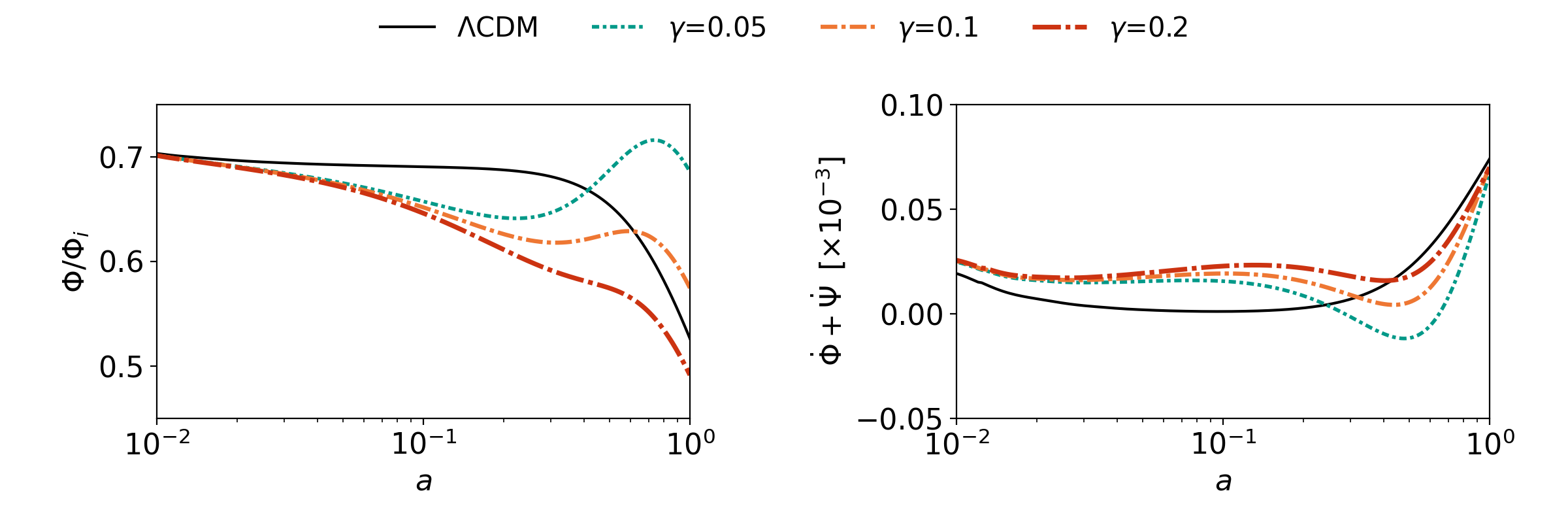}
    \caption{
    $\Lambda$--nDGP model: varying $\gm$, fixing $\al=+0.7$ and $\bt=1$, choosing Fourier modes near ${k\sim 10^{-2}} \,h/\mathrm{Mpc}$ (each line is for a particular Fourier mode within ${k=9.6\pm 0.1\times 10^{-3}} \,h/\mathrm{Mpc}$).
    \textit{Left panel:} evolution of $\Phi$ with scale factor (normalized with the initial value $\Phi_i$). 
    \textit{Right panel:} Evolution of $\dot\Phi+\dot\Psi$ with scale factor.}
    \label{fig:L_g_ap0d7_PhiPsi}
\end{figure}

Let us now discuss the phenomenology at linear scales of the $\Lambda$-nDGP model. We recall that this model has two parameters: $\al$ and $\gm$.
 We show the results in Figs.~\ref{fig:Perturb_L_a}, \ref{fig:L_a_PhiPsi}, \ref{fig:Perturb_L_g} and \ref{fig:L_g_ap0d7_PhiPsi}.
Specifically, we study the following sets of parameters to show their phenomenology: 
$\alpha=0.3,\, 0.7$  with $\gm=0.1$; and $\gamma=0.005,\, 0.1\,, 0.2$ with $\al=0.7$.

\paragraph*{TT power spectrum:}
In Fig.~\ref{fig:Perturb_L_a} top left panel, we can notice that a positive $\al$  suppresses the ISW tail w.r.t.~\LCDM\,(around $10 \%$ suppression with $\al=0.3$ and 20\% with $\al=0.7$). 
That is due to modified rate of evolution  of the  gravitational potentials at late time w.r.t.~\LCDM\, which changes the late--time ISW effect (see Fig.~\ref{fig:L_a_PhiPsi} right panel). 
There is also a  shift toward higher $\ell$ in the TT power spectrum due to the modified background, so the BAO happen at different angular scales. The amplitude of the peaks is not modified because the effective cosmological gravitational coupling at background has $\beta=1$.
In Fig.~\ref{fig:Perturb_L_g} left panels we now see the impact of varying $\gamma$. 
We can notice that depending on the value of $\gm$ we can have either enhancement or suppression of the ISW tail w.r.t. \LCDM~($\sim 20\%$ enhancement for $\gm=0.2$, and suppression for $\gm=0.1,0.05$ with deviations at smallest $\ell$ as large as $\sim 20\%$ and $40\%$, respectively). 
At high--$\ell$, a horizontal shift of the acoustic peaks towards higher $\ell$ is present for all the values of $\gamma$. This is an effect of the modified background introduced by $\alpha=0.7$. 

\paragraph*{Lensing power spectrum:} In the middle panels of Fig.~\ref{fig:Perturb_L_a}, 
when varying $\alpha$ we see a suppressed lensing power spectrum. The larger suppression corresponds to the higher value of $\alpha$. We can expect this phenomenology since in Fig.~\ref{fig:L_a_PhiPsi} left panel the gravitational potential is suppressed w.r.t.~\LCDM\, for most of the time evolution. 
In Fig.~\ref{fig:Perturb_L_g} middle panels, when fixing $\alpha$ and letting $\gamma$ to vary, we notice that the lensing power spectrum can be either enhanced or suppressed. We see that this might be expected by looking at the time evolution of the potential in Fig.~\ref{fig:L_g_ap0d7_PhiPsi} left panel. 
In detail, it is possible to realize an enhanced $C_\ell^{\phi\phi}$ for the smaller value of $\gamma$, with deviation from the \LCDM\, scenario which reaches the 40\%. On the contrary the higher value of $\gamma$ can realize a suppression of power ($\sim 20\%$).

\paragraph*{Matter power spectrum:} In Fig.~\ref{fig:Perturb_L_a} right panel, when varying $\alpha$ we see an enhancement of the matter power spectra, with relative difference w.r.t.~\LCDM\, around $\sim 5-10\%$ , depending on the value of $\alpha$. The higher its value, the larger is the difference.
This is indeed what we expected since at present time $\mu(z=0)>1$ (see Fig.~\ref{fig:mu_L} left panel).
When we let $\gamma$ to vary, instead, we notice that the larger the value of this parameter is, the less clustered the structures are (see Fig.~\ref{fig:Perturb_L_g} right panel). Given the fixed value of $\alpha$ that we chose, we can see that the resulting matter power spectra are all enhanced w.r.t. \LCDM\, except the one for the higher value $\gamma=0.2$. However, let us note that selecting a smaller value for $\alpha$ would push the $P(k)$ down.

We can conclude that there is some  degeneracy between $\alpha$ and $\gamma$.  We can also notice that the smaller value of $\gamma$ can realize a suppressed ISW tail which in general is favoured by CMB data \cite{Peirone:2019aua,Frusciante:2019puu,Atayde:2021pgb}, as well as an excess in the lensing power also expected in CMB data  \cite{Planck:2018vyg}, unfortunately it shows an enhanced matter power spectra which is not supported by weak lensing measurements and redshift surveys \cite{DiValentino:2020vvd}. However this needs further investigation  with actual data since the role of $\alpha$ is not so trivial. A way to break this degeneracy indeed is to consider the background data combined with CMB data which we expect would strongly constrain $\alpha$ give its role in the background equation.

\section{Conclusion}\label{Sec:Conclusion}

In this work we investigated the phenomenology of a logarithmic form of $f(Q)$ gravity resembling DGP--like behaviours, which is characterized by 3 parameters $\al, \bt, \gm$. 
We extended the analysis on this model by investigating not only the background cosmological evolution, as previously done, but considering also the cosmological perturbations at linear level. Additionally, while previous works include an extra dark energy component (a cosmological constant) at background level, in our work, because this logarithmic $f(Q)$ can present self--accelerated behaviour, we do not include any additional dark energy source.
We called this case sLog model. A sub-case of the sLog model is the sDGP model which, with the choice $\beta=1$ and $\al<0$, has the same background equation of the sDGP branch, hence the name. 
On the other hand, a choice of the parameters $\beta=1$ and $\al>0$ requires an additional dark energy component, resembling the nDGP branch: we include a cosmological constant and refer to this case  as the $\Lambda$--nDGP model. 
We found that the $\beta$ parameter in the sLog model determines the evolution at early time, which allows us to impose early dark energy constraints to restrict its viable parameter space. Furthermore, as $\bt$ defines the effective cosmological gravitational coupling it can be related to BBN constraint on the helium abundance. In Eq.~\eqref{eq:constraint-beta-Exact} we showed the combined constraints on this parameter.

In the second part of this work, we have studied the impact of the modification of the gravitational interaction on cosmological observables such as the CMB TT power spectrum, lensing potential auto-correlation power spectrum, and matter power spectrum. In order to do that we have created a new patch to the Einstein--Boltzmann code \texttt{MGCAMB}, where the logarithmic $f(Q)$ model has been implemented. 
In our analysis we found that this logarithmic $f(Q)$ model can have a rich and non-trivial phenomenology at linear level: 
i) the early time physics is controlled by the $\bt$ parameter which is present only in the sLog model; 
ii) degeneracy among the parameters is present ($\beta$ and $\gamma$ in the sLog model and $\al$ and $\gamma$ in the $\Lambda$--nDGP one) which we expect can be broken using a combination of background and CMB data because these data can strongly constrain either $\beta$ in the sLog model and $\al$ in the $\Lambda$--nDGP since, contrary to $\gm$, they enter in the background evolution and shift the high--$\ell$ peaks of the CMB TT power spectrum; 
iii) we also found that in the sDGP case it is possible to have a phenomenology such that the lensing auto-correlation power spectrum  is enhanced and the matter power spectrum is suppressed for the same choice of parameters. These features are of interest in view of understanding the excess of lensing in CMB data and the $\sigma_8$ tension; 
iv) in the $\Lambda$--nDGP model an  enhanced lensing auto-correlation power spectrum can be obtained but at the same time the matter power spectrum is enhanced as well;
v) in the $\Lambda$--nDGP model it is possible to have a suppressed ISW tail which is a feature preferred by CMB data.

We conclude that the Logarithmic form of the $f(Q)$--gravity has peculiar features that can be of interest in solving some of the observational problems of the \LCDM~model. Therefore, the model deserves further investigation in comparing  the dynamics of linear perturbations with observations.

\acknowledgments

We thank F.S.N. Lobo for useful discussion.
T.B.G. is supported by the Funda\c{c}\~{a}o para a Ci\^{e}ncia e a Tecnologia (FCT) through the Fellowship PRT/BD/153354/2021 (IDPASC PT-CERN Grant) and POCH/FSE, and research grants UIDB/04434/2020 and UIDP/04434/2020.
L.A. is supported by the Funda\c{c}\~{a}o para a Ci\^{e}ncia e a Tecnologia (FCT) through the Fellowship grant with ref. number 2022.11152.BD, and research grants UIDB/04434/2020 and UIDP/04434/2020.
N.F. is supported by the Italian Ministry of University and Research (MUR) through the Rita Levi Montalcini project ``Tests of gravity on cosmic scales" with reference PGR19ILFGP.
The authors also acknowledge the FCT project with ref. number PTDC/FIS-AST/0054/2021 and  the COST Action CosmoVerse, CA21136, supported by COST (European Cooperation in Science and Technology).

\appendix

\section{Full derivation of the Logarithmic Lagrangian of $f(Q)$-gravity} 
\label{appdx:fQderivation}

In this Appendix we review the proposal made in Ref.~\cite{Ayuso:2021vtj} to select the form of the $f(Q)$-function. Here the purpose is to show that another constant parameter needs to be considered  which is not present in the original work. For our investigation this parameter is important because it can shape the cosmological observables at linear perturbation level and impact the stability space, while it does not enter directly in the background dynamics.

Let us start with Eq.~\eqref{eq:Fried}, 
which can be cast in the form
\be\label{eq:Fried-boQ-appdx}
    Qf_Q-\frac12f = \ka^2 b(Q) \,,
\ee
where $b(Q)$ represents the energy density, $\rho$ as a function of $Q$.  The equation can be integrated, which has the
 following solution:
\be\label{eq:SOLfQ-boQ-appdx}
    f(Q) = \sqrt{Q} \int \frac{\ka^2 b(Q)}{Q^{3/2}} dQ\,.
\ee
A choice of $b(Q)\propto Q$ gives rise to the usual $H^2$ term in the Friedmann equation, and  adding another term such as $b(Q)\propto \sqrt{Q}$ gives rise to an additional $H$ term, present in modifications to GR, such as the DGP models. Therefore, in Ref.~\cite{Ayuso:2021vtj} it has been proposed 
\be\label{eq:boQ-DGP-appdx}
    b(Q) = \frac{1}{\ka^2}\left(\al'\sqrt{Q} +\bt Q\right) \,,
\ee
where $\al'$ and $\bt$ are constant parameters \footnote{
Here we are denoting one of the parameters with a prime ($\al'$) because later on we will be redefining this parameter ($\al'\rightarrow \al$) such that we will be working with a dimensionless parameter $\al$. While $\bt$ is already dimensionless.
}. 
We can now use this $b(Q)$ in Eq.~\eqref{eq:SOLfQ-boQ-appdx}. 

Let us now show, in detail, the integration between an arbitrary constant $Q_1$ and a variable $Q$, 
\bea
    \label{eq:int-fQ-deriv-appdx}
    &f(Q) &= \sqrt{Q} \int_{Q_1}^{Q} \frac{\left(\al'\sqrt{Q'} +\bt Q'\right)}{Q'^{3/2} } dQ' \nonumber\\
    & &= \sqrt{Q} \left[\al' \ln\left(Q'\right)+2\bt\sqrt{Q'}\right]_{Q_1}^{Q} \nonumber\\
    & &= \al' \sqrt{Q} \left[\ln\left(\frac{Q}{Q_1}\right) -2\frac{\bt}{\al'}\sqrt{Q_1}\right] +2\bt Q \nonumber\\
    & &= \al' \sqrt{Q} \ln\left(\frac{Q}{Q_1 \exp\left(2\frac{\bt}{\al'}\sqrt{Q_1}\right)} \right)  +2\bt Q \nonumber\\
    & &= \al' \sqrt{Q} \ln\left(\gm' Q\right) +2\bt Q \, ,
\eea
where, in the last step, we have defined the arbitrary constant ${\gm'\equiv \left[Q_1 \exp\left(2\frac{\bt}{\al'}\sqrt{Q_1}\right)\right]^{-1}}$.
As we have noted in Footnote~\ref{fn:gammaAyuso}, in Ref.~\cite{Ayuso:2021vtj} the parameter $\gm'$ is not considered and, as such, it is set to one. That is because in their analysis they are interested in studying the background dynamics which is not affected by this parameter [see Eq.~\eqref{eq:Fried-DGP-1}]. In our analysis this parameter enters the perturbations equations, and, therefore, we include it.
Furthermore, since we will be interested in doing numerical analysis, it will be useful to work with dimensionless parameters; to this end, we redefine parameters $\al'$ and $\gm'
$ in the main text [see Eq.~\eqref{eq:parms_nodimension}].

\section {\texttt{MGCAMB} implementation}
\label{appdx:MGCAMB}

In this Appendix, we  describe in detail the implementation of the different cases of the Logarithmic $f(Q)$ model in \texttt{MGCAMB}.

We have made two implementations, within the QSA approach, with the $(\mu,\eta)$ parametrization:
\begin{itemize}
    \item sLog: modifies both the background and linear perturbation equations, with $\al$ constrained by flatness. This includes the sDGP model by setting $\bt=1$.
    
    \item $\Lambda$-nDGP: modifies both the background and linear perturbation equations, with DE including a cosmological constant component.
\end{itemize}

Hereafter we will work in conformal time, $\tau$.

\subsection{Background evolution}

At background level, \texttt{MGCAMB} requires that we specify the Hubble evolution, the density and the pressure of the dark energy fluid as they  appear in the Friedmann equations:
\begin{eqnarray}
    &&\Hc^2=\frac{\ka^2 a^2}{3}\left(\rho+\rho_{DE} \right)\,,\\
     &&\Hc'= - \frac{\ka^2 a^2}{2}  (\rho+p+\rho_{DE}+p_{DE})  +\Hc^2      \,,
\end{eqnarray}
where $\Hc= aH$ is the the conformal Hubble parameter and a prime stands for a derivative w.r.t. conformal time $\tau$ (for which $d\tau = a^{-1}dt$). 

In detail for the different cases: 
\begin{itemize}
    \item sLog: 
    we implement the following DE energy density and pressure [cf. Eqs.~\eqref{eq:rhoQ-Log} and~\eqref{eq:pQ-Log}]: 
    \begin{eqnarray}
        &\rho_{DE}&\equiv \rho_Q =\frac{1}{\ka^2a^2}3\l[\l(1-\bt\r)\Hc^2 -\al a H_0 \Hc \r] \,, \\
        \label{eq:rhoDE-Exact}  
        &p_{DE} &\equiv p_Q =\frac{1}{\ka^2a^2}\l[-\l(1 -\bt\r)\l(2\Hc' +\Hc^2\r) +\al a H_0 \l(\dfrac{\Hc'}{\Hc} +2\Hc \r) \r] \,, 
        \label{eq:pDE-Exact}
    \end{eqnarray}
    using the Friedmann solution of this model [cf. Eq.~\eqref{eq:H-Exact}]:
    \bea
        && \Hc=\frac{H_0}{2\bt}\l[ -\al a +\sqrt{ \al^2 a^2 +4\bt \l(\frac{\Omm}{a} +\frac{\Omr}{a^2}\r)}\r] \,,\\
        && \Hc' = \frac{H_0  \Hc}{2\bt} \l( -\al a + \frac{\al^2 a^2 - 2\bt \l( \frac{\Omm}{a} + 2 \frac{\Omr}{a^2}\r) } {\sqrt{ \al^2 a^2 + 4\bt \l(\frac{\Omm}{a}+\frac{\Omr}{a^2}\r) } }\r) \,,
    \eea
    where $\al =  \Omm+\Omr-\bt$.
    
    \item sDGP: sLog implementation, fixing $\bt=1$.
    
    \item $\Lambda$-nDGP: 
    we implement the following DE energy density and pressure : 
    \begin{eqnarray}
        &\rho_{DE}&\equiv \rho_Q + \rho_\Lambda = \frac{3}{\ka^2 a^2} \l[\l(1-\bt\r) \Hc^2 -\al a H_0 \Hc  + H_0^2 \OmL a^2\r] \,, \\
        \label{eq:rhoDE-Exact-appendix}  
        &p_{DE} &\equiv p_Q + p_\Lambda=\frac{1}{\ka^2a^2}\l[-\l(1 -\bt\r)\l(2\Hc' +\Hc^2\r) +\al a H_0 \l(\dfrac{\Hc'}{\Hc} +2\Hc \r) -3 H_0 \OmL a^2\r] \,, 
        \label{eq:pDE-Exact-appendix}
    \end{eqnarray}
    using the Friedmann solution of this model:
    \bea
        && \Hc=\frac{H_0}{2\bt}\l[ -\al a +\sqrt{ \al^2 a^2 +4\bt \l(\frac{\Omm}{a} +\frac{\Omr}{a^2} + \OmL a^2 \r)}\r] \,,\\
        && \Hc' = \frac{H_0  \Hc}{2\bt} \l( -\al a + \frac{\al^2 a^2 - 2\bt \l( \frac{\Omm}{a} + 2 \frac{\Omr}{a^2} -2\OmL a^2 \r) } {\sqrt{ \al^2 a^2 + 4\bt \l(\frac{\Omm}{a}+\frac{\Omr}{a^2} + \OmL a^2\r) } }\r) \,,
    \eea
where $\OmL=\al+\bt-\Omm-\Omr$, and setting $\bt=1$.
    
\end{itemize}

\subsection{Evolution of the perturbations}

\texttt{MGCAMB} implements the formalism described in Sec.~\ref{Sec:linearpheno} at linear perturbations level. 
As such it requires an expression for $\mu$ and $\eta$ and their time derivatives as function of the Hubble rate and matter densities. 
We recall that $\eta=1$ for $f(Q)$-gravity in the QSA.
For the coupling function $\mu$ ($=1/f_Q$), we implement the following:
\be
    \label{eq:mu-conformal}
    \mu (\tau )= \frac{1}{\frac{\al  H_0 a}{2 \Hc} \left(\frac{1}{2} \log \left(\frac{\gm  \Hc ^2}{H_0^2 a^2}\right)+1\right)+\bt }\,,
\ee

\be
    \label{eq:muprime-conformal}
    \mu'(\tau) = - \mu^2 \frac{\al  H_0 a \left(\Hc^2-\Hc^\prime\right) \log \left(\frac{\gm  \Hc^2}{H_0^2 a^2}\right)}{4 \Hc^2}  \,.
\ee

With the following considerations for each case we study:
\begin{itemize}
    \item sLog: $\al = \Omm+\Omr-\bt$.

    \item sDGP: sLog implementation, fixing $\bt=1$.
    
    \item $\Lambda$-nDGP: $\al>0$ and $\bt=1$.

\end{itemize}


\bibliography{sample}

\begin{thebibliography}{89}%
\makeatletter
\providecommand \@ifxundefined [1]{%
 \@ifx{#1\undefined}
}%
\providecommand \@ifnum [1]{%
 \ifnum #1\expandafter \@firstoftwo
 \else \expandafter \@secondoftwo
 \fi
}%
\providecommand \@ifx [1]{%
 \ifx #1\expandafter \@firstoftwo
 \else \expandafter \@secondoftwo
 \fi
}%
\providecommand \natexlab [1]{#1}%
\providecommand \enquote  [1]{``#1''}%
\providecommand \bibnamefont  [1]{#1}%
\providecommand \bibfnamefont [1]{#1}%
\providecommand \citenamefont [1]{#1}%
\providecommand \href@noop [0]{\@secondoftwo}%
\providecommand \href [0]{\begingroup \@sanitize@url \@href}%
\providecommand \@href[1]{\@@startlink{#1}\@@href}%
\providecommand \@@href[1]{\endgroup#1\@@endlink}%
\providecommand \@sanitize@url [0]{\catcode `\\12\catcode `\$12\catcode
  `\&12\catcode `\#12\catcode `\^12\catcode `\_12\catcode `\%12\relax}%
\providecommand \@@startlink[1]{}%
\providecommand \@@endlink[0]{}%
\providecommand \url  [0]{\begingroup\@sanitize@url \@url }%
\providecommand \@url [1]{\endgroup\@href {#1}{\urlprefix }}%
\providecommand \urlprefix  [0]{URL }%
\providecommand \Eprint [0]{\href }%
\providecommand \doibase [0]{http://dx.doi.org/}%
\providecommand \selectlanguage [0]{\@gobble}%
\providecommand \bibinfo  [0]{\@secondoftwo}%
\providecommand \bibfield  [0]{\@secondoftwo}%
\providecommand \translation [1]{[#1]}%
\providecommand \BibitemOpen [0]{}%
\providecommand \bibitemStop [0]{}%
\providecommand \bibitemNoStop [0]{.\EOS\space}%
\providecommand \EOS [0]{\spacefactor3000\relax}%
\providecommand \BibitemShut  [1]{\csname bibitem#1\endcsname}%
\let\auto@bib@innerbib\@empty
\bibitem [{\citenamefont {Beltr\'an~Jim\'enez}\ \emph
  {et~al.}(2018{\natexlab{a}})\citenamefont {Beltr\'an~Jim\'enez},
  \citenamefont {Heisenberg},\ and\ \citenamefont
  {Koivisto}}]{BeltranJimenez:2017tkd}%
  \BibitemOpen
  \bibfield  {author} {\bibinfo {author} {\bibfnamefont {J.}~\bibnamefont
  {Beltr\'an~Jim\'enez}}, \bibinfo {author} {\bibfnamefont {L.}~\bibnamefont
  {Heisenberg}}, \ and\ \bibinfo {author} {\bibfnamefont {T.}~\bibnamefont
  {Koivisto}},\ }\href {\doibase 10.1103/PhysRevD.98.044048} {\bibfield
  {journal} {\bibinfo  {journal} {Phys. Rev. D}\ }\textbf {\bibinfo {volume}
  {98}},\ \bibinfo {pages} {044048} (\bibinfo {year} {2018}{\natexlab{a}})},\
  \Eprint {http://arxiv.org/abs/1710.03116} {arXiv:1710.03116 [gr-qc]}
  \BibitemShut {NoStop}%
\bibitem [{\citenamefont {Beltr\'an~Jim\'enez}\ \emph
  {et~al.}(2018{\natexlab{b}})\citenamefont {Beltr\'an~Jim\'enez},
  \citenamefont {Heisenberg},\ and\ \citenamefont
  {Koivisto}}]{BeltranJimenez:2018vdo}%
  \BibitemOpen
  \bibfield  {author} {\bibinfo {author} {\bibfnamefont {J.}~\bibnamefont
  {Beltr\'an~Jim\'enez}}, \bibinfo {author} {\bibfnamefont {L.}~\bibnamefont
  {Heisenberg}}, \ and\ \bibinfo {author} {\bibfnamefont {T.~S.}\ \bibnamefont
  {Koivisto}},\ }\href {\doibase 10.1088/1475-7516/2018/08/039} {\bibfield
  {journal} {\bibinfo  {journal} {JCAP}\ }\textbf {\bibinfo {volume} {08}},\
  \bibinfo {pages} {039} (\bibinfo {year} {2018}{\natexlab{b}})},\ \Eprint
  {http://arxiv.org/abs/1803.10185} {arXiv:1803.10185 [gr-qc]} \BibitemShut
  {NoStop}%
\bibitem [{\citenamefont {Maluf}(2013)}]{Maluf:2013gaa}%
  \BibitemOpen
  \bibfield  {author} {\bibinfo {author} {\bibfnamefont {J.~W.}\ \bibnamefont
  {Maluf}},\ }\href {\doibase 10.1002/andp.201200272} {\bibfield  {journal}
  {\bibinfo  {journal} {Annalen Phys.}\ }\textbf {\bibinfo {volume} {525}},\
  \bibinfo {pages} {339} (\bibinfo {year} {2013})},\ \Eprint
  {http://arxiv.org/abs/1303.3897} {arXiv:1303.3897 [gr-qc]} \BibitemShut
  {NoStop}%
\bibitem [{\citenamefont {Nester}\ and\ \citenamefont
  {Yo}(1999)}]{Nester:1998mp}%
  \BibitemOpen
  \bibfield  {author} {\bibinfo {author} {\bibfnamefont {J.~M.}\ \bibnamefont
  {Nester}}\ and\ \bibinfo {author} {\bibfnamefont {H.-J.}\ \bibnamefont
  {Yo}},\ }\href@noop {} {\bibfield  {journal} {\bibinfo  {journal} {Chin. J.
  Phys.}\ }\textbf {\bibinfo {volume} {37}},\ \bibinfo {pages} {113} (\bibinfo
  {year} {1999})},\ \Eprint {http://arxiv.org/abs/gr-qc/9809049}
  {arXiv:gr-qc/9809049} \BibitemShut {NoStop}%
\bibitem [{\citenamefont {Beltr\'an~Jim\'enez}\ \emph
  {et~al.}(2020{\natexlab{a}})\citenamefont {Beltr\'an~Jim\'enez},
  \citenamefont {Heisenberg}, \citenamefont {Koivisto},\ and\ \citenamefont
  {Pekar}}]{BeltranJimenez:2019tme}%
  \BibitemOpen
  \bibfield  {author} {\bibinfo {author} {\bibfnamefont {J.}~\bibnamefont
  {Beltr\'an~Jim\'enez}}, \bibinfo {author} {\bibfnamefont {L.}~\bibnamefont
  {Heisenberg}}, \bibinfo {author} {\bibfnamefont {T.~S.}\ \bibnamefont
  {Koivisto}}, \ and\ \bibinfo {author} {\bibfnamefont {S.}~\bibnamefont
  {Pekar}},\ }\href {\doibase 10.1103/PhysRevD.101.103507} {\bibfield
  {journal} {\bibinfo  {journal} {Phys. Rev. D}\ }\textbf {\bibinfo {volume}
  {101}},\ \bibinfo {pages} {103507} (\bibinfo {year} {2020}{\natexlab{a}})},\
  \Eprint {http://arxiv.org/abs/1906.10027} {arXiv:1906.10027 [gr-qc]}
  \BibitemShut {NoStop}%
\bibitem [{\citenamefont {Weinberg}(1989)}]{Weinberg:1988cp}%
  \BibitemOpen
  \bibfield  {author} {\bibinfo {author} {\bibfnamefont {S.}~\bibnamefont
  {Weinberg}},\ }\href {\doibase 10.1103/RevModPhys.61.1} {\bibfield  {journal}
  {\bibinfo  {journal} {Rev. Mod. Phys.}\ }\textbf {\bibinfo {volume} {61}},\
  \bibinfo {pages} {1} (\bibinfo {year} {1989})}\BibitemShut {NoStop}%
\bibitem [{\citenamefont {Carroll}(2001)}]{Carroll:2000fy}%
  \BibitemOpen
  \bibfield  {author} {\bibinfo {author} {\bibfnamefont {S.~M.}\ \bibnamefont
  {Carroll}},\ }\href {\doibase 10.12942/lrr-2001-1} {\bibfield  {journal}
  {\bibinfo  {journal} {Living Rev. Rel.}\ }\textbf {\bibinfo {volume} {4}},\
  \bibinfo {pages} {1} (\bibinfo {year} {2001})},\ \Eprint
  {http://arxiv.org/abs/astro-ph/0004075} {arXiv:astro-ph/0004075} \BibitemShut
  {NoStop}%
\bibitem [{\citenamefont {Velten}\ \emph {et~al.}(2014)\citenamefont {Velten},
  \citenamefont {vom Marttens},\ and\ \citenamefont
  {Zimdahl}}]{Velten:2014nra}%
  \BibitemOpen
  \bibfield  {author} {\bibinfo {author} {\bibfnamefont {H.~E.~S.}\
  \bibnamefont {Velten}}, \bibinfo {author} {\bibfnamefont {R.~F.}\
  \bibnamefont {vom Marttens}}, \ and\ \bibinfo {author} {\bibfnamefont
  {W.}~\bibnamefont {Zimdahl}},\ }\href {\doibase
  10.1140/epjc/s10052-014-3160-4} {\bibfield  {journal} {\bibinfo  {journal}
  {Eur. Phys. J. C}\ }\textbf {\bibinfo {volume} {74}},\ \bibinfo {pages}
  {3160} (\bibinfo {year} {2014})},\ \Eprint {http://arxiv.org/abs/1410.2509}
  {arXiv:1410.2509 [astro-ph.CO]} \BibitemShut {NoStop}%
\bibitem [{\citenamefont {Joyce}\ \emph {et~al.}(2015)\citenamefont {Joyce},
  \citenamefont {Jain}, \citenamefont {Khoury},\ and\ \citenamefont
  {Trodden}}]{Joyce:2014kja}%
  \BibitemOpen
  \bibfield  {author} {\bibinfo {author} {\bibfnamefont {A.}~\bibnamefont
  {Joyce}}, \bibinfo {author} {\bibfnamefont {B.}~\bibnamefont {Jain}},
  \bibinfo {author} {\bibfnamefont {J.}~\bibnamefont {Khoury}}, \ and\ \bibinfo
  {author} {\bibfnamefont {M.}~\bibnamefont {Trodden}},\ }\href {\doibase
  10.1016/j.physrep.2014.12.002} {\bibfield  {journal} {\bibinfo  {journal}
  {Phys. Rept.}\ }\textbf {\bibinfo {volume} {568}},\ \bibinfo {pages} {1}
  (\bibinfo {year} {2015})},\ \Eprint {http://arxiv.org/abs/1407.0059}
  {arXiv:1407.0059 [astro-ph.CO]} \BibitemShut {NoStop}%
\bibitem [{\citenamefont {Riess}\ \emph {et~al.}(2019)\citenamefont {Riess},
  \citenamefont {Casertano}, \citenamefont {Yuan}, \citenamefont {Macri},\ and\
  \citenamefont {Scolnic}}]{Riess:2019cxk}%
  \BibitemOpen
  \bibfield  {author} {\bibinfo {author} {\bibfnamefont {A.~G.}\ \bibnamefont
  {Riess}}, \bibinfo {author} {\bibfnamefont {S.}~\bibnamefont {Casertano}},
  \bibinfo {author} {\bibfnamefont {W.}~\bibnamefont {Yuan}}, \bibinfo {author}
  {\bibfnamefont {L.~M.}\ \bibnamefont {Macri}}, \ and\ \bibinfo {author}
  {\bibfnamefont {D.}~\bibnamefont {Scolnic}},\ }\href {\doibase
  10.3847/1538-4357/ab1422} {\bibfield  {journal} {\bibinfo  {journal}
  {Astrophys. J.}\ }\textbf {\bibinfo {volume} {876}},\ \bibinfo {pages} {85}
  (\bibinfo {year} {2019})},\ \Eprint {http://arxiv.org/abs/1903.07603}
  {arXiv:1903.07603 [astro-ph.CO]} \BibitemShut {NoStop}%
\bibitem [{\citenamefont {Wong}\ \emph {et~al.}(2020)\citenamefont {Wong} \emph
  {et~al.}}]{Wong:2019kwg}%
  \BibitemOpen
  \bibfield  {author} {\bibinfo {author} {\bibfnamefont {K.~C.}\ \bibnamefont
  {Wong}} \emph {et~al.},\ }\href {\doibase 10.1093/mnras/stz3094} {\bibfield
  {journal} {\bibinfo  {journal} {Mon. Not. Roy. Astron. Soc.}\ }\textbf
  {\bibinfo {volume} {498}},\ \bibinfo {pages} {1420} (\bibinfo {year}
  {2020})},\ \Eprint {http://arxiv.org/abs/1907.04869} {arXiv:1907.04869
  [astro-ph.CO]} \BibitemShut {NoStop}%
\bibitem [{\citenamefont {Freedman}\ \emph {et~al.}(2019)\citenamefont
  {Freedman} \emph {et~al.}}]{Freedman:2019jwv}%
  \BibitemOpen
  \bibfield  {author} {\bibinfo {author} {\bibfnamefont {W.~L.}\ \bibnamefont
  {Freedman}} \emph {et~al.},\ }\href {\doibase 10.3847/1538-4357/ab2f73}
  {\bibfield  {journal} {\bibinfo  {journal} {Astrophys. J.}\ }\textbf
  {\bibinfo {volume} {882}},\ \bibinfo {pages} {34} (\bibinfo {year} {2019})},\
  \Eprint {http://arxiv.org/abs/1907.05922} {arXiv:1907.05922 [astro-ph.CO]}
  \BibitemShut {NoStop}%
\bibitem [{\citenamefont {Di~Valentino}\ \emph
  {et~al.}(2021{\natexlab{a}})\citenamefont {Di~Valentino} \emph
  {et~al.}}]{DiValentino:2020zio}%
  \BibitemOpen
  \bibfield  {author} {\bibinfo {author} {\bibfnamefont {E.}~\bibnamefont
  {Di~Valentino}} \emph {et~al.},\ }\href {\doibase
  10.1016/j.astropartphys.2021.102605} {\bibfield  {journal} {\bibinfo
  {journal} {Astropart. Phys.}\ }\textbf {\bibinfo {volume} {131}},\ \bibinfo
  {pages} {102605} (\bibinfo {year} {2021}{\natexlab{a}})},\ \Eprint
  {http://arxiv.org/abs/2008.11284} {arXiv:2008.11284 [astro-ph.CO]}
  \BibitemShut {NoStop}%
\bibitem [{\citenamefont {Kuijken}\ \emph {et~al.}(2015)\citenamefont {Kuijken}
  \emph {et~al.}}]{Kuijken:2015vca}%
  \BibitemOpen
  \bibfield  {author} {\bibinfo {author} {\bibfnamefont {K.}~\bibnamefont
  {Kuijken}} \emph {et~al.},\ }\href {\doibase 10.1093/mnras/stv2140}
  {\bibfield  {journal} {\bibinfo  {journal} {Mon. Not. Roy. Astron. Soc.}\
  }\textbf {\bibinfo {volume} {454}},\ \bibinfo {pages} {3500} (\bibinfo {year}
  {2015})},\ \Eprint {http://arxiv.org/abs/1507.00738} {arXiv:1507.00738
  [astro-ph.CO]} \BibitemShut {NoStop}%
\bibitem [{\citenamefont {de~Jong}\ \emph {et~al.}(2015)\citenamefont {de~Jong}
  \emph {et~al.}}]{deJong:2015wca}%
  \BibitemOpen
  \bibfield  {author} {\bibinfo {author} {\bibfnamefont {J.~T.~A.}\
  \bibnamefont {de~Jong}} \emph {et~al.},\ }\href {\doibase
  10.1051/0004-6361/201526601} {\bibfield  {journal} {\bibinfo  {journal}
  {Astron. Astrophys.}\ }\textbf {\bibinfo {volume} {582}},\ \bibinfo {pages}
  {A62} (\bibinfo {year} {2015})},\ \Eprint {http://arxiv.org/abs/1507.00742}
  {arXiv:1507.00742 [astro-ph.CO]} \BibitemShut {NoStop}%
\bibitem [{\citenamefont {Hildebrandt}\ \emph {et~al.}(2017)\citenamefont
  {Hildebrandt} \emph {et~al.}}]{Hildebrandt:2016iqg}%
  \BibitemOpen
  \bibfield  {author} {\bibinfo {author} {\bibfnamefont {H.}~\bibnamefont
  {Hildebrandt}} \emph {et~al.},\ }\href {\doibase 10.1093/mnras/stw2805}
  {\bibfield  {journal} {\bibinfo  {journal} {Mon. Not. Roy. Astron. Soc.}\
  }\textbf {\bibinfo {volume} {465}},\ \bibinfo {pages} {1454} (\bibinfo {year}
  {2017})},\ \Eprint {http://arxiv.org/abs/1606.05338} {arXiv:1606.05338
  [astro-ph.CO]} \BibitemShut {NoStop}%
\bibitem [{\citenamefont {Di~Valentino}\ \emph
  {et~al.}(2021{\natexlab{b}})\citenamefont {Di~Valentino} \emph
  {et~al.}}]{DiValentino:2020vvd}%
  \BibitemOpen
  \bibfield  {author} {\bibinfo {author} {\bibfnamefont {E.}~\bibnamefont
  {Di~Valentino}} \emph {et~al.},\ }\href {\doibase
  10.1016/j.astropartphys.2021.102604} {\bibfield  {journal} {\bibinfo
  {journal} {Astropart. Phys.}\ }\textbf {\bibinfo {volume} {131}},\ \bibinfo
  {pages} {102604} (\bibinfo {year} {2021}{\natexlab{b}})},\ \Eprint
  {http://arxiv.org/abs/2008.11285} {arXiv:2008.11285 [astro-ph.CO]}
  \BibitemShut {NoStop}%
\bibitem [{\citenamefont {Clifton}\ \emph {et~al.}(2012)\citenamefont
  {Clifton}, \citenamefont {Ferreira}, \citenamefont {Padilla},\ and\
  \citenamefont {Skordis}}]{Clifton:2011jh}%
  \BibitemOpen
  \bibfield  {author} {\bibinfo {author} {\bibfnamefont {T.}~\bibnamefont
  {Clifton}}, \bibinfo {author} {\bibfnamefont {P.~G.}\ \bibnamefont
  {Ferreira}}, \bibinfo {author} {\bibfnamefont {A.}~\bibnamefont {Padilla}}, \
  and\ \bibinfo {author} {\bibfnamefont {C.}~\bibnamefont {Skordis}},\ }\href
  {\doibase 10.1016/j.physrep.2012.01.001} {\bibfield  {journal} {\bibinfo
  {journal} {Phys. Rept.}\ }\textbf {\bibinfo {volume} {513}},\ \bibinfo
  {pages} {1} (\bibinfo {year} {2012})},\ \Eprint
  {http://arxiv.org/abs/1106.2476} {arXiv:1106.2476 [astro-ph.CO]} \BibitemShut
  {NoStop}%
\bibitem [{\citenamefont {Capozziello}\ and\ \citenamefont
  {De~Laurentis}(2011)}]{Capozziello:2011et}%
  \BibitemOpen
  \bibfield  {author} {\bibinfo {author} {\bibfnamefont {S.}~\bibnamefont
  {Capozziello}}\ and\ \bibinfo {author} {\bibfnamefont {M.}~\bibnamefont
  {De~Laurentis}},\ }\href {\doibase 10.1016/j.physrep.2011.09.003} {\bibfield
  {journal} {\bibinfo  {journal} {Phys. Rept.}\ }\textbf {\bibinfo {volume}
  {509}},\ \bibinfo {pages} {167} (\bibinfo {year} {2011})},\ \Eprint
  {http://arxiv.org/abs/1108.6266} {arXiv:1108.6266 [gr-qc]} \BibitemShut
  {NoStop}%
\bibitem [{\citenamefont {Avelino}\ \emph {et~al.}(2016)\citenamefont {Avelino}
  \emph {et~al.}}]{Avelino:2016lpj}%
  \BibitemOpen
  \bibfield  {author} {\bibinfo {author} {\bibfnamefont {P.}~\bibnamefont
  {Avelino}} \emph {et~al.},\ }\href {\doibase 10.3390/sym8080070} {\bibfield
  {journal} {\bibinfo  {journal} {Symmetry}\ }\textbf {\bibinfo {volume} {8}},\
  \bibinfo {pages} {70} (\bibinfo {year} {2016})},\ \Eprint
  {http://arxiv.org/abs/1607.02979} {arXiv:1607.02979 [astro-ph.CO]}
  \BibitemShut {NoStop}%
\bibitem [{\citenamefont {Saridakis}\ \emph {et~al.}(2021)\citenamefont
  {Saridakis} \emph {et~al.}}]{CANTATA:2021ktz}%
  \BibitemOpen
  \bibfield  {author} {\bibinfo {author} {\bibfnamefont {E.~N.}\ \bibnamefont
  {Saridakis}} \emph {et~al.} (\bibinfo {collaboration} {CANTATA}),\
  }\href@noop {} {\  (\bibinfo {year} {2021})},\ \Eprint
  {http://arxiv.org/abs/2105.12582} {arXiv:2105.12582 [gr-qc]} \BibitemShut
  {NoStop}%
\bibitem [{\citenamefont {Sotiriou}\ and\ \citenamefont
  {Faraoni}(2010)}]{Sotiriou:2008rp}%
  \BibitemOpen
  \bibfield  {author} {\bibinfo {author} {\bibfnamefont {T.~P.}\ \bibnamefont
  {Sotiriou}}\ and\ \bibinfo {author} {\bibfnamefont {V.}~\bibnamefont
  {Faraoni}},\ }\href {\doibase 10.1103/RevModPhys.82.451} {\bibfield
  {journal} {\bibinfo  {journal} {Rev. Mod. Phys.}\ }\textbf {\bibinfo {volume}
  {82}},\ \bibinfo {pages} {451} (\bibinfo {year} {2010})},\ \Eprint
  {http://arxiv.org/abs/0805.1726} {arXiv:0805.1726 [gr-qc]} \BibitemShut
  {NoStop}%
\bibitem [{\citenamefont {De~Felice}\ and\ \citenamefont
  {Tsujikawa}(2010)}]{DeFelice:2010aj}%
  \BibitemOpen
  \bibfield  {author} {\bibinfo {author} {\bibfnamefont {A.}~\bibnamefont
  {De~Felice}}\ and\ \bibinfo {author} {\bibfnamefont {S.}~\bibnamefont
  {Tsujikawa}},\ }\href {\doibase 10.12942/lrr-2010-3} {\bibfield  {journal}
  {\bibinfo  {journal} {Living Rev. Rel.}\ }\textbf {\bibinfo {volume} {13}},\
  \bibinfo {pages} {3} (\bibinfo {year} {2010})},\ \Eprint
  {http://arxiv.org/abs/1002.4928} {arXiv:1002.4928 [gr-qc]} \BibitemShut
  {NoStop}%
\bibitem [{\citenamefont {Ferraro}\ and\ \citenamefont
  {Fiorini}(2007)}]{Ferraro:2006jd}%
  \BibitemOpen
  \bibfield  {author} {\bibinfo {author} {\bibfnamefont {R.}~\bibnamefont
  {Ferraro}}\ and\ \bibinfo {author} {\bibfnamefont {F.}~\bibnamefont
  {Fiorini}},\ }\href {\doibase 10.1103/PhysRevD.75.084031} {\bibfield
  {journal} {\bibinfo  {journal} {Phys. Rev. D}\ }\textbf {\bibinfo {volume}
  {75}},\ \bibinfo {pages} {084031} (\bibinfo {year} {2007})},\ \Eprint
  {http://arxiv.org/abs/gr-qc/0610067} {arXiv:gr-qc/0610067} \BibitemShut
  {NoStop}%
\bibitem [{\citenamefont {Ferraro}\ and\ \citenamefont
  {Fiorini}(2008)}]{Ferraro:2008ey}%
  \BibitemOpen
  \bibfield  {author} {\bibinfo {author} {\bibfnamefont {R.}~\bibnamefont
  {Ferraro}}\ and\ \bibinfo {author} {\bibfnamefont {F.}~\bibnamefont
  {Fiorini}},\ }\href {\doibase 10.1103/PhysRevD.78.124019} {\bibfield
  {journal} {\bibinfo  {journal} {Phys. Rev. D}\ }\textbf {\bibinfo {volume}
  {78}},\ \bibinfo {pages} {124019} (\bibinfo {year} {2008})},\ \Eprint
  {http://arxiv.org/abs/0812.1981} {arXiv:0812.1981 [gr-qc]} \BibitemShut
  {NoStop}%
\bibitem [{\citenamefont {Bengochea}\ and\ \citenamefont
  {Ferraro}(2009)}]{Bengochea:2008gz}%
  \BibitemOpen
  \bibfield  {author} {\bibinfo {author} {\bibfnamefont {G.~R.}\ \bibnamefont
  {Bengochea}}\ and\ \bibinfo {author} {\bibfnamefont {R.}~\bibnamefont
  {Ferraro}},\ }\href {\doibase 10.1103/PhysRevD.79.124019} {\bibfield
  {journal} {\bibinfo  {journal} {Phys. Rev. D}\ }\textbf {\bibinfo {volume}
  {79}},\ \bibinfo {pages} {124019} (\bibinfo {year} {2009})},\ \Eprint
  {http://arxiv.org/abs/0812.1205} {arXiv:0812.1205 [astro-ph]} \BibitemShut
  {NoStop}%
\bibitem [{\citenamefont {Cai}\ \emph {et~al.}(2016)\citenamefont {Cai},
  \citenamefont {Capozziello}, \citenamefont {De~Laurentis},\ and\
  \citenamefont {Saridakis}}]{Cai:2015emx}%
  \BibitemOpen
  \bibfield  {author} {\bibinfo {author} {\bibfnamefont {Y.-F.}\ \bibnamefont
  {Cai}}, \bibinfo {author} {\bibfnamefont {S.}~\bibnamefont {Capozziello}},
  \bibinfo {author} {\bibfnamefont {M.}~\bibnamefont {De~Laurentis}}, \ and\
  \bibinfo {author} {\bibfnamefont {E.~N.}\ \bibnamefont {Saridakis}},\ }\href
  {\doibase 10.1088/0034-4885/79/10/106901} {\bibfield  {journal} {\bibinfo
  {journal} {Rept. Prog. Phys.}\ }\textbf {\bibinfo {volume} {79}},\ \bibinfo
  {pages} {106901} (\bibinfo {year} {2016})},\ \Eprint
  {http://arxiv.org/abs/1511.07586} {arXiv:1511.07586 [gr-qc]} \BibitemShut
  {NoStop}%
\bibitem [{\citenamefont {Lazkoz}\ \emph {et~al.}(2019)\citenamefont {Lazkoz},
  \citenamefont {Lobo}, \citenamefont {Ortiz-Ba\~nos},\ and\ \citenamefont
  {Salzano}}]{Lazkoz:2019sjl}%
  \BibitemOpen
  \bibfield  {author} {\bibinfo {author} {\bibfnamefont {R.}~\bibnamefont
  {Lazkoz}}, \bibinfo {author} {\bibfnamefont {F.~S.~N.}\ \bibnamefont {Lobo}},
  \bibinfo {author} {\bibfnamefont {M.}~\bibnamefont {Ortiz-Ba\~nos}}, \ and\
  \bibinfo {author} {\bibfnamefont {V.}~\bibnamefont {Salzano}},\ }\href
  {\doibase 10.1103/PhysRevD.100.104027} {\bibfield  {journal} {\bibinfo
  {journal} {Phys. Rev. D}\ }\textbf {\bibinfo {volume} {100}},\ \bibinfo
  {pages} {104027} (\bibinfo {year} {2019})},\ \Eprint
  {http://arxiv.org/abs/1907.13219} {arXiv:1907.13219 [gr-qc]} \BibitemShut
  {NoStop}%
\bibitem [{\citenamefont {Dialektopoulos}\ \emph {et~al.}(2019)\citenamefont
  {Dialektopoulos}, \citenamefont {Koivisto},\ and\ \citenamefont
  {Capozziello}}]{Dialektopoulos:2019mtr}%
  \BibitemOpen
  \bibfield  {author} {\bibinfo {author} {\bibfnamefont {K.~F.}\ \bibnamefont
  {Dialektopoulos}}, \bibinfo {author} {\bibfnamefont {T.~S.}\ \bibnamefont
  {Koivisto}}, \ and\ \bibinfo {author} {\bibfnamefont {S.}~\bibnamefont
  {Capozziello}},\ }\href {\doibase 10.1140/epjc/s10052-019-7106-8} {\bibfield
  {journal} {\bibinfo  {journal} {Eur. Phys. J. C}\ }\textbf {\bibinfo {volume}
  {79}},\ \bibinfo {pages} {606} (\bibinfo {year} {2019})},\ \Eprint
  {http://arxiv.org/abs/1905.09019} {arXiv:1905.09019 [gr-qc]} \BibitemShut
  {NoStop}%
\bibitem [{\citenamefont {Lu}\ \emph {et~al.}(2019)\citenamefont {Lu},
  \citenamefont {Zhao},\ and\ \citenamefont {Chee}}]{Lu:2019hra}%
  \BibitemOpen
  \bibfield  {author} {\bibinfo {author} {\bibfnamefont {J.}~\bibnamefont
  {Lu}}, \bibinfo {author} {\bibfnamefont {X.}~\bibnamefont {Zhao}}, \ and\
  \bibinfo {author} {\bibfnamefont {G.}~\bibnamefont {Chee}},\ }\href {\doibase
  10.1140/epjc/s10052-019-7038-3} {\bibfield  {journal} {\bibinfo  {journal}
  {Eur. Phys. J. C}\ }\textbf {\bibinfo {volume} {79}},\ \bibinfo {pages} {530}
  (\bibinfo {year} {2019})},\ \Eprint {http://arxiv.org/abs/1906.08920}
  {arXiv:1906.08920 [gr-qc]} \BibitemShut {NoStop}%
\bibitem [{\citenamefont {Barros}\ \emph {et~al.}(2020)\citenamefont {Barros},
  \citenamefont {Barreiro}, \citenamefont {Koivisto},\ and\ \citenamefont
  {Nunes}}]{Barros:2020bgg}%
  \BibitemOpen
  \bibfield  {author} {\bibinfo {author} {\bibfnamefont {B.~J.}\ \bibnamefont
  {Barros}}, \bibinfo {author} {\bibfnamefont {T.}~\bibnamefont {Barreiro}},
  \bibinfo {author} {\bibfnamefont {T.}~\bibnamefont {Koivisto}}, \ and\
  \bibinfo {author} {\bibfnamefont {N.~J.}\ \bibnamefont {Nunes}},\ }\href
  {\doibase 10.1016/j.dark.2020.100616} {\bibfield  {journal} {\bibinfo
  {journal} {Phys. Dark Univ.}\ }\textbf {\bibinfo {volume} {30}},\ \bibinfo
  {pages} {100616} (\bibinfo {year} {2020})},\ \Eprint
  {http://arxiv.org/abs/2004.07867} {arXiv:2004.07867 [gr-qc]} \BibitemShut
  {NoStop}%
\bibitem [{\citenamefont {Esposito}\ \emph {et~al.}(2022)\citenamefont
  {Esposito}, \citenamefont {Carloni}, \citenamefont {Cianci},\ and\
  \citenamefont {Vignolo}}]{Esposito:2021ect}%
  \BibitemOpen
  \bibfield  {author} {\bibinfo {author} {\bibfnamefont {F.}~\bibnamefont
  {Esposito}}, \bibinfo {author} {\bibfnamefont {S.}~\bibnamefont {Carloni}},
  \bibinfo {author} {\bibfnamefont {R.}~\bibnamefont {Cianci}}, \ and\ \bibinfo
  {author} {\bibfnamefont {S.}~\bibnamefont {Vignolo}},\ }\href {\doibase
  10.1103/PhysRevD.105.084061} {\bibfield  {journal} {\bibinfo  {journal}
  {Phys. Rev. D}\ }\textbf {\bibinfo {volume} {105}},\ \bibinfo {pages}
  {084061} (\bibinfo {year} {2022})},\ \Eprint
  {http://arxiv.org/abs/2107.14522} {arXiv:2107.14522 [gr-qc]} \BibitemShut
  {NoStop}%
\bibitem [{\citenamefont {Frusciante}(2021)}]{Frusciante:2021sio}%
  \BibitemOpen
  \bibfield  {author} {\bibinfo {author} {\bibfnamefont {N.}~\bibnamefont
  {Frusciante}},\ }\href {\doibase 10.1103/PhysRevD.103.044021} {\bibfield
  {journal} {\bibinfo  {journal} {Phys. Rev. D}\ }\textbf {\bibinfo {volume}
  {103}},\ \bibinfo {pages} {044021} (\bibinfo {year} {2021})},\ \Eprint
  {http://arxiv.org/abs/2101.09242} {arXiv:2101.09242 [astro-ph.CO]}
  \BibitemShut {NoStop}%
\bibitem [{\citenamefont {Anagnostopoulos}\ \emph {et~al.}(2021)\citenamefont
  {Anagnostopoulos}, \citenamefont {Basilakos},\ and\ \citenamefont
  {Saridakis}}]{Anagnostopoulos:2021ydo}%
  \BibitemOpen
  \bibfield  {author} {\bibinfo {author} {\bibfnamefont {F.~K.}\ \bibnamefont
  {Anagnostopoulos}}, \bibinfo {author} {\bibfnamefont {S.}~\bibnamefont
  {Basilakos}}, \ and\ \bibinfo {author} {\bibfnamefont {E.~N.}\ \bibnamefont
  {Saridakis}},\ }\href {\doibase 10.1016/j.physletb.2021.136634} {\bibfield
  {journal} {\bibinfo  {journal} {Phys. Lett. B}\ }\textbf {\bibinfo {volume}
  {822}},\ \bibinfo {pages} {136634} (\bibinfo {year} {2021})},\ \Eprint
  {http://arxiv.org/abs/2104.15123} {arXiv:2104.15123 [gr-qc]} \BibitemShut
  {NoStop}%
\bibitem [{\citenamefont {Atayde}\ and\ \citenamefont
  {Frusciante}(2021)}]{Atayde:2021pgb}%
  \BibitemOpen
  \bibfield  {author} {\bibinfo {author} {\bibfnamefont {L.}~\bibnamefont
  {Atayde}}\ and\ \bibinfo {author} {\bibfnamefont {N.}~\bibnamefont
  {Frusciante}},\ }\href {\doibase 10.1103/PhysRevD.104.064052} {\bibfield
  {journal} {\bibinfo  {journal} {Phys. Rev. D}\ }\textbf {\bibinfo {volume}
  {104}},\ \bibinfo {pages} {064052} (\bibinfo {year} {2021})},\ \Eprint
  {http://arxiv.org/abs/2108.10832} {arXiv:2108.10832 [astro-ph.CO]}
  \BibitemShut {NoStop}%
\bibitem [{\citenamefont {Albuquerque}\ and\ \citenamefont
  {Frusciante}(2022)}]{Albuquerque:2022eac}%
  \BibitemOpen
  \bibfield  {author} {\bibinfo {author} {\bibfnamefont {I.~S.}\ \bibnamefont
  {Albuquerque}}\ and\ \bibinfo {author} {\bibfnamefont {N.}~\bibnamefont
  {Frusciante}},\ }\href {\doibase 10.1016/j.dark.2022.100980} {\bibfield
  {journal} {\bibinfo  {journal} {Phys. Dark Univ.}\ }\textbf {\bibinfo
  {volume} {35}},\ \bibinfo {pages} {100980} (\bibinfo {year} {2022})},\
  \Eprint {http://arxiv.org/abs/2202.04637} {arXiv:2202.04637 [astro-ph.CO]}
  \BibitemShut {NoStop}%
\bibitem [{\citenamefont {Ferreira}\ \emph {et~al.}(2022)\citenamefont
  {Ferreira}, \citenamefont {Barreiro}, \citenamefont {Mimoso},\ and\
  \citenamefont {Nunes}}]{Ferreira:2022jcd}%
  \BibitemOpen
  \bibfield  {author} {\bibinfo {author} {\bibfnamefont {J.}~\bibnamefont
  {Ferreira}}, \bibinfo {author} {\bibfnamefont {T.}~\bibnamefont {Barreiro}},
  \bibinfo {author} {\bibfnamefont {J.}~\bibnamefont {Mimoso}}, \ and\ \bibinfo
  {author} {\bibfnamefont {N.~J.}\ \bibnamefont {Nunes}},\ }\href {\doibase
  10.1103/PhysRevD.105.123531} {\bibfield  {journal} {\bibinfo  {journal}
  {Phys. Rev. D}\ }\textbf {\bibinfo {volume} {105}},\ \bibinfo {pages}
  {123531} (\bibinfo {year} {2022})},\ \Eprint
  {http://arxiv.org/abs/2203.13788} {arXiv:2203.13788 [astro-ph.CO]}
  \BibitemShut {NoStop}%
\bibitem [{\citenamefont {Anagnostopoulos}\ \emph {et~al.}(2023)\citenamefont
  {Anagnostopoulos}, \citenamefont {Gakis}, \citenamefont {Saridakis},\ and\
  \citenamefont {Basilakos}}]{Anagnostopoulos:2022gej}%
  \BibitemOpen
  \bibfield  {author} {\bibinfo {author} {\bibfnamefont {F.~K.}\ \bibnamefont
  {Anagnostopoulos}}, \bibinfo {author} {\bibfnamefont {V.}~\bibnamefont
  {Gakis}}, \bibinfo {author} {\bibfnamefont {E.~N.}\ \bibnamefont
  {Saridakis}}, \ and\ \bibinfo {author} {\bibfnamefont {S.}~\bibnamefont
  {Basilakos}},\ }\href {\doibase 10.1140/epjc/s10052-023-11190-x} {\bibfield
  {journal} {\bibinfo  {journal} {Eur. Phys. J. C}\ }\textbf {\bibinfo {volume}
  {83}},\ \bibinfo {pages} {58} (\bibinfo {year} {2023})},\ \Eprint
  {http://arxiv.org/abs/2205.11445} {arXiv:2205.11445 [gr-qc]} \BibitemShut
  {NoStop}%
\bibitem [{\citenamefont {Khyllep}\ \emph {et~al.}(2023)\citenamefont
  {Khyllep}, \citenamefont {Dutta}, \citenamefont {Saridakis},\ and\
  \citenamefont {Yesmakhanova}}]{Khyllep:2022spx}%
  \BibitemOpen
  \bibfield  {author} {\bibinfo {author} {\bibfnamefont {W.}~\bibnamefont
  {Khyllep}}, \bibinfo {author} {\bibfnamefont {J.}~\bibnamefont {Dutta}},
  \bibinfo {author} {\bibfnamefont {E.~N.}\ \bibnamefont {Saridakis}}, \ and\
  \bibinfo {author} {\bibfnamefont {K.}~\bibnamefont {Yesmakhanova}},\ }\href
  {\doibase 10.1103/PhysRevD.107.044022} {\bibfield  {journal} {\bibinfo
  {journal} {Phys. Rev. D}\ }\textbf {\bibinfo {volume} {107}},\ \bibinfo
  {pages} {044022} (\bibinfo {year} {2023})},\ \Eprint
  {http://arxiv.org/abs/2207.02610} {arXiv:2207.02610 [gr-qc]} \BibitemShut
  {NoStop}%
\bibitem [{\citenamefont {Atayde}\ and\ \citenamefont
  {Frusciante}(2023)}]{Atayde:2023aoj}%
  \BibitemOpen
  \bibfield  {author} {\bibinfo {author} {\bibfnamefont {L.}~\bibnamefont
  {Atayde}}\ and\ \bibinfo {author} {\bibfnamefont {N.}~\bibnamefont
  {Frusciante}},\ }\href {\doibase 10.1103/PhysRevD.107.124048} {\bibfield
  {journal} {\bibinfo  {journal} {Phys. Rev. D}\ }\textbf {\bibinfo {volume}
  {107}},\ \bibinfo {pages} {124048} (\bibinfo {year} {2023})},\ \Eprint
  {http://arxiv.org/abs/2306.03015} {arXiv:2306.03015 [astro-ph.CO]}
  \BibitemShut {NoStop}%
\bibitem [{\citenamefont {Ferreira}\ \emph {et~al.}(2023)\citenamefont
  {Ferreira}, \citenamefont {Barreiro}, \citenamefont {Mimoso},\ and\
  \citenamefont {Nunes}}]{Ferreira:2023awf}%
  \BibitemOpen
  \bibfield  {author} {\bibinfo {author} {\bibfnamefont {J.}~\bibnamefont
  {Ferreira}}, \bibinfo {author} {\bibfnamefont {T.}~\bibnamefont {Barreiro}},
  \bibinfo {author} {\bibfnamefont {J.~P.}\ \bibnamefont {Mimoso}}, \ and\
  \bibinfo {author} {\bibfnamefont {N.~J.}\ \bibnamefont {Nunes}},\ }\href
  {\doibase 10.1103/PhysRevD.108.063521} {\bibfield  {journal} {\bibinfo
  {journal} {Phys. Rev. D}\ }\textbf {\bibinfo {volume} {108}},\ \bibinfo
  {pages} {063521} (\bibinfo {year} {2023})},\ \Eprint
  {http://arxiv.org/abs/2306.10176} {arXiv:2306.10176 [astro-ph.CO]}
  \BibitemShut {NoStop}%
\bibitem [{\citenamefont {Shabani}\ \emph {et~al.}(2023)\citenamefont
  {Shabani}, \citenamefont {De}, \citenamefont {Loo},\ and\ \citenamefont
  {Saridakis}}]{Shabani:2023xfn}%
  \BibitemOpen
  \bibfield  {author} {\bibinfo {author} {\bibfnamefont {H.}~\bibnamefont
  {Shabani}}, \bibinfo {author} {\bibfnamefont {A.}~\bibnamefont {De}},
  \bibinfo {author} {\bibfnamefont {T.-H.}\ \bibnamefont {Loo}}, \ and\
  \bibinfo {author} {\bibfnamefont {E.~N.}\ \bibnamefont {Saridakis}},\
  }\href@noop {} {\  (\bibinfo {year} {2023})},\ \Eprint
  {http://arxiv.org/abs/2306.13324} {arXiv:2306.13324 [gr-qc]} \BibitemShut
  {NoStop}%
\bibitem [{\citenamefont {De}\ \emph {et~al.}(2023)\citenamefont {De},
  \citenamefont {Loo},\ and\ \citenamefont {Saridakis}}]{De:2023xua}%
  \BibitemOpen
  \bibfield  {author} {\bibinfo {author} {\bibfnamefont {A.}~\bibnamefont
  {De}}, \bibinfo {author} {\bibfnamefont {T.-H.}\ \bibnamefont {Loo}}, \ and\
  \bibinfo {author} {\bibfnamefont {E.~N.}\ \bibnamefont {Saridakis}},\
  }\href@noop {} {\  (\bibinfo {year} {2023})},\ \Eprint
  {http://arxiv.org/abs/2308.00652} {arXiv:2308.00652 [gr-qc]} \BibitemShut
  {NoStop}%
\bibitem [{\citenamefont {Ayuso}\ \emph {et~al.}(2022)\citenamefont {Ayuso},
  \citenamefont {Lazkoz},\ and\ \citenamefont {Mimoso}}]{Ayuso:2021vtj}%
  \BibitemOpen
  \bibfield  {author} {\bibinfo {author} {\bibfnamefont {I.}~\bibnamefont
  {Ayuso}}, \bibinfo {author} {\bibfnamefont {R.}~\bibnamefont {Lazkoz}}, \
  and\ \bibinfo {author} {\bibfnamefont {J.~P.}\ \bibnamefont {Mimoso}},\
  }\href {\doibase 10.1103/PhysRevD.105.083534} {\bibfield  {journal} {\bibinfo
   {journal} {Phys. Rev. D}\ }\textbf {\bibinfo {volume} {105}},\ \bibinfo
  {pages} {083534} (\bibinfo {year} {2022})},\ \Eprint
  {http://arxiv.org/abs/2111.05061} {arXiv:2111.05061 [astro-ph.CO]}
  \BibitemShut {NoStop}%
\bibitem [{\citenamefont {Dvali}\ \emph {et~al.}(2000)\citenamefont {Dvali},
  \citenamefont {Gabadadze},\ and\ \citenamefont {Porrati}}]{Dvali:2000hr}%
  \BibitemOpen
  \bibfield  {author} {\bibinfo {author} {\bibfnamefont {G.}~\bibnamefont
  {Dvali}}, \bibinfo {author} {\bibfnamefont {G.}~\bibnamefont {Gabadadze}}, \
  and\ \bibinfo {author} {\bibfnamefont {M.}~\bibnamefont {Porrati}},\ }\href
  {\doibase 10.1016/S0370-2693(00)00669-9} {\bibfield  {journal} {\bibinfo
  {journal} {Phys. Lett. B}\ }\textbf {\bibinfo {volume} {485}},\ \bibinfo
  {pages} {208} (\bibinfo {year} {2000})},\ \Eprint
  {http://arxiv.org/abs/hep-th/0005016} {arXiv:hep-th/0005016} \BibitemShut
  {NoStop}%
\bibitem [{\citenamefont {Deffayet}(2001)}]{Deffayet:2000uy}%
  \BibitemOpen
  \bibfield  {author} {\bibinfo {author} {\bibfnamefont {C.}~\bibnamefont
  {Deffayet}},\ }\href {\doibase 10.1016/S0370-2693(01)00160-5} {\bibfield
  {journal} {\bibinfo  {journal} {Phys. Lett. B}\ }\textbf {\bibinfo {volume}
  {502}},\ \bibinfo {pages} {199} (\bibinfo {year} {2001})},\ \Eprint
  {http://arxiv.org/abs/hep-th/0010186} {arXiv:hep-th/0010186} \BibitemShut
  {NoStop}%
\bibitem [{\citenamefont {Luty}\ \emph {et~al.}(2003)\citenamefont {Luty},
  \citenamefont {Porrati},\ and\ \citenamefont {Rattazzi}}]{Luty:2003vm}%
  \BibitemOpen
  \bibfield  {author} {\bibinfo {author} {\bibfnamefont {M.~A.}\ \bibnamefont
  {Luty}}, \bibinfo {author} {\bibfnamefont {M.}~\bibnamefont {Porrati}}, \
  and\ \bibinfo {author} {\bibfnamefont {R.}~\bibnamefont {Rattazzi}},\ }\href
  {\doibase 10.1088/1126-6708/2003/09/029} {\bibfield  {journal} {\bibinfo
  {journal} {JHEP}\ }\textbf {\bibinfo {volume} {09}},\ \bibinfo {pages} {029}
  (\bibinfo {year} {2003})},\ \Eprint {http://arxiv.org/abs/hep-th/0303116}
  {arXiv:hep-th/0303116} \BibitemShut {NoStop}%
\bibitem [{\citenamefont {Nicolis}\ and\ \citenamefont
  {Rattazzi}(2004)}]{Nicolis:2004qq}%
  \BibitemOpen
  \bibfield  {author} {\bibinfo {author} {\bibfnamefont {A.}~\bibnamefont
  {Nicolis}}\ and\ \bibinfo {author} {\bibfnamefont {R.}~\bibnamefont
  {Rattazzi}},\ }\href {\doibase 10.1088/1126-6708/2004/06/059} {\bibfield
  {journal} {\bibinfo  {journal} {JHEP}\ }\textbf {\bibinfo {volume} {06}},\
  \bibinfo {pages} {059} (\bibinfo {year} {2004})},\ \Eprint
  {http://arxiv.org/abs/hep-th/0404159} {arXiv:hep-th/0404159} \BibitemShut
  {NoStop}%
\bibitem [{\citenamefont {Charmousis}\ \emph {et~al.}(2006)\citenamefont
  {Charmousis}, \citenamefont {Gregory}, \citenamefont {Kaloper},\ and\
  \citenamefont {Padilla}}]{Charmousis:2006pn}%
  \BibitemOpen
  \bibfield  {author} {\bibinfo {author} {\bibfnamefont {C.}~\bibnamefont
  {Charmousis}}, \bibinfo {author} {\bibfnamefont {R.}~\bibnamefont {Gregory}},
  \bibinfo {author} {\bibfnamefont {N.}~\bibnamefont {Kaloper}}, \ and\
  \bibinfo {author} {\bibfnamefont {A.}~\bibnamefont {Padilla}},\ }\href
  {\doibase 10.1088/1126-6708/2006/10/066} {\bibfield  {journal} {\bibinfo
  {journal} {JHEP}\ }\textbf {\bibinfo {volume} {10}},\ \bibinfo {pages} {066}
  (\bibinfo {year} {2006})},\ \Eprint {http://arxiv.org/abs/hep-th/0604086}
  {arXiv:hep-th/0604086} \BibitemShut {NoStop}%
\bibitem [{\citenamefont {Fairbairn}\ and\ \citenamefont
  {Goobar}(2006)}]{Fairbairn:2005ue}%
  \BibitemOpen
  \bibfield  {author} {\bibinfo {author} {\bibfnamefont {M.}~\bibnamefont
  {Fairbairn}}\ and\ \bibinfo {author} {\bibfnamefont {A.}~\bibnamefont
  {Goobar}},\ }\href {\doibase 10.1016/j.physletb.2006.07.048} {\bibfield
  {journal} {\bibinfo  {journal} {Phys. Lett. B}\ }\textbf {\bibinfo {volume}
  {642}},\ \bibinfo {pages} {432} (\bibinfo {year} {2006})},\ \Eprint
  {http://arxiv.org/abs/astro-ph/0511029} {arXiv:astro-ph/0511029} \BibitemShut
  {NoStop}%
\bibitem [{\citenamefont {Maartens}\ and\ \citenamefont
  {Majerotto}(2006)}]{Maartens:2006yt}%
  \BibitemOpen
  \bibfield  {author} {\bibinfo {author} {\bibfnamefont {R.}~\bibnamefont
  {Maartens}}\ and\ \bibinfo {author} {\bibfnamefont {E.}~\bibnamefont
  {Majerotto}},\ }\href {\doibase 10.1103/PhysRevD.74.023004} {\bibfield
  {journal} {\bibinfo  {journal} {Phys. Rev. D}\ }\textbf {\bibinfo {volume}
  {74}},\ \bibinfo {pages} {023004} (\bibinfo {year} {2006})},\ \Eprint
  {http://arxiv.org/abs/astro-ph/0603353} {arXiv:astro-ph/0603353} \BibitemShut
  {NoStop}%
\bibitem [{\citenamefont {Fang}\ \emph {et~al.}(2008)\citenamefont {Fang},
  \citenamefont {Wang}, \citenamefont {Hu}, \citenamefont {Haiman},
  \citenamefont {Hui},\ and\ \citenamefont {May}}]{Fang:2008kc}%
  \BibitemOpen
  \bibfield  {author} {\bibinfo {author} {\bibfnamefont {W.}~\bibnamefont
  {Fang}}, \bibinfo {author} {\bibfnamefont {S.}~\bibnamefont {Wang}}, \bibinfo
  {author} {\bibfnamefont {W.}~\bibnamefont {Hu}}, \bibinfo {author}
  {\bibfnamefont {Z.}~\bibnamefont {Haiman}}, \bibinfo {author} {\bibfnamefont
  {L.}~\bibnamefont {Hui}}, \ and\ \bibinfo {author} {\bibfnamefont
  {M.}~\bibnamefont {May}},\ }\href {\doibase 10.1103/PhysRevD.78.103509}
  {\bibfield  {journal} {\bibinfo  {journal} {Phys. Rev. D}\ }\textbf {\bibinfo
  {volume} {78}},\ \bibinfo {pages} {103509} (\bibinfo {year} {2008})},\
  \Eprint {http://arxiv.org/abs/0808.2208} {arXiv:0808.2208 [astro-ph]}
  \BibitemShut {NoStop}%
\bibitem [{\citenamefont {Ade}\ \emph {et~al.}(2016)\citenamefont {Ade} \emph
  {et~al.}}]{Planck:2015bue}%
  \BibitemOpen
  \bibfield  {author} {\bibinfo {author} {\bibfnamefont {P.~A.~R.}\
  \bibnamefont {Ade}} \emph {et~al.} (\bibinfo {collaboration} {Planck}),\
  }\href {\doibase 10.1051/0004-6361/201525814} {\bibfield  {journal} {\bibinfo
   {journal} {Astron. Astrophys.}\ }\textbf {\bibinfo {volume} {594}},\
  \bibinfo {pages} {A14} (\bibinfo {year} {2016})},\ \Eprint
  {http://arxiv.org/abs/1502.01590} {arXiv:1502.01590 [astro-ph.CO]}
  \BibitemShut {NoStop}%
\bibitem [{\citenamefont {Hu}\ and\ \citenamefont {White}(1996)}]{Hu:1996vq}%
  \BibitemOpen
  \bibfield  {author} {\bibinfo {author} {\bibfnamefont {W.}~\bibnamefont
  {Hu}}\ and\ \bibinfo {author} {\bibfnamefont {M.~J.}\ \bibnamefont {White}},\
  }\href {\doibase 10.1086/177951} {\bibfield  {journal} {\bibinfo  {journal}
  {Astrophys. J.}\ }\textbf {\bibinfo {volume} {471}},\ \bibinfo {pages} {30}
  (\bibinfo {year} {1996})},\ \Eprint {http://arxiv.org/abs/astro-ph/9602019}
  {arXiv:astro-ph/9602019} \BibitemShut {NoStop}%
\bibitem [{\citenamefont {Sachs}\ and\ \citenamefont
  {Wolfe}(1967)}]{Sachs:1967er}%
  \BibitemOpen
  \bibfield  {author} {\bibinfo {author} {\bibfnamefont {R.~K.}\ \bibnamefont
  {Sachs}}\ and\ \bibinfo {author} {\bibfnamefont {A.~M.}\ \bibnamefont
  {Wolfe}},\ }\href {\doibase 10.1007/s10714-007-0448-9} {\bibfield  {journal}
  {\bibinfo  {journal} {Astrophys. J.}\ }\textbf {\bibinfo {volume} {147}},\
  \bibinfo {pages} {73} (\bibinfo {year} {1967})}\BibitemShut {NoStop}%
\bibitem [{\citenamefont {Kofman}\ and\ \citenamefont
  {Starobinsky}(1985)}]{Kofman:1985fp}%
  \BibitemOpen
  \bibfield  {author} {\bibinfo {author} {\bibfnamefont {L.}~\bibnamefont
  {Kofman}}\ and\ \bibinfo {author} {\bibfnamefont {A.~A.}\ \bibnamefont
  {Starobinsky}},\ }\href@noop {} {\bibfield  {journal} {\bibinfo  {journal}
  {Sov. Astron. Lett.}\ }\textbf {\bibinfo {volume} {11}},\ \bibinfo {pages}
  {271} (\bibinfo {year} {1985})}\BibitemShut {NoStop}%
\bibitem [{\citenamefont {Acquaviva}\ and\ \citenamefont
  {Baccigalupi}(2006)}]{Acquaviva:2005xz}%
  \BibitemOpen
  \bibfield  {author} {\bibinfo {author} {\bibfnamefont {V.}~\bibnamefont
  {Acquaviva}}\ and\ \bibinfo {author} {\bibfnamefont {C.}~\bibnamefont
  {Baccigalupi}},\ }\href {\doibase 10.1103/PhysRevD.74.103510} {\bibfield
  {journal} {\bibinfo  {journal} {Phys. Rev. D}\ }\textbf {\bibinfo {volume}
  {74}},\ \bibinfo {pages} {103510} (\bibinfo {year} {2006})},\ \Eprint
  {http://arxiv.org/abs/astro-ph/0507644} {arXiv:astro-ph/0507644} \BibitemShut
  {NoStop}%
\bibitem [{\citenamefont {Carbone}\ \emph {et~al.}(2013)\citenamefont
  {Carbone}, \citenamefont {Baldi}, \citenamefont {Pettorino},\ and\
  \citenamefont {Baccigalupi}}]{Carbone:2013dna}%
  \BibitemOpen
  \bibfield  {author} {\bibinfo {author} {\bibfnamefont {C.}~\bibnamefont
  {Carbone}}, \bibinfo {author} {\bibfnamefont {M.}~\bibnamefont {Baldi}},
  \bibinfo {author} {\bibfnamefont {V.}~\bibnamefont {Pettorino}}, \ and\
  \bibinfo {author} {\bibfnamefont {C.}~\bibnamefont {Baccigalupi}},\ }\href
  {\doibase 10.1088/1475-7516/2013/09/004} {\bibfield  {journal} {\bibinfo
  {journal} {JCAP}\ }\textbf {\bibinfo {volume} {09}},\ \bibinfo {pages} {004}
  (\bibinfo {year} {2013})},\ \Eprint {http://arxiv.org/abs/1305.0829}
  {arXiv:1305.0829 [astro-ph.CO]} \BibitemShut {NoStop}%
\bibitem [{\citenamefont {Peebles}(1984)}]{Peebles:1984ge}%
  \BibitemOpen
  \bibfield  {author} {\bibinfo {author} {\bibfnamefont {P.~J.~E.}\
  \bibnamefont {Peebles}},\ }\href {\doibase 10.1086/162425} {\bibfield
  {journal} {\bibinfo  {journal} {Astrophys. J.}\ }\textbf {\bibinfo {volume}
  {284}},\ \bibinfo {pages} {439} (\bibinfo {year} {1984})}\BibitemShut
  {NoStop}%
\bibitem [{\citenamefont {{Barrow}}\ and\ \citenamefont
  {{Saich}}(1993)}]{1993MNRAS.262..717B}%
  \BibitemOpen
  \bibfield  {author} {\bibinfo {author} {\bibfnamefont {J.~D.}\ \bibnamefont
  {{Barrow}}}\ and\ \bibinfo {author} {\bibfnamefont {P.}~\bibnamefont
  {{Saich}}},\ }\href {\doibase 10.1093/mnras/262.3.717} {\bibfield  {journal}
  {\bibinfo  {journal} {MNRAS}\ }\textbf {\bibinfo {volume} {262}},\ \bibinfo
  {pages} {717} (\bibinfo {year} {1993})}\BibitemShut {NoStop}%
\bibitem [{\citenamefont {Zhao}\ \emph {et~al.}(2009)\citenamefont {Zhao},
  \citenamefont {Pogosian}, \citenamefont {Silvestri},\ and\ \citenamefont
  {Zylberberg}}]{Zhao:2008bn}%
  \BibitemOpen
  \bibfield  {author} {\bibinfo {author} {\bibfnamefont {G.-B.}\ \bibnamefont
  {Zhao}}, \bibinfo {author} {\bibfnamefont {L.}~\bibnamefont {Pogosian}},
  \bibinfo {author} {\bibfnamefont {A.}~\bibnamefont {Silvestri}}, \ and\
  \bibinfo {author} {\bibfnamefont {J.}~\bibnamefont {Zylberberg}},\ }\href
  {\doibase 10.1103/PhysRevD.79.083513} {\bibfield  {journal} {\bibinfo
  {journal} {Phys. Rev. D}\ }\textbf {\bibinfo {volume} {79}},\ \bibinfo
  {pages} {083513} (\bibinfo {year} {2009})},\ \Eprint
  {http://arxiv.org/abs/0809.3791} {arXiv:0809.3791 [astro-ph]} \BibitemShut
  {NoStop}%
\bibitem [{\citenamefont {Hojjati}\ \emph {et~al.}(2011)\citenamefont
  {Hojjati}, \citenamefont {Pogosian},\ and\ \citenamefont
  {Zhao}}]{Hojjati:2011ix}%
  \BibitemOpen
  \bibfield  {author} {\bibinfo {author} {\bibfnamefont {A.}~\bibnamefont
  {Hojjati}}, \bibinfo {author} {\bibfnamefont {L.}~\bibnamefont {Pogosian}}, \
  and\ \bibinfo {author} {\bibfnamefont {G.-B.}\ \bibnamefont {Zhao}},\ }\href
  {\doibase 10.1088/1475-7516/2011/08/005} {\bibfield  {journal} {\bibinfo
  {journal} {JCAP}\ }\textbf {\bibinfo {volume} {08}},\ \bibinfo {pages} {005}
  (\bibinfo {year} {2011})},\ \Eprint {http://arxiv.org/abs/1106.4543}
  {arXiv:1106.4543 [astro-ph.CO]} \BibitemShut {NoStop}%
\bibitem [{\citenamefont {Zucca}\ \emph {et~al.}(2019)\citenamefont {Zucca},
  \citenamefont {Pogosian}, \citenamefont {Silvestri},\ and\ \citenamefont
  {Zhao}}]{Zucca:2019xhg}%
  \BibitemOpen
  \bibfield  {author} {\bibinfo {author} {\bibfnamefont {A.}~\bibnamefont
  {Zucca}}, \bibinfo {author} {\bibfnamefont {L.}~\bibnamefont {Pogosian}},
  \bibinfo {author} {\bibfnamefont {A.}~\bibnamefont {Silvestri}}, \ and\
  \bibinfo {author} {\bibfnamefont {G.-B.}\ \bibnamefont {Zhao}},\ }\href
  {\doibase 10.1088/1475-7516/2019/05/001} {\bibfield  {journal} {\bibinfo
  {journal} {JCAP}\ }\textbf {\bibinfo {volume} {05}},\ \bibinfo {pages} {001}
  (\bibinfo {year} {2019})},\ \Eprint {http://arxiv.org/abs/1901.05956}
  {arXiv:1901.05956 [astro-ph.CO]} \BibitemShut {NoStop}%
\bibitem [{\citenamefont {Wang}\ \emph {et~al.}(2023)\citenamefont {Wang},
  \citenamefont {Mirpoorian}, \citenamefont {Pogosian}, \citenamefont
  {Silvestri},\ and\ \citenamefont {Zhao}}]{Wang:2023tjj}%
  \BibitemOpen
  \bibfield  {author} {\bibinfo {author} {\bibfnamefont {Z.}~\bibnamefont
  {Wang}}, \bibinfo {author} {\bibfnamefont {S.~H.}\ \bibnamefont
  {Mirpoorian}}, \bibinfo {author} {\bibfnamefont {L.}~\bibnamefont
  {Pogosian}}, \bibinfo {author} {\bibfnamefont {A.}~\bibnamefont {Silvestri}},
  \ and\ \bibinfo {author} {\bibfnamefont {G.-B.}\ \bibnamefont {Zhao}},\
  }\href {\doibase 10.1088/1475-7516/2023/08/038} {\bibfield  {journal}
  {\bibinfo  {journal} {JCAP}\ }\textbf {\bibinfo {volume} {08}},\ \bibinfo
  {pages} {038} (\bibinfo {year} {2023})},\ \Eprint
  {http://arxiv.org/abs/2305.05667} {arXiv:2305.05667 [astro-ph.CO]}
  \BibitemShut {NoStop}%
\bibitem [{\citenamefont {Gomes}\ \emph {et~al.}(2023)\citenamefont {Gomes},
  \citenamefont {Beltr\'an~Jim\'enez}, \citenamefont {Cano},\ and\
  \citenamefont {Koivisto}}]{Gomes:2023tur}%
  \BibitemOpen
  \bibfield  {author} {\bibinfo {author} {\bibfnamefont {D.~A.}\ \bibnamefont
  {Gomes}}, \bibinfo {author} {\bibfnamefont {J.}~\bibnamefont
  {Beltr\'an~Jim\'enez}}, \bibinfo {author} {\bibfnamefont {A.~J.}\
  \bibnamefont {Cano}}, \ and\ \bibinfo {author} {\bibfnamefont {T.~S.}\
  \bibnamefont {Koivisto}},\ }\href@noop {} {\  (\bibinfo {year} {2023})},\
  \Eprint {http://arxiv.org/abs/2311.04201} {arXiv:2311.04201 [gr-qc]}
  \BibitemShut {NoStop}%
\bibitem [{\citenamefont {Heisenberg}\ \emph {et~al.}(2023)\citenamefont
  {Heisenberg}, \citenamefont {Hohmann},\ and\ \citenamefont
  {Kuhn}}]{Heisenberg:2023wgk}%
  \BibitemOpen
  \bibfield  {author} {\bibinfo {author} {\bibfnamefont {L.}~\bibnamefont
  {Heisenberg}}, \bibinfo {author} {\bibfnamefont {M.}~\bibnamefont {Hohmann}},
  \ and\ \bibinfo {author} {\bibfnamefont {S.}~\bibnamefont {Kuhn}},\
  }\href@noop {} {\  (\bibinfo {year} {2023})},\ \Eprint
  {http://arxiv.org/abs/2311.05495} {arXiv:2311.05495 [gr-qc]} \BibitemShut
  {NoStop}%
\bibitem [{\citenamefont {Deffayet}\ and\ \citenamefont
  {Rombouts}(2005)}]{Deffayet:2005ys}%
  \BibitemOpen
  \bibfield  {author} {\bibinfo {author} {\bibfnamefont {C.}~\bibnamefont
  {Deffayet}}\ and\ \bibinfo {author} {\bibfnamefont {J.-W.}\ \bibnamefont
  {Rombouts}},\ }\href {\doibase 10.1103/PhysRevD.72.044003} {\bibfield
  {journal} {\bibinfo  {journal} {Phys. Rev. D}\ }\textbf {\bibinfo {volume}
  {72}},\ \bibinfo {pages} {044003} (\bibinfo {year} {2005})},\ \Eprint
  {http://arxiv.org/abs/gr-qc/0505134} {arXiv:gr-qc/0505134} \BibitemShut
  {NoStop}%
\bibitem [{\citenamefont {Mukohyama}(2010)}]{Mukohyama:2010xz}%
  \BibitemOpen
  \bibfield  {author} {\bibinfo {author} {\bibfnamefont {S.}~\bibnamefont
  {Mukohyama}},\ }\href {\doibase 10.1088/0264-9381/27/22/223101} {\bibfield
  {journal} {\bibinfo  {journal} {Class. Quant. Grav.}\ }\textbf {\bibinfo
  {volume} {27}},\ \bibinfo {pages} {223101} (\bibinfo {year} {2010})},\
  \Eprint {http://arxiv.org/abs/1007.5199} {arXiv:1007.5199 [hep-th]}
  \BibitemShut {NoStop}%
\bibitem [{\citenamefont {Gabadadze}\ \emph {et~al.}(2019)\citenamefont
  {Gabadadze}, \citenamefont {Older},\ and\ \citenamefont
  {Pirtskhalava}}]{Gabadadze:2019lld}%
  \BibitemOpen
  \bibfield  {author} {\bibinfo {author} {\bibfnamefont {G.}~\bibnamefont
  {Gabadadze}}, \bibinfo {author} {\bibfnamefont {D.}~\bibnamefont {Older}}, \
  and\ \bibinfo {author} {\bibfnamefont {D.}~\bibnamefont {Pirtskhalava}},\
  }\href {\doibase 10.1103/PhysRevD.100.124017} {\bibfield  {journal} {\bibinfo
   {journal} {Phys. Rev. D}\ }\textbf {\bibinfo {volume} {100}},\ \bibinfo
  {pages} {124017} (\bibinfo {year} {2019})},\ \Eprint
  {http://arxiv.org/abs/1907.13491} {arXiv:1907.13491 [hep-th]} \BibitemShut
  {NoStop}%
\bibitem [{\citenamefont {Hu}\ \emph {et~al.}(2023)\citenamefont {Hu},
  \citenamefont {Zhao}, \citenamefont {Ren}, \citenamefont {Wang},
  \citenamefont {Saridakis},\ and\ \citenamefont {Cai}}]{Hu:2023juh}%
  \BibitemOpen
  \bibfield  {author} {\bibinfo {author} {\bibfnamefont {Y.-M.}\ \bibnamefont
  {Hu}}, \bibinfo {author} {\bibfnamefont {Y.}~\bibnamefont {Zhao}}, \bibinfo
  {author} {\bibfnamefont {X.}~\bibnamefont {Ren}}, \bibinfo {author}
  {\bibfnamefont {B.}~\bibnamefont {Wang}}, \bibinfo {author} {\bibfnamefont
  {E.~N.}\ \bibnamefont {Saridakis}}, \ and\ \bibinfo {author} {\bibfnamefont
  {Y.-F.}\ \bibnamefont {Cai}},\ }\href {\doibase
  10.1088/1475-7516/2023/07/060} {\bibfield  {journal} {\bibinfo  {journal}
  {JCAP}\ }\textbf {\bibinfo {volume} {07}},\ \bibinfo {pages} {060} (\bibinfo
  {year} {2023})},\ \Eprint {http://arxiv.org/abs/2302.03545} {arXiv:2302.03545
  [gr-qc]} \BibitemShut {NoStop}%
\bibitem [{\citenamefont {Ayuso}\ \emph {et~al.}(2021)\citenamefont {Ayuso},
  \citenamefont {Lazkoz},\ and\ \citenamefont {Salzano}}]{Ayuso:2020dcu}%
  \BibitemOpen
  \bibfield  {author} {\bibinfo {author} {\bibfnamefont {I.}~\bibnamefont
  {Ayuso}}, \bibinfo {author} {\bibfnamefont {R.}~\bibnamefont {Lazkoz}}, \
  and\ \bibinfo {author} {\bibfnamefont {V.}~\bibnamefont {Salzano}},\ }\href
  {\doibase 10.1103/PhysRevD.103.063505} {\bibfield  {journal} {\bibinfo
  {journal} {Phys. Rev. D}\ }\textbf {\bibinfo {volume} {103}},\ \bibinfo
  {pages} {063505} (\bibinfo {year} {2021})},\ \Eprint
  {http://arxiv.org/abs/2012.00046} {arXiv:2012.00046 [astro-ph.CO]}
  \BibitemShut {NoStop}%
\bibitem [{\citenamefont {Lazkoz}\ \emph {et~al.}(2006)\citenamefont {Lazkoz},
  \citenamefont {Maartens},\ and\ \citenamefont {Majerotto}}]{Lazkoz:2006gp}%
  \BibitemOpen
  \bibfield  {author} {\bibinfo {author} {\bibfnamefont {R.}~\bibnamefont
  {Lazkoz}}, \bibinfo {author} {\bibfnamefont {R.}~\bibnamefont {Maartens}}, \
  and\ \bibinfo {author} {\bibfnamefont {E.}~\bibnamefont {Majerotto}},\ }\href
  {\doibase 10.1103/PhysRevD.74.083510} {\bibfield  {journal} {\bibinfo
  {journal} {Phys. Rev. D}\ }\textbf {\bibinfo {volume} {74}},\ \bibinfo
  {pages} {083510} (\bibinfo {year} {2006})},\ \Eprint
  {http://arxiv.org/abs/astro-ph/0605701} {arXiv:astro-ph/0605701} \BibitemShut
  {NoStop}%
\bibitem [{\citenamefont {Olive}\ and\ \citenamefont
  {Steigman}(1995)}]{Olive:1994fe}%
  \BibitemOpen
  \bibfield  {author} {\bibinfo {author} {\bibfnamefont {K.~A.}\ \bibnamefont
  {Olive}}\ and\ \bibinfo {author} {\bibfnamefont {G.}~\bibnamefont
  {Steigman}},\ }\href {\doibase 10.1086/192134} {\bibfield  {journal}
  {\bibinfo  {journal} {Astrophys. J. Suppl.}\ }\textbf {\bibinfo {volume}
  {97}},\ \bibinfo {pages} {49} (\bibinfo {year} {1995})},\ \Eprint
  {http://arxiv.org/abs/astro-ph/9405022} {arXiv:astro-ph/9405022} \BibitemShut
  {NoStop}%
\bibitem [{\citenamefont {Izotov}\ and\ \citenamefont
  {Thuan}(1998)}]{Izotov:1998mj}%
  \BibitemOpen
  \bibfield  {author} {\bibinfo {author} {\bibfnamefont {Y.~I.}\ \bibnamefont
  {Izotov}}\ and\ \bibinfo {author} {\bibfnamefont {T.~X.}\ \bibnamefont
  {Thuan}},\ }\href {\doibase 10.1086/305698} {\bibfield  {journal} {\bibinfo
  {journal} {Astrophys. J.}\ }\textbf {\bibinfo {volume} {500}},\ \bibinfo
  {pages} {188} (\bibinfo {year} {1998})}\BibitemShut {NoStop}%
\bibitem [{\citenamefont {O'Meara}\ \emph {et~al.}(2001)\citenamefont
  {O'Meara}, \citenamefont {Tytler}, \citenamefont {Kirkman}, \citenamefont
  {Suzuki}, \citenamefont {Prochaska}, \citenamefont {Lubin},\ and\
  \citenamefont {Wolfe}}]{OMeara:2000tmq}%
  \BibitemOpen
  \bibfield  {author} {\bibinfo {author} {\bibfnamefont {J.~M.}\ \bibnamefont
  {O'Meara}}, \bibinfo {author} {\bibfnamefont {D.}~\bibnamefont {Tytler}},
  \bibinfo {author} {\bibfnamefont {D.}~\bibnamefont {Kirkman}}, \bibinfo
  {author} {\bibfnamefont {N.}~\bibnamefont {Suzuki}}, \bibinfo {author}
  {\bibfnamefont {J.~X.}\ \bibnamefont {Prochaska}}, \bibinfo {author}
  {\bibfnamefont {D.}~\bibnamefont {Lubin}}, \ and\ \bibinfo {author}
  {\bibfnamefont {A.~M.}\ \bibnamefont {Wolfe}},\ }\href {\doibase
  10.1086/320579} {\bibfield  {journal} {\bibinfo  {journal} {Astrophys. J.}\
  }\textbf {\bibinfo {volume} {552}},\ \bibinfo {pages} {718} (\bibinfo {year}
  {2001})},\ \Eprint {http://arxiv.org/abs/astro-ph/0011179}
  {arXiv:astro-ph/0011179} \BibitemShut {NoStop}%
\bibitem [{\citenamefont {Chen}\ \emph {et~al.}(2001)\citenamefont {Chen},
  \citenamefont {Scherrer},\ and\ \citenamefont {Steigman}}]{Chen:2000xxa}%
  \BibitemOpen
  \bibfield  {author} {\bibinfo {author} {\bibfnamefont {X.-l.}\ \bibnamefont
  {Chen}}, \bibinfo {author} {\bibfnamefont {R.~J.}\ \bibnamefont {Scherrer}},
  \ and\ \bibinfo {author} {\bibfnamefont {G.}~\bibnamefont {Steigman}},\
  }\href {\doibase 10.1103/PhysRevD.63.123504} {\bibfield  {journal} {\bibinfo
  {journal} {Phys. Rev. D}\ }\textbf {\bibinfo {volume} {63}},\ \bibinfo
  {pages} {123504} (\bibinfo {year} {2001})},\ \Eprint
  {http://arxiv.org/abs/astro-ph/0011531} {arXiv:astro-ph/0011531} \BibitemShut
  {NoStop}%
\bibitem [{\citenamefont {Amendola}\ \emph {et~al.}(2007)\citenamefont
  {Amendola}, \citenamefont {Gannouji}, \citenamefont {Polarski},\ and\
  \citenamefont {Tsujikawa}}]{Amendola:2006we}%
  \BibitemOpen
  \bibfield  {author} {\bibinfo {author} {\bibfnamefont {L.}~\bibnamefont
  {Amendola}}, \bibinfo {author} {\bibfnamefont {R.}~\bibnamefont {Gannouji}},
  \bibinfo {author} {\bibfnamefont {D.}~\bibnamefont {Polarski}}, \ and\
  \bibinfo {author} {\bibfnamefont {S.}~\bibnamefont {Tsujikawa}},\ }\href
  {\doibase 10.1103/PhysRevD.75.083504} {\bibfield  {journal} {\bibinfo
  {journal} {Phys. Rev. D}\ }\textbf {\bibinfo {volume} {75}},\ \bibinfo
  {pages} {083504} (\bibinfo {year} {2007})},\ \Eprint
  {http://arxiv.org/abs/gr-qc/0612180} {arXiv:gr-qc/0612180} \BibitemShut
  {NoStop}%
\bibitem [{\citenamefont {Frusciante}\ \emph {et~al.}(2014)\citenamefont
  {Frusciante}, \citenamefont {Raveri},\ and\ \citenamefont
  {Silvestri}}]{Frusciante:2013zop}%
  \BibitemOpen
  \bibfield  {author} {\bibinfo {author} {\bibfnamefont {N.}~\bibnamefont
  {Frusciante}}, \bibinfo {author} {\bibfnamefont {M.}~\bibnamefont {Raveri}},
  \ and\ \bibinfo {author} {\bibfnamefont {A.}~\bibnamefont {Silvestri}},\
  }\href {\doibase 10.1088/1475-7516/2014/02/026} {\bibfield  {journal}
  {\bibinfo  {journal} {JCAP}\ }\textbf {\bibinfo {volume} {1402}},\ \bibinfo
  {pages} {026} (\bibinfo {year} {2014})},\ \Eprint
  {http://arxiv.org/abs/1310.6026} {arXiv:1310.6026 [astro-ph.CO]} \BibitemShut
  {NoStop}%
\bibitem [{\citenamefont {Frusciante}\ \emph {et~al.}(2016)\citenamefont
  {Frusciante}, \citenamefont {Raveri}, \citenamefont {Vernieri}, \citenamefont
  {Hu},\ and\ \citenamefont {Silvestri}}]{Frusciante:2015maa}%
  \BibitemOpen
  \bibfield  {author} {\bibinfo {author} {\bibfnamefont {N.}~\bibnamefont
  {Frusciante}}, \bibinfo {author} {\bibfnamefont {M.}~\bibnamefont {Raveri}},
  \bibinfo {author} {\bibfnamefont {D.}~\bibnamefont {Vernieri}}, \bibinfo
  {author} {\bibfnamefont {B.}~\bibnamefont {Hu}}, \ and\ \bibinfo {author}
  {\bibfnamefont {A.}~\bibnamefont {Silvestri}},\ }\href {\doibase
  10.1016/j.dark.2016.03.002} {\bibfield  {journal} {\bibinfo  {journal} {Phys.
  Dark Univ.}\ }\textbf {\bibinfo {volume} {13}},\ \bibinfo {pages} {7}
  (\bibinfo {year} {2016})},\ \Eprint {http://arxiv.org/abs/1508.01787}
  {arXiv:1508.01787 [astro-ph.CO]} \BibitemShut {NoStop}%
\bibitem [{\citenamefont {Amendola}\ \emph {et~al.}(2008)\citenamefont
  {Amendola}, \citenamefont {Kunz},\ and\ \citenamefont
  {Sapone}}]{Amendola:2007rr}%
  \BibitemOpen
  \bibfield  {author} {\bibinfo {author} {\bibfnamefont {L.}~\bibnamefont
  {Amendola}}, \bibinfo {author} {\bibfnamefont {M.}~\bibnamefont {Kunz}}, \
  and\ \bibinfo {author} {\bibfnamefont {D.}~\bibnamefont {Sapone}},\ }\href
  {\doibase 10.1088/1475-7516/2008/04/013} {\bibfield  {journal} {\bibinfo
  {journal} {JCAP}\ }\textbf {\bibinfo {volume} {04}},\ \bibinfo {pages} {013}
  (\bibinfo {year} {2008})},\ \Eprint {http://arxiv.org/abs/0704.2421}
  {arXiv:0704.2421 [astro-ph]} \BibitemShut {NoStop}%
\bibitem [{\citenamefont {Bean}\ and\ \citenamefont
  {Tangmatitham}(2010)}]{PhysRevD.81.083534}%
  \BibitemOpen
  \bibfield  {author} {\bibinfo {author} {\bibfnamefont {R.}~\bibnamefont
  {Bean}}\ and\ \bibinfo {author} {\bibfnamefont {M.}~\bibnamefont
  {Tangmatitham}},\ }\href {\doibase 10.1103/PhysRevD.81.083534} {\bibfield
  {journal} {\bibinfo  {journal} {Phys. Rev. D}\ }\textbf {\bibinfo {volume}
  {81}},\ \bibinfo {pages} {083534} (\bibinfo {year} {2010})}\BibitemShut
  {NoStop}%
\bibitem [{\citenamefont {Silvestri}\ \emph {et~al.}(2013)\citenamefont
  {Silvestri}, \citenamefont {Pogosian},\ and\ \citenamefont
  {Buniy}}]{Silvestri:2013ne}%
  \BibitemOpen
  \bibfield  {author} {\bibinfo {author} {\bibfnamefont {A.}~\bibnamefont
  {Silvestri}}, \bibinfo {author} {\bibfnamefont {L.}~\bibnamefont {Pogosian}},
  \ and\ \bibinfo {author} {\bibfnamefont {R.~V.}\ \bibnamefont {Buniy}},\
  }\href {\doibase 10.1103/PhysRevD.87.104015} {\bibfield  {journal} {\bibinfo
  {journal} {Phys. Rev. D}\ }\textbf {\bibinfo {volume} {87}},\ \bibinfo
  {pages} {104015} (\bibinfo {year} {2013})},\ \Eprint
  {http://arxiv.org/abs/1302.1193} {arXiv:1302.1193 [astro-ph.CO]} \BibitemShut
  {NoStop}%
\bibitem [{\citenamefont {{Pogosian}}\ \emph {et~al.}(2010)\citenamefont
  {{Pogosian}}, \citenamefont {{Silvestri}}, \citenamefont {{Koyama}},\ and\
  \citenamefont {{Zhao}}}]{2010PhRvD..81j4023P}%
  \BibitemOpen
  \bibfield  {author} {\bibinfo {author} {\bibfnamefont {L.}~\bibnamefont
  {{Pogosian}}}, \bibinfo {author} {\bibfnamefont {A.}~\bibnamefont
  {{Silvestri}}}, \bibinfo {author} {\bibfnamefont {K.}~\bibnamefont
  {{Koyama}}}, \ and\ \bibinfo {author} {\bibfnamefont {G.-B.}\ \bibnamefont
  {{Zhao}}},\ }\href {\doibase 10.1103/PhysRevD.81.104023} {\bibfield
  {journal} {\bibinfo  {journal} {Phys. Rev. D}\ }\textbf {\bibinfo {volume}
  {81}},\ \bibinfo {eid} {104023} (\bibinfo {year} {2010})},\ \Eprint
  {http://arxiv.org/abs/1002.2382} {arXiv:1002.2382 [astro-ph.CO]} \BibitemShut
  {NoStop}%
\bibitem [{\citenamefont {Amendola}\ \emph {et~al.}(2020)\citenamefont
  {Amendola}, \citenamefont {Bettoni}, \citenamefont {Pinho},\ and\
  \citenamefont {Casas}}]{Amendola:2019laa}%
  \BibitemOpen
  \bibfield  {author} {\bibinfo {author} {\bibfnamefont {L.}~\bibnamefont
  {Amendola}}, \bibinfo {author} {\bibfnamefont {D.}~\bibnamefont {Bettoni}},
  \bibinfo {author} {\bibfnamefont {A.~M.}\ \bibnamefont {Pinho}}, \ and\
  \bibinfo {author} {\bibfnamefont {S.}~\bibnamefont {Casas}},\ }\href
  {\doibase 10.3390/universe6020020} {\bibfield  {journal} {\bibinfo  {journal}
  {Universe}\ }\textbf {\bibinfo {volume} {6}},\ \bibinfo {pages} {20}
  (\bibinfo {year} {2020})},\ \Eprint {http://arxiv.org/abs/1902.06978}
  {arXiv:1902.06978 [astro-ph.CO]} \BibitemShut {NoStop}%
\bibitem [{\citenamefont {Beltr\'an~Jim\'enez}\ \emph
  {et~al.}(2020{\natexlab{b}})\citenamefont {Beltr\'an~Jim\'enez},
  \citenamefont {Heisenberg}, \citenamefont {Koivisto},\ and\ \citenamefont
  {Pekar}}]{Jimenez:2019ovq}%
  \BibitemOpen
  \bibfield  {author} {\bibinfo {author} {\bibfnamefont {J.}~\bibnamefont
  {Beltr\'an~Jim\'enez}}, \bibinfo {author} {\bibfnamefont {L.}~\bibnamefont
  {Heisenberg}}, \bibinfo {author} {\bibfnamefont {T.~S.}\ \bibnamefont
  {Koivisto}}, \ and\ \bibinfo {author} {\bibfnamefont {S.}~\bibnamefont
  {Pekar}},\ }\href {\doibase 10.1103/PhysRevD.101.103507} {\bibfield
  {journal} {\bibinfo  {journal} {Phys. Rev. D}\ }\textbf {\bibinfo {volume}
  {101}},\ \bibinfo {pages} {103507} (\bibinfo {year} {2020}{\natexlab{b}})},\
  \Eprint {http://arxiv.org/abs/1906.10027} {arXiv:1906.10027 [gr-qc]}
  \BibitemShut {NoStop}%
\bibitem [{\citenamefont {Sawicki}\ and\ \citenamefont
  {Bellini}(2015)}]{Sawicki:2015zya}%
  \BibitemOpen
  \bibfield  {author} {\bibinfo {author} {\bibfnamefont {I.}~\bibnamefont
  {Sawicki}}\ and\ \bibinfo {author} {\bibfnamefont {E.}~\bibnamefont
  {Bellini}},\ }\href {\doibase 10.1103/PhysRevD.92.084061} {\bibfield
  {journal} {\bibinfo  {journal} {Phys. Rev.}\ }\textbf {\bibinfo {volume}
  {D92}},\ \bibinfo {pages} {084061} (\bibinfo {year} {2015})},\ \Eprint
  {http://arxiv.org/abs/1503.06831} {arXiv:1503.06831 [astro-ph.CO]}
  \BibitemShut {NoStop}%
\bibitem [{\citenamefont {Aghanim}\ \emph {et~al.}(2020)\citenamefont {Aghanim}
  \emph {et~al.}}]{Planck:2018vyg}%
  \BibitemOpen
  \bibfield  {author} {\bibinfo {author} {\bibfnamefont {N.}~\bibnamefont
  {Aghanim}} \emph {et~al.} (\bibinfo {collaboration} {Planck}),\ }\href
  {\doibase 10.1051/0004-6361/201833910} {\bibfield  {journal} {\bibinfo
  {journal} {Astron. Astrophys.}\ }\textbf {\bibinfo {volume} {641}},\ \bibinfo
  {pages} {A6} (\bibinfo {year} {2020})},\ \bibinfo {note} {[Erratum:
  Astron.Astrophys. 652, C4 (2021)]},\ \Eprint
  {http://arxiv.org/abs/1807.06209} {arXiv:1807.06209 [astro-ph.CO]}
  \BibitemShut {NoStop}%
\bibitem [{\citenamefont {Peirone}\ \emph {et~al.}(2019)\citenamefont
  {Peirone}, \citenamefont {Benevento}, \citenamefont {Frusciante},\ and\
  \citenamefont {Tsujikawa}}]{Peirone:2019aua}%
  \BibitemOpen
  \bibfield  {author} {\bibinfo {author} {\bibfnamefont {S.}~\bibnamefont
  {Peirone}}, \bibinfo {author} {\bibfnamefont {G.}~\bibnamefont {Benevento}},
  \bibinfo {author} {\bibfnamefont {N.}~\bibnamefont {Frusciante}}, \ and\
  \bibinfo {author} {\bibfnamefont {S.}~\bibnamefont {Tsujikawa}},\ }\href
  {\doibase 10.1103/PhysRevD.100.063540} {\bibfield  {journal} {\bibinfo
  {journal} {Phys. Rev. D}\ }\textbf {\bibinfo {volume} {100}},\ \bibinfo
  {pages} {063540} (\bibinfo {year} {2019})},\ \Eprint
  {http://arxiv.org/abs/1905.05166} {arXiv:1905.05166 [astro-ph.CO]}
  \BibitemShut {NoStop}%
\bibitem [{\citenamefont {Frusciante}\ \emph {et~al.}(2020)\citenamefont
  {Frusciante}, \citenamefont {Peirone}, \citenamefont {Atayde},\ and\
  \citenamefont {De~Felice}}]{Frusciante:2019puu}%
  \BibitemOpen
  \bibfield  {author} {\bibinfo {author} {\bibfnamefont {N.}~\bibnamefont
  {Frusciante}}, \bibinfo {author} {\bibfnamefont {S.}~\bibnamefont {Peirone}},
  \bibinfo {author} {\bibfnamefont {L.}~\bibnamefont {Atayde}}, \ and\ \bibinfo
  {author} {\bibfnamefont {A.}~\bibnamefont {De~Felice}},\ }\href {\doibase
  10.1103/PhysRevD.101.064001} {\bibfield  {journal} {\bibinfo  {journal}
  {Phys. Rev. D}\ }\textbf {\bibinfo {volume} {101}},\ \bibinfo {pages}
  {064001} (\bibinfo {year} {2020})},\ \Eprint
  {http://arxiv.org/abs/1912.07586} {arXiv:1912.07586 [astro-ph.CO]}
  \BibitemShut {NoStop}%
\end{thebibliography}%

\end{document}